\documentclass{article}
\usepackage{geometry}
\usepackage[utf8]{inputenc}
\usepackage{authblk}
\usepackage{graphics}
\usepackage{bm}
\usepackage{amsmath}
\usepackage{extarrows}
\usepackage{amssymb}
\usepackage{enumerate}
\usepackage{multirow}
\usepackage{bbm}
\usepackage{framed}  
\usepackage{scalerel}
\usepackage{algorithm}
\usepackage{algorithmic}
\usepackage{color}
\usepackage{booktabs}
\usepackage{chngcntr}
\counterwithin*{section}{part}
\usepackage{hyperref}

\newcommand{\mr}{\mathrm} 
\newcommand{\bb}{\mathbbm}

\newcommand{\BE}{\begin{equation}}
\newcommand{\EE}{\end{equation}}
\newcommand{\BS}{\begin{subequations}}
\newcommand{\ES}{\end{subequations}}
\renewcommand{\bf}{\bm}
\newtheorem{theorem}{Theorem}
\newtheorem{proposition}{Proposition}

\newtheorem{definition}{Definition}

\newtheorem{lemma}{Lemma}

\title{Generalized Memory Approximate Message Passing}
\author[1]{Feiyan Tian}
\author[$*$2]{Lei Liu}
\author[1]{Xiaoming Chen\thanks{Corresponding authors: leiliu@jaist.ac.jp,  chen\_xiaoming@zju.edu.cn}}

\affil[1]{\small College of Information Science and Electronic Engineering, Zhejiang University, Hangzhou 310027, China.}
\affil[2]{\small School of Information Science, Japan Institute of Science and Technology (JAIST), Nomi 923-1292, Japan.}

\date{}
\begin{document}

\maketitle

\begin{abstract}
Generalized approximate message passing (GAMP) is a promising technique for unknown signal reconstruction of generalized linear models (GLM). However, it requires that the transformation matrix has independent and identically distributed (IID) entries. In this context, generalized vector AMP (GVAMP) is proposed for general unitarily-invariant transformation matrices but it has a high-complexity matrix inverse. To this end, we propose a universal generalized memory AMP (GMAMP) framework including the existing orthogonal AMP/VAMP, GVAMP, and memory AMP (MAMP) as special instances. Due to the characteristics that local processors are all memory, GMAMP requires stricter orthogonality to guarantee the asymptotic IID Gaussianity and state evolution. To satisfy such orthogonality, local orthogonal memory estimators are established. The GMAMP framework provides a principle toward building new advanced AMP-type algorithms. As an example, we construct a Bayes-optimal GMAMP (BO-GMAMP), which uses a low-complexity memory linear estimator to suppress the linear interference, and thus its complexity is comparable to GAMP. Furthermore, we prove that for unitarily-invariant transformation matrices, BO-GMAMP achieves the replica minimum (i.e., Bayes-optimal) MSE if it has a unique fixed point.

This paper is divided into two parts that can be read independently: The first part (main text) presents the universal framework of GMAMP, discusses the construction of BO-GMAMP and its properties. The second part (supplementary information) provides the additional explanatory for our main text including extended technical description of results, full details of mathematical models as well as all the proofs. The source code of this work is publicly available at \href{https://sites.google.com/site/leihomepage/research}{sites.google.com/site/leihomepage/research}.
\end{abstract}

\tableofcontents

\subsection*{Notations}
Bold upper (lower) letters denote matrices (column vectors). $(\cdot)^{\rm T}$ and $(\cdot)^{\rm H}$ denote transpose and conjugate transpose, respectively. ${\rm E}\{\cdot\}$ denotes expectation. $\mathcal{CN}(\bf{\mu},\bf{\Sigma})$ denotes the complex Gaussian distribution of a vector with mean vector $\bf{\mu}$ and covariance matrix $\bf{\Sigma}$. $|\!\langle\bf{A}_{M\times N}, \bf{B}_{M\times N} \rangle\!| \equiv \frac{1}{N}\bf{A}^{\rm H}_{M\times N}\bf{B}_{M\times N}$, $\|\cdot\|$ denotes the $\ell_2$-norm. ${\rm diag}\{\cdot\}$, ${\rm det}\{\cdot\}$ and ${\rm tr}(\cdot)$ respectively denote the diagonal vector, the determinant and the trace of a matrix.  ${\rm min}\{\cdot\}$ and ${\rm max}\{\cdot\}$ denotes the minimum and maximum value of a set. $\bf{I}$ and $\bf{0}$ are identity matrix and zero matrix or vector. $X\sim Y$ represents that $X$ follows the distribution $Y$, $\overset{\rm a.s.}{=}$ denotes almost sure equivalence. $\bf{A}$ matrix is said column-wise IIDG and row-wise joint-Gaussian (CIIDG-RJG) if its each column is IIDG and its each row is joint Gaussian.

 \newpage
\part{Main text}
\section{Introduction}
In signal processing fields, many applications can be formulated as unknown signal reconstruction problems of standard linear models (SLM), i.e., $\bf{y}=\bf{Ax}+\bf{n}$, where $\bf{y}$ is the measurement vector, $\bf{A}$ is the transformation matrix, $\bf{x}$ is the unknown signal vector, and $\bf{n}$ is an additive Gaussian noise vector. For solving such problems, approximate message passing (AMP) is a high-efficient approach by employing a low-complexity matched filter (MF) \cite{Donoho2009,Bayati2011} to suppress the linear interference. More importantly, it has been proved that AMP is Bayes-optimal \cite{Reeves2019,Barbier2020} through the asymptotic behavior analysis of AMP described by a simple set of state evolution (SE) equations \cite{Bayati2011}.The information-theoretical (i.e., capacity) optimality of AMP was proved in \cite{lei2021AMP}. However, AMP is viable only when the transformation matrix $\bf{A}$ comprises independent and identically distributed Gaussian (IIDG) entries \cite{Rangan2017,Vila2015}. As a result, AMP has limited application scenarios.

In this context, a variant of AMP was discovered in \cite{UTAMPa, UTAMPb} based on a unitary transformation, called UTAMP. It performs well for difficult (e.g. correlated) matrices $\bf{A}$. Independently, orthogonal/vector AMP (OAMP/VAMP) was proposed for SLM problems with general transformation matrices \cite{Ma2017,Rangan2019}. The SE for OAMP/VAMP was conjectured for LUIS in \cite{Ma2017} and rigorously proved in \cite{Rangan2019,Takeuchi2020}. Specifically, OAMP/VAMP has a two-estimator architecture, i.e., a linear estimator (LE) and a non-linear estimator (NLE). Via iterations between the two estimators, OAMP/VAMP can achieve the Bayes optimality when the compression rate of system is larger than a certain value \cite{Ma2017, Tulino2013, Barbier2018,Takeda2006}. The information-theoretical (i.e., capacity) optimality of OAMP/VAMP was proved in \cite{lei2021OAMP}. To guarantee the local optimality of LE, OMAP/VAMP usually adopts a linear minimum mean square error (LMMSE) estimator, resulting in high computational complexity, especially in large-scale systems. To avoid the high-complexity LMMSE estimator in LE, the singular-value decomposition (SVD) estimator was utilized in \cite{Rangan2019}. Yet, the complexity of SVD is not much lower than LMMSE. In \cite{Ma2017}, the LMMSE estimator is replaced by a simple MF, but the performance of OAMP/VAMP deteriorated sharply. 
Additionally, a low-complexity convolutional AMP (CAMP) approach was proposed for unitarily invariant transformation matrix  \cite{Takeuchi2021}. However, the CAMP has a relatively low convergence speed and even fails to converge, particularly for matrices with high condition numbers. Similarly, an AMP  was constructed in \cite{Opper2016} to solve the Thouless-Anderson-Palmer equations for Ising models with invariant random matrices. The results in \cite{Opper2016} were rigorously justified via SE in \cite{Fan2020arxiv}. To this end, a memory AMP (MAMP) was recently designed by replacing the LMMSE estimator with an MF in the presence of memory, where the introduced memory can assist the MF in converging the locally optimal LMMSE \cite{lei2020mamp,lei2021mamp}. Hence, MAMP can achieve the Bayes optimality with low complexity and unitarily invariant transformation matrix \cite{lei2020mamp,lei2021mamp}.

On the other hand, SLM is not so accurate in practical applications. For instance, in wireless communications, the received signal may suffer a nonlinear impact from the receiver. Consequently, the received signal is given by $\bf{y}=Q(\bf{Ax})$, which does not follow the SLM as $Q(\cdot)$ may be a non-linear function. Indeed, it is a generalized linear model (GLM) \cite{Rangan2010}. To alleviate the nonlinear impact, a generalized AMP (GAMP) was proposed by introducing a maximum posterior estimator \cite{Rangan2010}. Similar to AMP, GAMP has low complexity, but is only limited to IIDG transformation matrices. To address the limitation of GAMP, the generalized VAMP (GVAMP) was applied for GLM with general unitarily-invariant transformation matrix \cite{Schniter2016}. The SE  and Bayes optimality (predicted by Replica method) of GVAMP were proved in \cite{pandit2020}. Similar to OAMP/VAMP, GVAMP has a high computational complexity. In this context, it is desired to design a low-complexity, high-accurate, and wide-applicable AMP framework.

In this paper, we propose a novel generalized MAMP (GMAMP) framework to solve the GLM problem in the presence of memory terms. The proposed GMAMP framework is universal that includes the existing OAMP/VAMP, GVAMP and MAMP as special instances. The asymptotic IID Gaussianity of GMAMP is guaranteed under a new orthogonality principle, i.e., the current output estimation error of each local estimator is orthogonal to all the memory input estimation errors. To satisfy such orthogonality, we establish two kinds of orthogonal estimators, i.e., orthogonal memory LE (MLE) and orthogonal memory NLE (MNLE). The GMAMP framework provides new principle toward building new advanced AMP-type algorithms. As an example, we construct a Bayes-optimal GMAMP (BO-GMAMP). We prove that for unitarily-invariant transformation matrices, BO-GMAMP achieves the minimum (i.e., Bayes-optimal) MSE as predicted by the replica method if it has a unique fixed point. Hence, BO-GMAMP overcomes the IID transformation matrix limitation of GAMP. In addition, BO-GMAMP uses a low-complexity MLE to suppress the linear interference, and thus its complexity is comparable to GAMP. That is, BO-GMAMP avoids the high-complexity LMMSE in GVAMP. Finally, simulation results are provided to validate the accuracy of the theoretical analysis.

Simply, the key properties of these AMP-type algorithms including GAMP, GVAMP and the expected GMAMP are summarized in Table \ref{table:comp}.
\begin{table}[t]
\centering
\begin{tabular}{cccc}\label{table:comp}
Algorithms & Transform matrices & Complexity & Optimality \\ 
\midrule
GAMP & IID & Low & Bayes-optimal\\
GVAMP & General & High & Bayes-optimal\\
GMAMP(proposed) & General & Low & Bayes-optimal\\
\bottomrule
\end{tabular}
\caption{\footnotesize Comparison of GAMP, GVAMP and GMAMP}
\end{table}

\section{Preliminaries}

\subsection{Problem Model}

\begin{figure}[b]
\centering 
\includegraphics[scale=0.45]{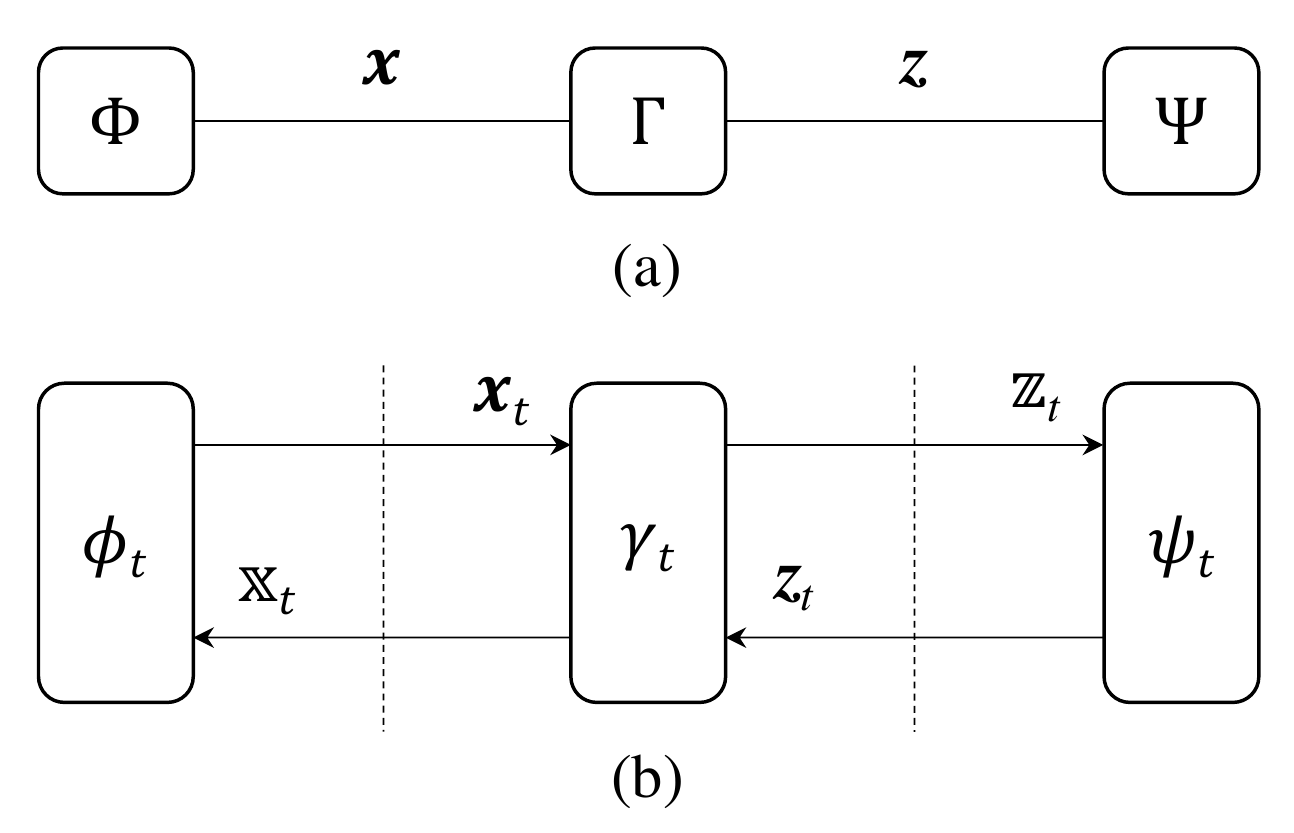}
\caption{\footnotesize Graphic illustrations for (a) a system model with three constraints $\Phi$, $\Gamma$ and $\Psi$, and (b) a non-memory iterative process (NMIP) involving three processors $\phi_t$, $\gamma_t$ and $\psi_t$ respectively corresponding to the constraints $\Phi$, $\Gamma$ and $\Psi$ in (a).}
\label{fig:system diagram0}
\end{figure}

Fig. \ref{fig:system diagram0}(a) illustrates an generalized linear model (GLM) consisted of three constraints:
\BS\label{Eqn:problem}
\begin{align} 
\Psi:\quad &\bf{y}=Q(\bf{z}),\\
\Gamma:\quad &\bf{z}=\bf{A}\bf{x},\\
\Phi:\quad &{x}_i \sim P_X(x),
\end{align} 
\ES
where $\bf{A}$ is an $M \times N$ matrix, $\bf{y}$ is an $M \times 1$ vector, $\bf{x}=\{x_n\}$ is an $N \times 1$ vector with independent and identically distributed (IID) entries, i.e., $x_n\sim P_X(x)$, and $Q(\cdot)$ is a symbol-by-symbol function (can be non-linear). It is assumed that the receiver knows $\bf{y}$, $\bf{A}$, $Q(\cdot)$, and the distributions of $\bf{x}$.

Our aim is to utilize the AMP-type iterative approach to find an MMSE estimation of $\bf{x}$, i.e., its MSE converges to
\BE
\label{Eqn:post_mean}
{\rm mmse}\{\bf{x}|\bf{y}, \bf{A}, \Psi, {\Gamma}, {\Phi}\} \equiv \tfrac{1}{N}{\mathrm{E}}\{\|\hat{\bf{x}}-{\bf{x}}\|^2\},
\EE
where $\hat{\bf{x}} \equiv {\mathrm{E}}\{\bf{x}|\bf{y}, \bf{A}, \Psi, {\Gamma}, {\Phi} \}$ is the \emph{a-posteriori} mean of $\bf{x}$.

\begin{definition}[Bayes Optimality]
An iterative approach is said to be Bayes optimal if its MSE converges to the MMSE of the system in \eqref{Eqn:problem}.
\end{definition}

\subsection{Assumptions}\label{Sec:Ass}

Let the singular value decomposition of $\bm{A}$ be $\bf{A}\!=\!\bf{U}\bf{\Sigma} \bf{V}^{\rm H}$, where $\bf{U}\!\in\! \mathbb{C}^{M\times M}$ and $\bf{V}\!\in\! \mathbb{C}^{N\times N}$ are unitary matrices, and $\bf{\Sigma} $ is a rectangular diagonal matrix. We assume that $\bf{A}$  is unitarily-invariant, i.e., $\bf{U}$, $\bf{V}$ and $\bf{\Sigma}$ are independent, and $\bf{U}$ and $\bf{V}$ are Haar distributed (or equivalently, isotropically random). Let $\lambda_t=\tfrac{1}{M}{\rm E}\{(\bf{A}\bf{A}^{\rm H})^t\}$ and ${\lambda}^\dag\equiv[ \lambda_{\max}+ \lambda_{\min}]/2$, where $\lambda_{\min}$ and $\lambda_{\max}$ denote the minimal and maximal eigenvalues of $\bf{A}\bf{A}^{\rm H}$, respectively. Without loss of generality, supposing that $\{\lambda_{\min}, \lambda_{\max}, \lambda_t\}$ are known. For specific random matrices such as IID Gaussian matrices, Wigner matrices and Wishart matrices, $\{\lambda_{\min}, \lambda_{\max}, \lambda_t\}$ is available \cite{Tulino2004}. Otherwise, some approximations of $\{\lambda_{\min}, \lambda_{\max}, \lambda_t\}$ in \cite{lei2020mamp} can be used if it is unavailable. 

\subsection{Overview of GVAMP} 

Fig. \ref{fig:system diagram0}(b) illustrates a \emph{non-memory iterative process (NMIP)}: Starting with $t=1$, 
\BS\label{Eqn:NMIP0}\begin{align}
     {\rm NLE:}\quad &\left[ \!\!\begin{array}{c}
         \bf{x}_{t}   \\
        \bf{z}_{t}
    \end{array} \!\!\right]   =\left[ \!\!\begin{array}{c}
        \phi_t(\bf{\bb{x}}_t)   \\
        \psi_t(\bf{\bb{z}}_t)
    \end{array} \!\!\right],\\
 {\rm LE:}\quad  & \left[ \!\!\begin{array}{c}
         \bf{\bb x}_{t + 1}   \\
        \bf{\bb z}_{t + 1}
    \end{array} \!\!\right]  = \gamma_t \big(\bf{x}_t, \bf{z}_t\big)  
    =\left[ \!\!\begin{array}{c}
        \gamma_t^{x} (\bf{x}_t, \bf{z}_t)  \\
        \gamma_t^{z} (\bf{x}_t, \bf{z}_t)
    \end{array} \!\!\right], 
\end{align}\ES
where $\gamma_t(\cdot)$, $\phi_t(\cdot)$ and $\psi_t(\cdot)$ process the  three  constraints $\Gamma$, $\Phi$ and $\Psi$ separately. We call \eqref{Eqn:NMIP0} NMIP since both $\gamma_t(\cdot)$ $\phi_t(\cdot)$ and $\psi_t(\cdot)$ are memoryless, depending only on their current inputs $(\bf{x}_t, \bf{z}_t)$, $\bf{\bb{x}}_t$ and $\bf{\bb{z}}_t$, respectively.

GVAMP \cite{Schniter2016} is an instance of NMIP. The derivation of GVAMP and its properties including orthogonality, asymptotic IID Gaussianity, Bayes optimality and state evolution are briefly introduced in  \textcolor{blue}{SI Appendix, section 1}.

\section{Generalized Memory AMP Framework}

In this part, we first introduce the memory iterative process (MIP) and the orthogonality for MIP. Then, a universal framework is constructed for low-complexity generalized memory AMP (GMAMP) using arbitrary-length memory. It is worth pointing out that this universal framework is also applied to  OAMP/VAMP, MAMP and GVAMP, i.e., they are instances of this universal framework.

\subsection{Memory Iterative Process and Orthogonality}

\begin{figure}[b]
\centering
\includegraphics[scale=0.28]{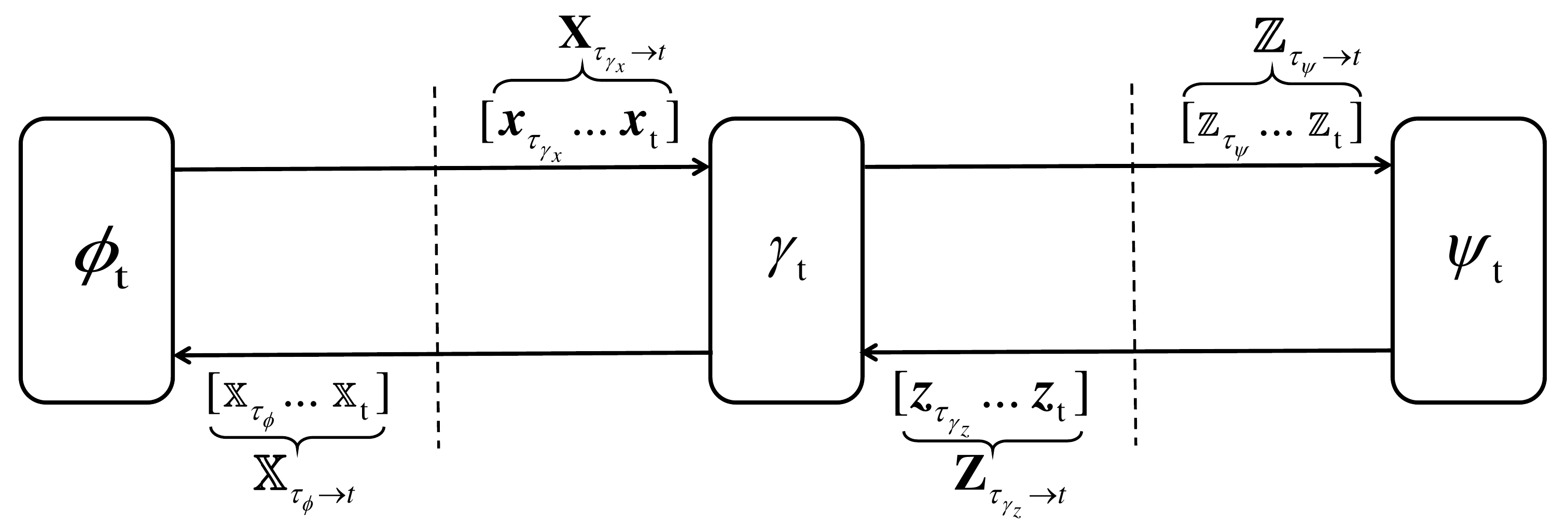}
\caption{\footnotesize Graphic illustration for a memory iterative process (MIP).}
\label{fig:system diagram2}
\end{figure}

\emph{Memory  Iterative Process (MIP):} Fig. \ref{fig:system diagram2} illustrates an MIP based on two memory non-linear estimators (MNLEs) and a memory linear estimator (MLE), defined as:  Starting with $t=1$, $   \bf{\bb{X}}_1 =\bf{0}$ and $\bf{\bb{Z}}_1 =\bf{0}$,
\BS\label{Eqn:MIP}\begin{alignat}{2}
{\rm NLE:}\qquad \left[ \!\!\begin{array}{c}
         \bf{x}_{t}   \\
        \bf{z}_{t}
    \end{array} \!\!\right]  & =\left[ \!\!\begin{array}{c}
        \phi_t(\bf{\bb{X}}_{\tau_{\phi}\rightarrow t})   \\
        \psi_t(\bf{\bb{Z}}_{\tau_{\psi}\rightarrow t})
    \end{array} \!\!\right],\\
{\rm MLE:} \quad  \left[ \!\!\begin{array}{c}
         \bf{\bb x}_{t + 1}   \\
        \bf{\bb z}_{t + 1}
    \end{array} \!\!\right] & = \gamma_t \big(\bf{X}_{\tau_{\gamma_x}\rightarrow t}, \bf{Z}_{\tau_{\gamma_z}\rightarrow t}\big) \nonumber\\ 
    &=\left[ \!\!\begin{array}{c}
        \gamma_t^{x} (\bf{X}_{\tau_{\gamma_x}\rightarrow t}, \bf{Z}_{\tau_{\gamma_z}\rightarrow t})  \\
        \gamma_t^{z} (\bf{X}_{\tau_{\gamma_x}\rightarrow t}, \bf{Z}_{\tau_{\gamma_z}\rightarrow t})
    \end{array} \!\!\right], 
\end{alignat}\ES	  
where $\bf{\bb{X}}_{\tau_{\phi}\rightarrow t}\!=\![\bf{\bb{x}}_{\tau_{\phi}} \cdots \bf{\bb{x}}_t]$, $\bf{\bb{Z}}_{\tau_{\psi}\rightarrow t}\!=\![\bf{\bb{z}}_{\tau_{\psi}} \cdots \bf{\bb{z}}_t]$, $\bf{X}_{\tau_{\gamma_x}\rightarrow t}\!=\![\bf{x}_{\tau_{\gamma_x}} \cdots \bf{x}_t]$, $\bf{Z}_{\tau_{\gamma_z}\rightarrow t}\!=\![\bf{z}_{\tau_{\gamma_z}}\cdots\bf{z}_t]$.  
$\tau_{\phi}$, $\tau_{\psi}$, $\tau_{\gamma_x}$ and $\tau_{\gamma_z}$ belong to $[1,t]$. $\gamma_t(\cdot)$, $\phi_t(\cdot)$ and $\psi_t(\cdot)$ process the  three  constraints $\Gamma$, $\Phi$ and $\Psi$ separately. Furthermore, we assume that $\phi_t(\cdot)$, $\psi_t(\cdot)$ are separable and Lipschitz-continuous functions \cite{Berthier2017}. Let
\BS\begin{align}
    \bf{\bb{x}}_t&=\bf{x}+\bf{\bb{f}}_t,&\bf{\bb{z}}_t&=\bf{z}+\bf{\bb{s}}_t,\\
    \bf{x}_t&=\bf{x}+\bf{f}_t,
    &
    \bf{z}_t&=\bf{z}+\bf{s}_t,
\end{align}\ES
where $\bf{\bb{f}}_t$, $\bf{f}_t$, $\bf{\bb{s}}_t$ and $\bf{s}_t$ indicate the estimation errors with zero mean and covariances: for $1\leq t'\leq t$,
\BS\label{err_defi}\begin{align}
    &\bb{v}^{x}_{t,t'}\equiv |\! \langle\bf{\bb{f}}_t, \bf{\bb{f}}_{t'} \rangle\!| ,&\bb{v}^{z}_{t,t'}\equiv |\! \langle \bf{\bb{s}}_t,  \bf{\bb{s}}_{t'}\rangle\!|,\\
    &v^x_{t,t'}\equiv |\! \langle\bf{f}_t, \bf{f}_{t'} \rangle\!|, &
    v^z_{t,t'}\equiv |\! \langle  \bf{s}_t,  \bf{s}_{t'}\rangle\!|.
\end{align}\ES

We call \eqref{Eqn:MIP} MIP since each local processor ($\gamma_t$, $\phi_t$ or $\psi_t$) is not only a function of the current input, but also some memories generated in previous iterations. Intuitively, MIP will degenerate into NMIP if $\{\tau_{\phi},\tau_{\psi},\tau_{\gamma_x},\tau_{\gamma_z}\}=t$.

\begin{definition}[Generalized memory AMP]
Generalized memory AMP (GMAMP) is a particular case of MIP when the following orthogonal constraints hold for $t\geq 1$:
\BS\label{Eqn:orth_GMAMP}\begin{align} 
    |\! \langle  \bf{\bb{f}}_t,  \bf{x} \rangle\!| & \overset{\rm a.s.}{=}0, &
    |\! \langle \bf{s}_t, \bf{z} \rangle\!|  &\overset{\rm a.s.}{=}0,\label{Eqn:orth_a}\\
    |\! \langle {\bf{f}_t}, \bf{\bb{F}}_{\tau_{\phi}\rightarrow t} \rangle\!|  &\overset{\rm a.s.}{=}\bf{0},&
    |\! \langle  \bf{s}_t,\bf{\bb{S}}_{\tau_{\psi}\rightarrow t}\rangle\!| &\overset{\rm a.s.}{=}\bf{0},\label{Eqn:orth_b}\\
    |\! \langle \bf{\bb{f}}_{t+1}, \bf{F}_{\tau_{\gamma_x}\rightarrow t}\rangle\!|   &\overset{\rm a.s.}{=}\bf{0},&
    |\! \langle \bf{\bb{s}}_{t+1}, \bf{S}_{\tau_{\gamma_z}\rightarrow t}\rangle\!|    &\overset{\rm a.s.}{=}\bf{0},\label{Eqn:orth_c}
\end{align}\ES
where $\bf{\bb{F}}_{\tau_{\phi}\rightarrow t}\!=\![\bf{\bb{f}}_{\tau_{\phi}}\cdots \bf{\bb{f}}_t],\bf{\bb{S}}_{\tau_{\psi}\rightarrow t}\!=\![\bf{\bb{s}}_{\tau_{\psi}}\cdots \bf{\bb{s}}_t],\bf{F}_{\tau_{\gamma_x}\rightarrow t}\!=\![\bf{f}_{\tau_{\gamma_x}}\cdots \bf{f}_t],\bf{S}_{\tau_{\gamma_z}\rightarrow t}\!=\![\bf{s}_{\tau_{\gamma_z}}\cdots \bf{s}_t]$. 
Specifically,  \eqref{Eqn:orth_a} shows that  $\bf{x}$ is orthogonal to the input estimation error of $\Phi_t$ and  $\bf{z}$ is orthogonal to the output estimation error of $\Psi_t$, and \eqref{Eqn:orth_b} and \eqref{Eqn:orth_c} show that current output estimation error of each local processor (NLE or MLE) is orthogonal to all input estimation errors, i.e., $\bf{f}_t, \bf{s}_t$, $\bf{\bb{f}}_{t+1}$ and $\bf{\bb{s}}_{t+1}$ are orthogonal to $\bf{\bb{F}}_{\tau_{\phi}\rightarrow t}$, $\bf{\bb{S}}_{\tau_{\psi}\rightarrow t}$,    $\bf{F}_{\tau_{\gamma_x}\rightarrow t}$ and  $\bf{S}_{\tau_{\gamma_z}\rightarrow t}$, respectively.
\end{definition}

The orthogonality in \eqref{Eqn:orth_a} is used to simplify the designs of $\phi_t(\cdot)$, $\psi_t(\cdot)$ and $\gamma_t(\cdot)$ in GMAMP. Given arbitrary $\hat{\bf{\bb{x}}}_t$ and $\hat{ \bf{z}}_t$, to satisfy \eqref{Eqn:orth_a}, we can construct   
\BS\label{norm_orth}\BE
    \bf{\bb{x}}_t = \bb{c}^x_t \hat{\bf{\bb{x}}}_t, \qquad   \bf{z}_t = c^z_t \hat{ \bf{z}}_t,
\EE
with
\BE
   \bb{c}^x_t  =  1/ |\! \langle \hat{\bf{\bb{x}}}_t , \bf{x}\rangle\!| , \qquad  c^z_t =  1/|\! \langle \hat{\bf{z}}_t , \bf{z}\rangle\!|.
\EE\ES 
Furthermore, to further improve the MSE performance of GMAMP, the following orthogonality is required for estimations $\bf{x}_t$ and $\bf{\bb{z}}_t$.
\BE
     |\! \langle   \bf{f}_t,   \bf{x}_t  \rangle\!| \overset{\rm a.s.}{=}0, \qquad \qquad
    |\! \langle   \bf{\bb{s}}_t,   \bf{\bb{z}}_t \rangle\!|  \overset{\rm a.s.}{=}0,\label{Eqn:orth_d}
\EE
which is a necessary model condition that minimizes the MSE of an estimation. Given arbitrary $\hat{\bf{x}}_t$ and $\hat{ \bf{\bb{z}}}_t$, to satisfy \eqref{Eqn:orth_d}, we can construct   
\BS\label{MMSE_orth}\BE
    \bf{x}_t = c^x_t \hat{\bf{x}}_t, \qquad   \bf{\bb{z}}_t = \bb{c}^z_t \hat{ \bf{\bb{z}}}_t,
\EE
with
\BE
   c^x_t = \arg\min_{c^x_t } \|\bf{x}_t -\bf{x}\|^2, \qquad  \bb{c}^z_t = \arg\min_{\bb{c}^z_t } \| \bf{\bb{z}}_t -\bf{z}\|^2.
\EE\ES
For more details, refer to \cite{lei2021OAMP}.

It should be emphasized that the step-by-step orthogonalization between current input and output estimation errors is not sufficient to guarantee the asymptotic IID Gaussianity for GMAMP. The following theorem shows a stricter orthogonality for the asymptotic IID Gaussianity (AIIDG) property of GMAMP. Specially, the strictest orthogonality requirement occurs with full memory, i.e., $\{\tau_{\gamma_x},\tau_{\gamma_z},\tau_{\phi},\tau_{\psi}\}=1$.

\begin{theorem} [Orthogonality and AIIDG]\label{Lem:IIDG_MIP}
Assume that $\bf{A}$ is unitarily invariant with $M, N\!\to\! \infty$,  $\{{\gamma}_t(\cdot)\}$ is Lipschitz-continuous \cite{Berthier2017} and   $\{\phi_t(\cdot), \psi_t(\cdot)\}$ are separable-and-Lipschitz-continuous \cite{Berthier2017}, Then, following generalized Stein's Lemma \cite{Stein1981}, the following orthogonality holds for GMAMP:
\BS\label{Eqn:error_orth_MIP}\begin{align} 
     |\! \langle  \bf{\bb{f}}_t,   \bf{x} \rangle\!|   &\overset{\rm a.s.}{=}0,&
     |\! \langle  \bf{s}_t,   \bf{z} \rangle\!|  &\overset{\rm a.s.}{=}0,\\
   % |\! \langle  \bf{f}_t ,\bf{x}_t \rangle\!|      &\overset{\rm a.s.}{=}0, &
    %|\! \langle  \bf{\bb{s}}_t, \bf{\bb{z}}_t \rangle\!|  &\overset{\rm a.s.}{=}0,\\
    |\! \langle  {\bf{f}_t},  \bf{\bb{F}}_{1\to t}\rangle\!|  &\overset{\rm a.s.}{=}\bf{0},&
    |\! \langle  {\bf{s}_t}, \bf{\bb{S}}_{1\to t}  \rangle\!| &\overset{\rm a.s.}{=}\bf{0},\\
    |\! \langle  \bf{\bb{f}}_{t+1},  \bf{F}_{1\to t}\rangle\!|  &\overset{\rm a.s.}{=}\bf{0},&
   |\! \langle  \bf{\bb{s}}_{t+1}, \bf{S}_{1\to t} \rangle\!|&\overset{\rm a.s.}{=}\bf{0}.
\end{align}\ES  
Since $\{\bf{\bb{X}}_{1\rightarrow t}, \bf{X}_{1\rightarrow t}\}$ and $\{\bf{\bb{Z}}_{1\rightarrow t}, \bf{Z}_{1\rightarrow t}\}$ correspond to independent Haar matrices $\bf{V}$ and $\bf{U}$, respectively. Therefore, similar to the AIIDGs of OAMP/VAMP \cite{Takeuchi2020, Rangan2019} and GVAMP \cite{Schniter2016, pandit2020}, the AIIDGs of $\bf{\bb{X}}_{1\rightarrow t}$ are guaranteed by the Haar matrix $\bf{V}$ and the orthogonality between $\bf{x}, \bf{\bb{F}}_{1\to t}$ and $\bf{F}_{1\to t}$ (see the equations on the left in \eqref{Eqn:error_orth_MIP}), and the AIIDGs of $\bf{Z}_{1\rightarrow t}$ are guaranteed by the Haar matrix $\bf{U}$ and the orthogonality between  $\bf{z}, \bf{\bb{S}}_{1\to t}$ and $\bf{S}_{1\to t}$ (see the equations on the right in \eqref{Eqn:error_orth_MIP}). Let $\bf{X}=\bf{x}\cdot\bf{1}^{\rm T}$ and $\bf{Z}=\bf{z}\cdot\bf{1}^{\rm T}$, where $\bf{1}$ is an all-ones vector with proper length. Then, under the orthogonality in \eqref{Eqn:error_orth_MIP},  we have \cite{Takeuchi2020, Rangan2019}:
$\forall 1\!\le \!t'\!\leq\! t$,  
\BS\label{Eqn:IIDG_MIP}  \begin{align}
&  v_{t,t'}^{x}   \! \overset{\rm a.s.}{=}\!    |\! \langle  \phi_t(\bf{X}\!\!+\!\bf{\bb{N}}_{\!\tau_{\phi}\!\to t}^{x})\!-\!\bf{x},\,  \phi_{t'}(\bf{X}\!\!+\!\bf{\bb{N}}_{\!\tau_{\phi}\!\to t'}^{x})\!-\!\bf{x}  \rangle\!|,\\ 
&\bb{v}_{t+1,{t'}\!+1}^{x} \!  \overset{\rm a.s.}{=} \!   |\! \langle  \gamma^x_t(\bf{X}_{\tau_{\gamma_x}\rightarrow t},  \bf{Z}\!+\!\bf{N}_{\!\tau_{\gamma_z}\!\to t}^{z})\!-\!\bf{x},  \gamma^x_{t'}(\bf{X}_{\tau_{\gamma_x}\rightarrow t}, \bf{Z}\!+\!\bf{N}_{\!\tau_{\gamma_z}\!\to t}^{z})\!-\!\bf{x}  \rangle\!|,\\
& \bb{v}_{t+1,{t'}\!+1}^{z}  \! \!  \overset{\rm a.s.}{=} \! \!|\! \langle  \gamma^z_t(\bf{X}_{\tau_{\gamma_x}\rightarrow t}, \bf{Z}\!+\!\bf{N}_{\!\tau_{\gamma_z}\!\to t}^{z})\!-\!\bf{z},  \gamma^z_{t'}(\bf{X}_{\tau_{\gamma_x}\rightarrow t}, \bf{Z}\!+\!\bf{N}_{\!\tau_{\gamma_z}\!\to t'}^{z})\!-\!\bf{z}  \rangle\!|, 
\end{align}\ES 
where $\bf{F}_{1\rightarrow t}$ and $\bf{\bb{S}}_{1\rightarrow t}$ are column-wise IID with zero mean, $\bf{\bb{N}}^x_{\tau_{\phi}\rightarrow t}\!=\![\bf{\bb{n}}^x_{\tau_{\phi}}\cdots \bf{\bb{n}}^x_t],  \bf{N}^z_{\tau_{\gamma_z}\rightarrow t}\!=\![\bf{n}^z_{\tau_{\gamma_z}}\cdots \bf{n}^z_t]$, and $\bf{\bb{N}}^x_{1\rightarrow t}$ and $\bf{N}^z_{1\rightarrow t}$ are respectively CIIDG-RJG with zero mean and covariances: for $1\leq t'\leq t$,
\begin{align}
    &{\rm E}\big\{|\! \langle \bf{\bb{n}}^x_t, \bf{\bb{n}}^x_{t'} \rangle\!|\big\}=\bb{v}^{x}_{t,t'}, 
    &{\rm E}\big\{|\!\langle  \bf{n}^z_t, \bf{n}^z_{t'}\rangle\!|\big\}=v^{z}_{t,t'}. 
\end{align} 
Meanwhile,  $\bf{\bb{N}}^x_{1\rightarrow t}$ is independent of $\bf{x}$, $\bf{N}^z_{1\rightarrow t}$ is independent of $\bf{z}$, $\bf{n}^x_t$ is independent of $\bf{x}_{t}$,  and $\{\bf{\bb{N}}^x_{1\rightarrow t}, \bf{F}_{1\rightarrow t},\bf{x}\}$ is independent of $\{\bf{N}^z_{1\rightarrow t}, \bf{\bb{S}}_{1\rightarrow t}, \bf{z}\}$. 
\end{theorem}

\subsection{Orthogonal MLE}
In this subsection, we give a construction of orthogonal MLE $\{\gamma_t^x(\cdot),\gamma_t^z(\cdot)\}$ to satisfy the orthogonality in \eqref{Eqn:error_orth_MIP}. For notational simplicity, we drop the subscript $t$.

\begin{definition}[Orthogonal MLE] 
Assume that $\bf{X}_{t}=[\bf{x}_{\tau_x}, \cdots\!, \bf{x}_t]$  and $\bf{Z}_t=[\bf{z}_{\tau_z}, \cdots\!, \bf{z}_t]$. An orthogonal MLE is defined as  
\BS\label{Eqn:orth_MLE}\begin{align}
     \left[ \!\!\!\begin{array}{c}
         \bf{\bb{x}}_t   \\
        \bf{\bb{z}}_t
    \end{array} \!\!\!\right] \!\!=\! \!\left[ \!\!\!\begin{array}{c}
         {\gamma}_x(\bf{X}_t;\bf{Z}_t)  \\
        {\gamma}_z(\bf{X}_t;\bf{Z}_t)
    \end{array} \!\!\!\right]  
    \! \!= \!\!\left[\! \!\!\begin{array}{c}
    \textstyle\sum_{i={\tau_z}}^t \!\bf{\mathcal{Q}}_i \bf{z}_i \!- \!\!\textstyle\sum_{i={\tau_x}}^t\! \bf{\mathcal{P}}_i \bf{x}_i \vspace{2mm}\\
    \textstyle\sum_{i={\tau_z}}^t \!\bf{\mathcal{G}}_i \bf{z}_i \!-\!\! \textstyle\sum_{i={\tau_x}}^t \!\bf{\mathcal{H}}_i \bf{A}\bf{x}_i 
    \end{array} \!\!\!\right]\!\!,
\end{align}
where $\bf{\mathcal{Q}}_i\bf{A}$ and $\bf{\mathcal{P}}_i$ are polynomials in $\bf{A}^{\rm H}\bf{A}$,  $\bf{\mathcal{G}}_i$ and $\bf{\mathcal{H}}_i$  are polynomials in $\bf{A}\bf{A}^{\rm H}$, and
\begin{align} 
   & {\rm tr} \big\{ \bf{\mathcal{P}}_i \big\} =0, \quad i={\tau_x},\dots,t, \\
   & {\rm tr} \big\{ \bf{\mathcal{G}}_i \big\} =0, \quad i={\tau_z},\dots,t,\\
   &\tfrac{1}{N}{\rm tr}\big\{\textstyle\sum_{i={\tau_z}}^t\!\bf{\mathcal{Q}}_i\bf{A} \!-\!\! \textstyle\sum_{i={\tau_x}}^t \!\bf{\mathcal{P}}_i\big\} = 1.
\end{align} \ES 
Let $\bf{\bb{f}}_t={\gamma}_x(\bf{X}_t;\bf{Z}_t)  - \bf{x} $ and $\bf{\bb{s}}_t={\gamma}_z(\bf{X}_t;\bf{Z}_t)  - \bf{z} $, recall $\bf{f}_i=\bf{x}_i-\bf{x}$ and $\bf{s}_i=\bf{z}_i-\bf{z}$, $\forall i$, and assume\footnote{This assumption is guaranteed by the orthogonal MNLE (in the next subsection) and orthogonal MLE by induction.} that $\bf{F}_t=[\bf{f}_1 \cdots \bf{f}_t]$ and $\bf{S}_{t}=[\bf{s}_1 \cdots \bf{s}_t]$ are column-wise IID with zero mean, and $\bf{F}_t$, $\bf{S}_t$ and $\bf{z}$ are independent. Then, the MLE in \eqref{Eqn:orth_MLE} satisfies the following orthogonalization:
\BS\begin{align} 
    |\! \langle \bf{\bb{f}}_t, \bf{F}_t \rangle\!| &\overset{\rm a.s.}{=}  \bf{0},  \\ 
    |\! \langle  \bf{\bb{s}}_t, \bf{S}_t \rangle\!| &\overset{\rm a.s.}{=}  \bf{0}, \\ 
    |\! \langle \bf{\bb{f}}_t, \bf{x} \rangle\!| &\overset{\rm a.s.}{=}  0.
\end{align} \ES 
Furthermore, the orthogonality $|\! \langle \bf{\bb{s}}_t, \bf{\bb{z}}_t \rangle\!|\overset{\rm a.s.}{=}  0$ can be easily satisfied by certain scaling (see \eqref{MMSE_orth}).
\end{definition}

The lemma below constructs an orthogonal MLE based on a general MLE.

\begin{lemma} 
 Given a general MLE
\BS\label{Eqn:orth_MLE_build}\begin{align}
     \left[ \!\!\!\begin{array}{c}
        \hat{\bf{\bb{x}}}_t  \\
        \hat{\bf{\bb{z}}}_t
    \end{array} \!\!\!\right]  \!  \!=\! \! \left[ \!\!\!\begin{array}{c}
         \hat{\gamma}_x(\bf{X}_t;\bf{Z}_t)  \\
        \hat{\gamma}_z(\bf{X}_t;\bf{Z}_t)
    \end{array} \!\!\!\right]  
     \! \!= \!\!  \left[\! \!\!\begin{array}{c}
    \textstyle\sum_{i={\tau_z}}^t\! {\bf{Q}}_i \bf{z}_i \! -\!\! \textstyle\sum_{i={\tau_x}}^t\! {\bf{P}}_i \bf{x}_i \vspace{2mm}\\
    \textstyle\sum_{i={\tau_z}}^t \! \bf{G}_i \bf{z}_i \! -\!\! \textstyle\sum_{i={\tau_x}}^t \! \bf{H}_i \bf{A}\bf{x}_i 
     \end{array} \!\!\!\right]\!\!,
\end{align}
where $\bf{Q}_i\bf{A}$ and $\bf{P}_i$ are polynomials in $\bf{A}^{\rm H}\bf{A}$, $\bf{G}_i$ and $\bf{H}_i$ are polynomials in $\bf{A}\bf{A}^{\rm H}$, we can construct an orthogonal MLE by
\label{Eqn:OMLE}\begin{align}
     \left[ \!\!\begin{array}{c}
         \bf{\bb{x}}_t    \\
        \bf{\bb{z}}_t 
    \end{array} \!\!\right]\!  \!=\! \! \left[ \!\!\!\begin{array}{c}
        {\gamma}_x(\bf{X}_t;\bf{Z}_t)  \\
       {\gamma}_z(\bf{X}_t;\bf{Z}_t)
    \end{array} \!\!\!\right]  
     \! \!= \!\!   \left[ \!\!\begin{array}{c}
    \bb{c}^x_t\big(\hat{\bf{\bb{x}}}_t +  \bf{X}_t\bf{p}_t \big) \vspace{2mm}\\
    \bb{c}^z_t\big(\hat{\bf{\bb{z}}}_t- \bf{Z}_t\bf{g}_t \big)
    \end{array} \!\!\right],
\end{align}
where 
\begin{align}
   & \bf{p}_t=\big[\tfrac{1}{N}{\rm tr}  \{ {\bf{P}}_{\tau_x}\}\,\cdots\,  \tfrac{1}{N}{\rm tr}  \{ {\bf{P}}_t\}\big]^{\rm T},\\
   & \bf{g}_t=\big[\tfrac{1}{M}{\rm tr}  \{ {\bf{G}}_{\tau_z}\}\,\cdots\,  \tfrac{1}{M}{\rm tr}  \{ {\bf{G}}_t\}\big]^{\rm T}, 
\end{align}\ES 
and $\bb{c}^x_t$ and $\bb{c}^z_t$ are the scaling coefficients given in \eqref{norm_orth} and \eqref{MMSE_orth}, respectively. Specifically, 
\BE
 \bb{c}^x_t = {N} / {\rm tr} \big\{\textstyle\sum_{i=\tau_z }^t{\bf{Q}}_i \bf{A} - \textstyle\sum_{i=\tau_x}^t  {\bf{P}}_i \big\}.
\EE
\end{lemma} 

It is easy to verify that the MLE in \eqref{Eqn:orth_MLE_build} satisfies \eqref{Eqn:orth_MLE}, i.e., it is an orthogonal MLE.

\underline{\emph{Example:}} Let $\tau_x=\tau_z=t$, $\bf{Q}_t=\bf{A}^{\rm H} (\rho_t\bf{I}+\bf{A}^{\rm H}\bf{A})^{-1}$, $\bf{P}_t=\bf{Q}_t\bf{A}$, $\bf{G}_t=\bf{A}\bf{Q}_t$ and $\bf{H}_t=\bf{A}\bf{Q}_t +\bf{I}$. Then, the orthogonal MLE in \eqref{Eqn:OMLE} is degraded to the LMMSE-LE of GVAMP in \textcolor{blue}{SI Appendix, section 1}.

\subsection{Orthogonal MNLE}
In this subsection, we construct two orthogonal MNLEs $\phi_t(\cdot)$ and $\psi_t(\cdot)$ to satisfy the orthogonality in \eqref{Eqn:error_orth_MIP}. %Let's take $\phi_t(\cdot)$ as an example, and the same thing applies to $\psi_t(\cdot)$. 
For notational simplicity, we omit the subscript $t$.

\begin{definition}[Orthogonal MNLE]
Let $\bf{\bb{X}}_t=[\bf{\bb{x}}_{\tau}, \cdots\!, \bf{\bb{x}}_t]$,   $\bf{f}_t \!=\! {\phi}(\bf{\bb{X}}_t) -\bf{x}$,  and recall $\bf{\bb{f}}_i=\bf{\bb{x}}_i-\bf{x}$, $\forall i$. Define that $\bf{\bb{F}}_t =[\bf{\bb{f}}_{\tau}\cdots \bf{\bb{f}}_t]$. Then, the MNLE ${\phi}(\bf{\bb{X}}_t)$ is called orthogonal MNLE if 
 \BE\label{Eqn:Orth_MNLE} 
     |\! \langle \bf{f}_{t}, \bf{\bb{F}}_t\rangle\!|  \overset{\rm a.s.}{=} \bf{0}.
 \EE
\end{definition} 

 The lemma below constructs an orthogonal MNLE based on an arbitrary MNLE.

\begin{lemma}\label{Lem:MNLE}
   Assume\footnote{This assumption is guaranteed by the orthogonal MLE (in previous subsection)  and orthogonal MNLE by induction.} that $\bf{\bb{F}}_t$ is CIIDG-RJG with zero mean and   independent of $\bf{x}$. Given arbitrary separable-and-Lipschitz-continuous  $\hat{\phi}(\bf{\bb{X}}_t)$, we can construct an orthogonal MNLE by
    \BS\label{Eqn:OMNLE_build}\begin{align}
        \phi(\bf{\bb{X}}_t)&=  \varpi^*_t\big( \hat{\phi}(\bf{\bb{X}}_t) -    \bf{\bb{X}}_t \bf{\varpi}_t \big),\label{Eqn:OMNLE_build_a} 
    \end{align}
    where 
     \begin{align}
        \bf{\varpi}_t &=  (\bf{\bb{F}}_t^{\rm H} \bf{\bb{F}}_t)^{-1}\bf{\bb{F}}_t^{\rm H}\hat{\phi}(\bf{\bb{X}}_t),\label{Eqn:OMNLE_build_b} 
    \end{align}\ES
    and $\varpi^*_t$ is scaling coefficient given in \eqref{norm_orth} or \eqref{MMSE_orth}. 
\end{lemma}

See Appendix B in \cite{lei2020mamp} for further details. In general, $\varpi^*_t$ in \eqref{Eqn:OMNLE_build} is determined by minimizing the MSE of ${\phi}(\cdot)$.

\begin{lemma}\label{Lem:MNLE_div}
   Assume that $\bf{\bb{F}}_t$ is CIIDG-RJG with zero mean and is independent of $\bf{x}$. For arbitrary separable-and-Lipschitz-continuous and differentiable $\hat{\phi}(\bf{\bb{X}}_t)$, using generalized Stein's Lemma \cite{Stein1981}, we have
    \BS\begin{align}
        \bf{\bb{F}}_t^{\rm H}\hat{\phi}(\bf{\bb{X}}_t)  &\overset{\rm a.s.}{=}\bf{\bb{F}}_t^{\rm H} \bf{\bb{F}}_t  \big[{\rm E}\{\tfrac{\partial \hat{\phi} } {\partial  \bf{\bb{x}}_\tau}\} \,\cdots \, {\rm E}\{\tfrac{\partial \hat{\phi} } {\partial  \bf{\bb{x}}_t}\} \big]^{\rm T}.
    \end{align}\ES
   Then, $\bf{\varpi}_t$ in \eqref{Eqn:OMNLE_build} can be rewritten to
    \BS\begin{align}
        \bf{\varpi}_t&=   \big[{\rm E}\{\tfrac{\partial \hat{\phi} } {\partial  \bf{\bb{x}}_\tau}\} \,\cdots \, {\rm E}\{\tfrac{\partial \hat{\phi} } {\partial  \bf{\bb{x}}_t}\} \big]^{\rm T}.
    \end{align}\ES
\end{lemma}

Due to the symmetry of the problem, we can construct the orthogonal $\psi(\bf{\bb{Z}}_t)$ in a similar way.

\underline{\emph{Example:}} Let $\tau=t$, $\varpi^*_{x_t}=\frac{1}{1-\varpi_{x_t}}$, $\varpi^*_{z_t}=\frac{1}{1-\varpi_{z_t}}$,  $\hat{\phi}$ and $\hat{\psi}$ be MMSE estimators given by  $\hat{\phi}(\bf{\bb{x}}_t)\equiv \mr{E}\{\bf{x}|\bf{\bb{x}}_t,\Phi\}$ and $\hat{\psi}(\bf{\bb{z}}_t)\equiv \mr{E}\{\bf{z}|\bf{\bb{z}}_t,\Psi\}$. Then, $\varpi_{x_t}={\rm E}\{\frac{d \hat{\phi} }{d  \bf{\bb{x}}_t} \} \overset{\rm a.s.}{=} \frac{ \|\hat{\phi}(\bf{\bb{x}}_t)-\bf{x}\|^2}{\|\bf{\bb{x}}_t -\bf{x}\|^2}=\frac{v^x_{t,t}}{\bb{v}^x_{t,t}}$ and $\varpi_{z_t}={\rm E}\{\frac{d \hat{\psi} }{d  \bf{\bb{z}}_t}\} \overset{\rm a.s.}{=} \frac{ \|\hat{\psi}(\bf{\bb{z}}_t)-\bf{z}\|^2}{\|\bf{\bb{z}}_t -\bf{z}\|^2}=\frac{v^z_{t,t}}{\bb{v}^z_{t,t}}$. In this case, the orthogonal MNLE in \eqref{Eqn:OMNLE_build} is degraded to the MMSE-NLE of GVAMP in \textcolor{blue}{SI Appendix, section 1}.

\section{Bayes-optimal Generalized Memory AMP}
As mentioned, GVAMP has a high computational complexity due to LMMSE-LE, which costs $\mathcal{O}(M^2N+M^3)$ time complexity per iteration for matrix multiplication and matrix inversion. To reduce the complexity, we introduce  a Bayes-optimal memory LE to suppress the linear interference. However, the NLEs are the same as these in GVAMP since they are symbol-by-symbol with time complexity as low as $\mathcal{O}(N\!+\!M)$ per iteration. 

\BS\begin{alignat}{2}
{\rm NLE:}\quad &\left[ \!\!\begin{array}{c}
         \bf{x}_{t}   \\
        \bf{z}_{t}
    \end{array} \!\!\right]   =\left[ \!\!\begin{array}{c}
        \phi_t(\bf{\bb{x}}_t)   \\
        \psi_t(\bf{\bb{z}}_t)
    \end{array} \!\!\right],\\
{\rm MLE:} \quad \quad & (\bf{\bb x}_{t + 1}, \bf{\bb z}_{t + 1})  = \gamma_t \big([\bf{x}_1 \cdots \bf{x}_t], [\bf{z}_1\cdots\bf{z}_t]\big).
\end{alignat}\ES	  

\begin{figure}[ht]
\centering
\includegraphics[scale=0.4]{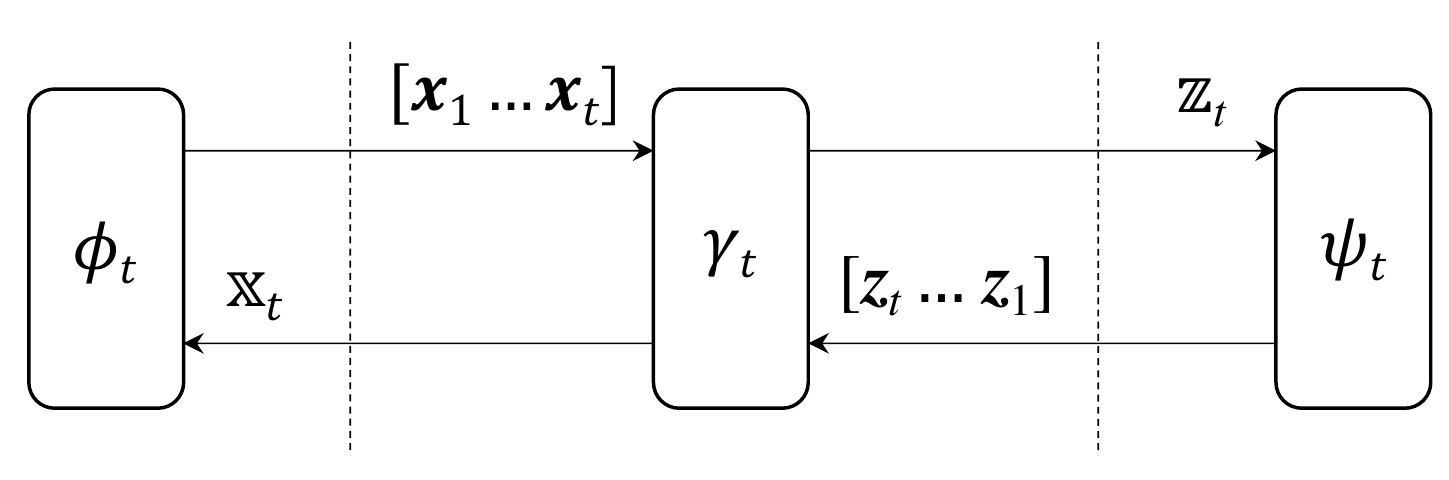}
\caption{\footnotesize Graphic illustration with a long-memory LE and two non-memory NLEs.}
\label{fig:system diagram3}
\end{figure}

Fig. \ref{fig:system diagram3} is a special case of the MIP in \eqref{Eqn:MIP} with $\tau_{\gamma_x}=\tau_{\gamma_z}=1$ and $\tau_{\phi}=\tau_{\psi}=t$. In this case, the orthogonality in \eqref{Eqn:orth_GMAMP} of GMAMP degenerates to 
\BS\label{Eqn:orth_BO-GMAMP}\begin{align} 
    |\! \langle  \bf{\bb{f}}_t,  \bf{x} \rangle\!| & \overset{\rm a.s.}{=}0, &
    |\! \langle \bf{s}_t, \bf{z} \rangle\!|  &\overset{\rm a.s.}{=}0,\\
    |\! \langle {\bf{f}_t}, \bf{\bb{f}}_t \rangle\!|  &\overset{\rm a.s.}{=}0,&
    |\! \langle  \bf{s}_t,\bf{\bb{s}}_t\rangle\!| &\overset{\rm a.s.}{=}0,\\
    |\! \langle \bf{\bb{f}}_{t+1}, \bf{F}_{1\rightarrow t}\rangle\!|   &\overset{\rm a.s.}{=}\bf{0},&
    |\! \langle \bf{\bb{s}}_{t+1}, \bf{S}_{1\rightarrow t}\rangle\!|    &\overset{\rm a.s.}{=}\bf{0}.
\end{align}\ES

Based on this theoretical framework, we design a specific Bayes-optimal GMAMP (BO-GMAMP) algorithm by exploiting the above constructions of orthogonal MLE and orthogonal MNLE, which guarantees not only Bayes optimality but also low implement complexity. Meanwhile, the main properties of BO-GMAMP including orthogonality and AIIDG are provided below. In addition, we derive the state evolution of BO-GMAMP and prove the Bayes optimality of BO-GMAMP via state evolution. 

\subsection{Bayes-optimal Generalized Memory AMP}
As an instance of GMAMP framework, BO-GMAMP is given in the following algorithm. 
\begin{framed}
 \emph{Bayes-Optimal Generalized Memory AMP:} Let $\bf{B}  = \lambda^\dag\bf{I} - \bf{A}\bf{A}^{\mr H}$. Consider a memory linear estimations: 
\begin{align}
    \hat{\bf{\bb{z}}}_t &= \theta_t\bf{B}\hat{\bf{\bb{z}}}_{t-1}+\xi_t(\bf{z}_t-\bf{Ax}_t),\\
    \hat{\bf{\bb{x}}}_t &= \bf{A}^{\rm H}\hat{\bf{\bb{z}}}_t. 
\end{align}
 A BO-GMAMP process is defined as: Starting with $t=1$ and $\hat{\bf{\bb{x}}}_{0}= {\bf{\bb{x}}}_{1}= \hat{\bf{\bb{z}}}_{1}= \bf{0}$, 
\BS\label{Eqn:GMAMP0}
\begin{alignat}{2}
&{\rm MNLE:} \;\; \left[ \!\!\begin{array}{c}
         \bf{x}_{t}   \\
        \bf{z}_{t}
    \end{array} \!\!\right]  
    =\left[ \!\!\begin{array}{c}
         \Bar{\phi}_t(\bf{\bb{x}}_t)   \\
        \Bar{\psi}_t(\bf{\bb{z}}_t)
    \end{array} \!\!\right]     =\left[ \!\!\begin{array}{c}
         \scaleto{\zeta}{8pt}_{t, l_t} \phi_t(\bf{\bb{x}}_t)+\textstyle\sum_{i=1}^{l_t-1} \scaleto{\zeta}{8pt}_{t,i}\bf{x}_{t-l_t+i}, \vspace{2mm} \\
        \varrho_{t, l_t} \psi_t(\bf{\bb{z}}_t)+\textstyle\sum_{i=1}^{l_t-1} \varrho_{t,i}\bf{z}_{t-l_t+i}
    \end{array} \!\!\right],\label{Eqn:NLE0}\\
&{\rm MLE:} \quad \left[ \!\!\begin{array}{c}
         \bf{\bb{x}}_{t+1}   \\
        \bf{\bb{z}}_{t+1}
    \end{array} \!\!\right]
    = \gamma_t \big(\bf{X}_{1\rightarrow t},\bf{Z}_{1\rightarrow t}\big)  
    =\left[ \!\!\!\begin{array}{c}
        \bb{c}_t^x \big(  \delta^{-1} \hat{\bf{\bb{x}}}_t  +  \textstyle\sum_{i=1}^t p_{t,i} \bf{x}_i  \big) \\
        \bb{c}_t^z \big[ \bf{A} \big( \hat{\bf{\bb{x}}}_t +\frac{\xi_t}{\theta_t}\bf{x}_t\big)   - \textstyle\sum_{i=1}^t p_{t,i} \bf{z}_i \big] 
    \end{array}\!\!\! \right], 
    \label{Eqn:MLE0}
\end{alignat}\ES	 
where $\phi_t(\cdot)$ and $\psi_t(\cdot)$ are separable and Lipschitz-continuous functions and the same as that in GVAMP (see \textcolor{blue}{SI Appendix, section 1} for further details).
\end{framed}

The following are some intuitive interpretations of the parameters in BO-GMAMP. The optimization of these parameters will be provided later.  
\begin{itemize}
    \item In MLE, all preceding messages $\{\bf{x}_i, \bf{z}_i\}$ are utilized to guarantee the orthogonality in \eqref{Eqn:error_orth_MIP}. 
    \item $\theta_t $, optimized in \eqref{Eqn:theta}, ensures that BO-GMAMP has a Bayes-optimal fixed point. $\xi_t$, optimized in \textcolor{blue}{SI Appendix, section 5}, accelerates the convergence of BO-GMAMP. 
    \item $\{\bb{c}_t^x\}$, given by $\bb{c}_t^x = 1/\textstyle\sum_{i=1}^t p_{t,i}$, and $\{\bb{c}_t^z\}$, optimized in \eqref{Eqn:alpha_opt0}, and $\{ p_{t, i}\}$ guarantee the orthogonality in \eqref{Eqn:error_orth_MIP} (see Theorem \ref{The:IIDG_GMAMP}).   
    \item  $\scaleto{\bf{\zeta}}{8pt}_t=[{\scaleto{\zeta}{8pt}}_{t,1}, \cdots, {\scaleto{\zeta}{8pt}}_{t,l_t}]^{\rm T}$ and $\bf{\varrho}_t=[{\varrho}_{t,1}, \cdots, {\varrho}_{t,l_t}]^{\rm T}$, optimized in \eqref{Eqn:zeta_varrho}, are damping vectors with $\textstyle\sum_{i=1}^{l_t} {\scaleto{\zeta}{8pt}}_{t,i} =1$ and $\textstyle\sum_{i=1}^{l_t} \varrho_{t,i} =1$. We set $l_t=\min\{L, t\}$, where $L$ is the maximum damping length. Damping guarantees and improves the convergence of BO-GMAMP. In general, we set $L\le3$.
\end{itemize}

As we can see, only matrix-vector multiplications are involved in each iteration. Thus, the time complexity of BO-GMAMP is as low as ${\cal O}(MN)$ per iteration, which is comparable to AMP.

\subsection{Orthogonality and AIIDG}

The following theorem is based on Theorem \ref{Lem:IIDG_MIP} and Proposition 1 in \textcolor{blue}{SI Appendix, section 2}.

\begin{theorem}\label{The:IIDG_GMAMP}
Assume that $\bf{A}$ is unitarily invariant with $M, N\!\to\! \infty$. The orthogonality in \eqref{Eqn:orth_BO-GMAMP} holds for BO-GMAMP. Therefore, following Theorem \ref{Lem:IIDG_MIP}, the asymptotic IID Gaussianity in \eqref{Eqn:IIDG_MIP} holds for BO-GMAMP \cite{Takeuchi2020, Rangan2019, pandit2020}.
\end{theorem}

See \textcolor{blue}{SI Appendix, section 2} for further details.

Using Theorem \ref{The:IIDG_GMAMP}, the performance of BO-GMAMP can be tracked by using the state evolution discussed in the following subsection.

\subsection{State Evolution}
Define the covariance vectors and covariance matrices as follows:
\BS\begin{alignat}{2}
   \bf{\bb{v}}^{x}_t &=[\bb{v}^{x}_{t,1},...,\bb{v}^{x}_{t,t}]^{\rm T}, \qquad &\bf{\bb{V}}^{x}_t &=[\bb{v}^{x}_{i,j}]_{t\times t},\\
    \bf{v}^x_t&=[v^x_{t,1},...,v^x_{t,t}]^{\rm T}, &\bf{V}^x_t &=[v^x_{i,j}]_{t\times t},\\
    \bf{\bb{v}}^{z}_t&=[\bb{v}^{z}_{t,1},...,\bb{v}^{z}_{t,t}]^{\rm T}, &\bf{\bb{V}}^{z}_t &=[\bb{v}^{z}_{i,j}]_{t\times t},\\
   \bf{v}^z_t &=[v^z_{t,1},...,v^z_{t,t}]^{\rm T}, &\bf{V}^z_t &=[v^z_{i,j}]_{t\times t}, 
\end{alignat}\ES
where $\bb{v}^{x}_{t,t'}$, $\bb{v}^{z}_{t,t'}$, $v^x_{t,t'}$ and $v^z_{t,t'}$ are defined in \eqref{err_defi}.

Hence, starting with $\bb{v}^{x}_1=1$ and $\bb{v}^{z}_1=\lambda_1$, the error covariance matrices of BO-GMAMP can be tracked by the following state evolution.
\BS\begin{align}
  {\rm NLE:} \qquad\;\, (\bf{v}_{t}^{x}, \bf{v}_{t}^{z}) & = \big(\Bar{\phi}_{\rm SE}(\bf{\bb{V}}_t^x), \Bar{\psi}_{\rm SE}(\bf{\bb{V}}_t^{z})\big),\\
    {\rm MLE:} \quad (\bf{\bb{v}}_{t+1}^{x},\bf{\bb{v}}_{t+1}^{z})&=\gamma_{\rm SE}(\bf{V}_t^{x},\bf{V}_t^{z}).
\end{align}\ES
where $\Bar{\phi}_{\rm SE}(\cdot)$, $\Bar{\psi}_{\rm SE}(\cdot)$ and $\gamma_{\rm SE}(\cdot)$ are given in \textcolor{blue}{SI Appendix, section 3}.

\subsection{Convergence and Bayes Optimality}  
It has been proved that GVAMP achieves the minimum (i.e., Bayes-optimal) MSE as predicted by the replica method if it has a unique fixed point  \cite{pandit2020,Reeves2017,Gabrie2018}. The following theorem gives the convergence and Bayes optimality of BO-GMAMP. 
 
\begin{theorem}\label{The:Conv_GVAMP}
Assume that $\bf{A}$ is unitarily-invariant with $M,N\to\infty$. The proposed BO-GMAMP converges to the same fixed point as GVAMP, i.e.,  BO-GMAMP achieves the minimum (i.e., Bayes-optimal) MSE as predicted by the replica method if it has a unique fixed point. 
\end{theorem}
 
See \textcolor{blue}{SI Appendix, section 4} for further details.

\subsection{Complexity Comparison}
The time and space complexity of BO-GMAMP are given below, where $T$ represents the maximum number of iterations and $L\ll T \ll N$.
\begin{itemize}
\item BO-GMAMP costs $\mathcal{O}(MNT)$ time complexity for matrix-vector multiplications $\{\bf{AA}^{\rm H}\hat{\bf{\bb{x}}}_t\}$ and $\{\bf{Ax}_t\}$, which is dominant for the case $T \ll N$, $\mathcal{O}\big((N+M)T^2\big)$ for $\{\sum\nolimits_{i=1}^t p_{ti}\bf{x}_i\}$ and $\{\sum\nolimits_{i=1}^t p_{ti}\bf{z}_i\}$, $\mathcal{O}(T^3)$ for calculating $\{\bf{\bb{v}}_{t,t}^{x}, \bf{\bb{v}}_{t,t}^{z}\}$, and $\mathcal{O}(L^3T)$ for calculation of $\{\bf{\zeta}_t, \bf{\varrho}_t\}$.

\item BO-GMAMP needs $\mathcal{O}(MN)$ space complexity to store $\bf{A}$, which is dominant for the case $T \ll N$, $\mathcal{O}\big((M+N)T\big)$ space for $\{\bf{x}_t,\bf{z}_t\}$.
\end{itemize}

The time and space complexity of BO-GMAMP, GVAMP and GAMP are compared in Table \ref{tab:my_label}.  In general, $L\ll T \ll N$. Hence, $L$ and $T$ are negligible and can be selectively ignored.
\newcommand{\tabincell}[2]{\begin{tabular}{@{}#1@{}}#2\end{tabular}}
 \begin{table}[b]%\vspace{-2mm}
\renewcommand{\arraystretch}{1.5} 
    \centering    \small
    \setlength{\tabcolsep}{1.4mm}{\begin{tabular}{|c||c|c|}
        \hline
        Algorithms & Time complexity & Space complexity \\
        \hline\hline
        GAMP & $\mathcal{O}(MN)$ & $\mathcal{O}(MN)$ \\
        \hline
         \tabincell{c}{GVAMP\vspace{-0.2cm}\\ (SVD)} & $\mathcal{O}\big(M^2N+MN\big)$ &  $\mathcal{O}(N^2+M^2+MN)$ \\
        \hline
         \tabincell{c}{GVAMP\vspace{-0.2cm}\\ (matrix inverse)} & $\mathcal{O}\big(M^2N+M^3\big)$ &  $\mathcal{O}(MN)$ \\ 
        \hline
        \tabincell{c}{ BO-GMAMP\vspace{-0.2cm}\\  (proposed)}  & $\mathcal{O}\big(MN+(N+M)T^2\big)$ & $\mathcal{O}\big(MN+(N+M)T\big)$\\
        \hline
    \end{tabular}}
    \caption{\footnotesize Time and Space Complexity Comparison}
    \label{tab:my_label}
\end{table}

\subsection{BO-GMAMP for linear $Q(\cdot)$}
When $Q(\cdot)$ in GLM is a linear function, the $\Psi$ constraint in \eqref{Eqn:problem} can be simplified to
\BE
\bf{y}=Q(\bf{Ax})=\bf{Ax}+\bf{n},
\EE
where $\bf{n}$ denotes the additive white Gaussian noise (AWGN). In this case, the $\{\bf{\bb{z}}_t\}$ and $\psi_t$ in BO-GMAMP, see Fig. \ref{fig:system diagram3}, are removed. In addition, the variables $\{\bf{z}_1\cdots\bf{z}_t\}$ are replaced by $\bf{y}$ and their covariance matrix is replaced by noise variance $\tfrac{1}{M}{\rm E}\{\bf{n}^{\rm H}\bf{n}\}=\sigma^2$. Then, BO-GMAMP degenerates into the standard MAMP in \cite{lei2020mamp, lei2021mamp}.

\section{Parameter Optimization}
In this section, we provide the optimized parameters $\{\bf{\zeta}_t,\bf{\varrho}_t,\theta_t,\xi_t,\bb{c}_t^z\}$ directly. For details of the solution to these optimized parameters, please refer to \textcolor{blue}{SI Appendix, section 5} and \cite{lei2020mamp}. Notice that the optimized $\{\bf{\zeta}_t,\bf{\varrho}_t,\theta_t,\xi_t,\bb{c}_t^z\}$ guarantee the convergence of BO-GMAMP as well as improve the convergence speed of BO-GMAMP.

\begin{itemize}
\item An optimal $\bf{\zeta}_t$ that minimizes $v_{t+1,t+1}^x$ is given by
    \BS\label{Eqn:zeta_varrho}\begin{align}
    \bf{\zeta}_t^{\rm opt}=
    \begin{cases}
    \dfrac{[\bf{V}_{t+1}^{\phi}]^{-1}\bf{1}}{\bf{1}^{\rm T}[\bf{V}_{t+1}^{\phi}]^{-1}\bf{1}},\hspace{0.9cm} {\rm if}\ {\rm det}(\bf{V}_{t+1}^{\phi})>0\vspace{2mm}\\
    [0,\cdots,0,1]^{\rm T},\hspace{0.9cm} {\rm otherwise}
    \end{cases}
    \end{align}     
    and an optimal $\bf{\varrho}_t$ that minimizes $v_{t+1,t+1}^z$ is given by
    \begin{align}
    \bf{\varrho}_t^{\rm opt}=
    \begin{cases}
    \dfrac{[\bf{V}_{t+1}^{\psi}]^{-1}\bf{1}}{\bf{1}^{\rm T}[\bf{V}_{t+1}^{\psi}]^{-1}\bf{1}},\hspace{0.9cm} {\rm if}\ {\rm det}(\bf{V}_{t+1}^{\psi})>0\vspace{2mm}\\
    [0,\cdots,0,1]^{\rm T},\hspace{0.9cm} {\rm otherwise}
    \end{cases},
    \end{align}\ES
where the error covariance matrix  $\bf{V}_{t}^{\phi}$  of $\{\bf{x}_{t-l_t+1},...,\bf{x}_{t-1},\phi_t(\bf{\bb{x}}_t)\}$ (inputs of $\Bar{\phi}_t$) and the error covariance matrix $\bf{V}_{t}^{\psi}$  of $\{\bf{z}_{t-l_t+1},...,\bf{z}_{t-1},\psi_t(\bf{\bb{z}}_t)\}$ (inputs of $\Bar{\psi}_t$) are defined in \textcolor{blue}{SI Appendix, section 3}.

\item An optimal $\theta_t$ that minimizes the spectral radius of $\bf{C}_t=\bf{I}-\theta_t(\rho_t\bf{I}+\bf{AA}^{\rm H})$ is given by
    \begin{align}\label{Eqn:theta}
    \theta_t^{\rm opt}=(\lambda^{\dagger}+\rho_t)^{-1},
    \end{align}
    where $\rho_t=v_{t,t}^z/v_{t,t}^x$ and $\lambda^{\dagger}=(\lambda_{\rm max}+\lambda_{\rm min})/2$ with $\lambda_{\rm min}$ and $\lambda_{\rm max}$ being the minimal and maximal eigenvalues of $\bf{AA}^{\rm H}$.
\item An optimal $\xi_t$ that minimizes $\bb{v}_{t+1,t+1}^x$ is given in \textcolor{blue}{SI Appendix, section 5}. Note that $\xi_t$ does not change the fixed point of BO-GMAMP, but the optimized $\xi_t$ can improve the convergence speed of BO-GMAMP.
    
\item An optimal $\bb{c}_t^z$ that minimizes $\bb{v}_{t+1,t+1}^{z}$ is given by
\BE\label{Eqn:alpha_opt0}
   \bb{c}_t^{z,\rm opt} = \frac{\beta_t w_0}{\beta_t^2 w_0+v^{\tilde{\bf{\bb{s}}}}_{t,t}},
\EE
where $\beta_t$, $w_0$ and $v^{\tilde{\bf{\bb{s}}}}_{t,t}$ are defined in \textcolor{blue}{SI Appendix, section 3}.
\end{itemize}

\section{Algorithm Summary}
We have summarized BO-GMAMP and the state evolution of BO-GMAMP to Algorithm 1 and Algorithm 2. Please see \textcolor{blue}{SI Appendix, section 6} for specific algorithm flowcharts.

\section{Simulation Results}
In this part, we conduct simulations to evaluate our proposed BO-GMAMP algorithm.
For comparison, we provide BO-GMAMP and another two baselines: GVAMP and our BO-GMAMP without optimized parameters. All simulation settings are detailed as follows.

We address a compressed sensing problem using the proposed BO-GMAMP. $\bf{x}$ is sparse and follows Bernoulli-Gaussian distribution, i.e., the elements of $\bf{x}$ are non-zero Gaussian with probability $\mu$ and zero with $1-\mu$, which is given by
\BE
 x_i\sim \left\{\begin{array}{ll}
      0,&  {\rm probability} = 1- \mu  \\
      {\cal N} (0, \mu^{-1}),&  {\rm probability} = \mu
 \end{array}\right..
\EE
A specific GLM is described as
\BE\label{Eqn:example}
\left\{
\begin{array}{l}
\bf{y}=Q(\bf{z})\equiv {\rm Clip}(\bf{z})+\bf{n}\\
\bf{z}=\bf{A}\bf{x}\\
\bf{x}_i \sim P_X(x)
\end{array}
\right.,
\EE
where ${\rm Clip}(\cdot)$ is a symbol-by-symbol function given by 
\BE
{\rm Clip}(z)=\left\{
\begin{array}{l}
\mathfrak{c},\ \ \ \ \rm{if}\ \ \it{z}\geq\mathfrak{c}\\
z,\ \ \ \ \rm{if}\ \ -\mathfrak{c}<\it{z}<\mathfrak{c}\\
-\mathfrak{c},\ \ \rm{if}\ \ \it{z}\leq-\mathfrak{c}
\end{array}
\right. 
\EE
with $\mathfrak{c}$ being the judgement threshold.

Additionally, let the SVD of $\bm{A}$ be $\bm{A}=\bm{U \Sigma V^{\rm H}}$. The system model in \eqref{Eqn:example} is rewritten as:
\BE
{\bf{z}} = {\bm{U \Sigma V}}^{\rm H}\bf{x}.
\EE
To reduce the calculation complexity of matrix multiplication in GVAMP, we approximate two large random unitary matrices by $\bf{U} \!=\! \bf{F}_1\bf{\Pi}_1$ and ${\bf{V}}^{\rm H} \!=\! {\bf{\Pi}_2 \bf{F}_2}$, where $\bf{\Pi}_1$, $\bf{\Pi}_2$ are random permutation matrices and $\bm{F}_1$, $\bf{F}_2$ are discrete Fourier transform (DFT) matrices with different dimensions ($M$ and $N$, respectively). Note that this transformation significantly improves the implementation speed of GVAMP. 
The singular values $\{d_i\}$ are generated as: $d_i/d_{i+1}=\kappa^{1/
J}$ for $i = 1,\ldots, J-1$ and $\sum_{i=1}^Jd_i^2=N$, where $J=\min\{M, N\}$. $\kappa\ge1$ denotes the condition number of $\bf{A}$, representing the stability of system. 

\begin{figure}[b!]
\centering
\includegraphics[width=.4\linewidth]{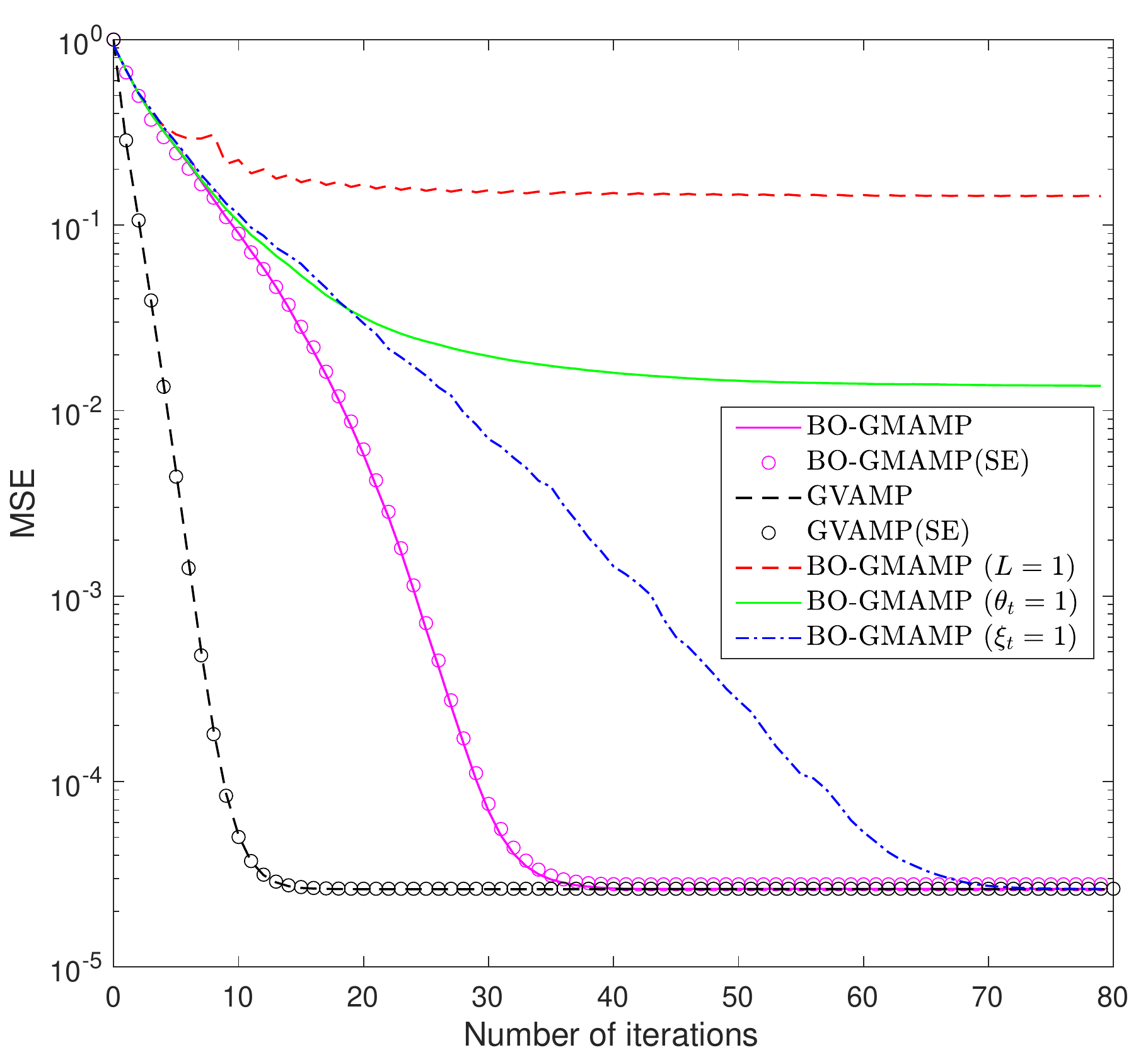}
\caption{\footnotesize MSE versus the number of iterations for GVAMP, BO-GMAMP and un-optimized BO-GMAMP."BO-GMAMP" with parameters $\mu=0.1$, $\mathfrak{c}=2$, $N=8192$, $\delta=0.5$, $\kappa=30$, ${\rm SNR}=40{\rm dB}$ and damping length $L=3$. "SE" denotes state evolution. The last three lines change only one parameter in brackets.}
\label{fig:sim1}
\end{figure}
As shown in Figure \ref{fig:sim1}, the proposed BO-GMAMP converges to the same fixed point with the high-complexity GVAMP, which validates the Bayes optimality of BO-GMAMP. Although GVAMP converges faster than BO-GMAMP, the computational complexity of GVAMP is intolerable, especially in large-scale systems (e.g., $N=8192$). Furthermore, the optimized parameters ($\{\theta_t,\xi_t\}$) and damping vectors (when damping length $L\neq1$) improve significantly the convergence speed of BO-GMAMP. Other Results are presented in \textcolor{blue}{SI Appendix, section 7}, including the influence of compression ratio $\delta$, condition number of system $\kappa$ and damping length $L$ on the performance of BO-GMAMP.

%\subsection{Simulation Settings}

\bibliographystyle{unsrt}
\bibliography{Reference}

\newpage

\part{Supplementary information}

\section{Overview of GVAMP}%1
\subsection{Non-memory Iterative Process and Orthogonality}
\begin{figure}[h]
\centering
\includegraphics[width=0.3\textwidth]{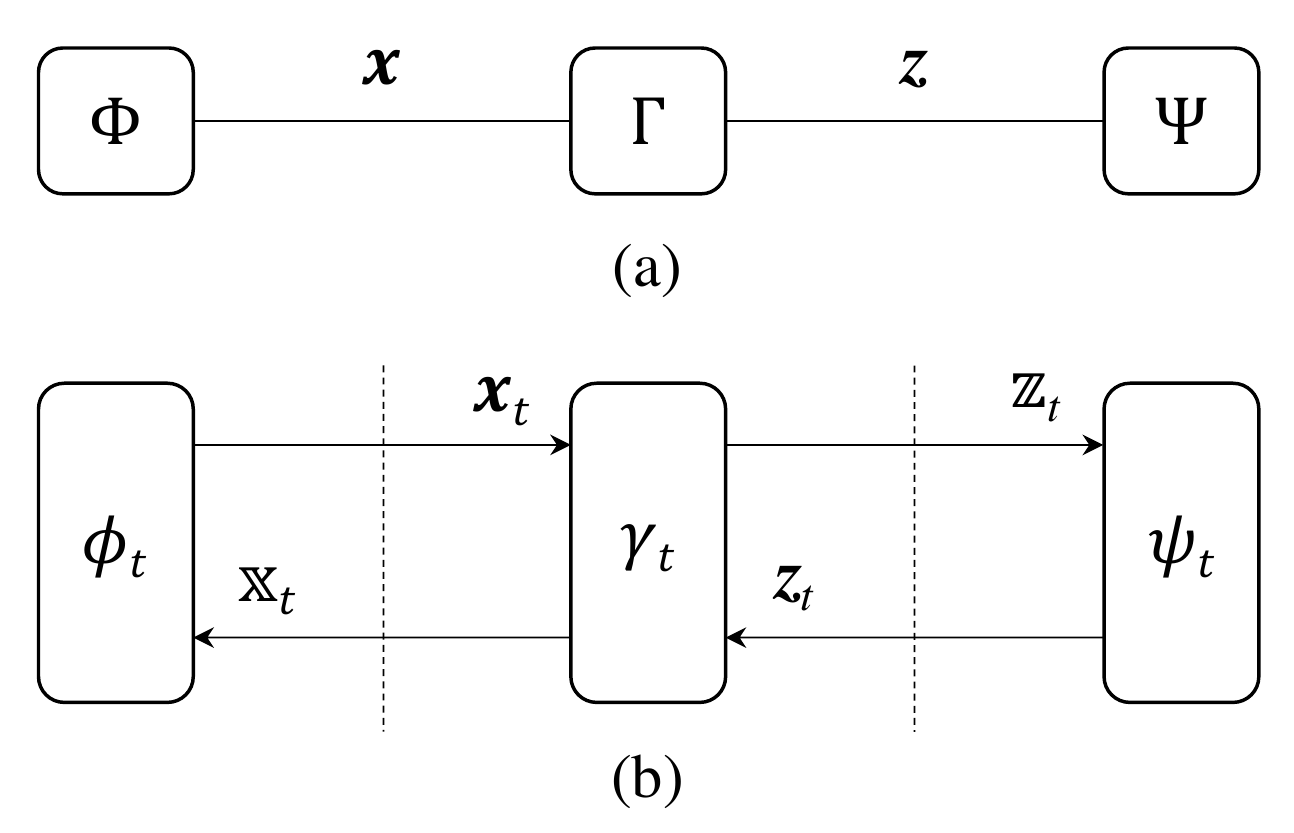}
\caption{\footnotesize Graphic illustrations for (a) a system model with three constraints $\Phi$, $\Gamma$ and $\Psi$, and (b) a non-memory iterative process (NMIP) involving three processors $\phi_t$, $\gamma_t$ and $\psi_t$ respectively corresponding to the constraints $\Phi$, $\Gamma$ and $\Psi$ in (a).}
\label{fig:system diagram1}
\end{figure}

{\emph{Non-memory Iterative Process (NMIP):}} Fig. \ref{fig:system diagram1}(b) illustrates an NMIP: Starting with $t=1$, 
\BS\label{Eqn:NMIP}\begin{align}
     {\rm NLE:}\quad &\left[ \!\!\begin{array}{c}
         \bf{x}_{t}   \\
        \bf{z}_{t}
    \end{array} \!\!\right]   =\left[ \!\!\begin{array}{c}
        \phi_t(\bf{\bb{x}}_t)   \\
        \psi_t(\bf{\bb{z}}_t)
    \end{array} \!\!\right],\\
 {\rm LE:}\quad  & \left[ \!\!\begin{array}{c}
         \bf{\bb{x}}_{t+1}   \\
        \bf{\bb{z}}_{t+1}
    \end{array} \!\!\right]   = \gamma_t(\bf{x}_t,\bf{z}_t) = \left[ \!\!\begin{array}{c}
       \gamma_t^x(\bf{x}_t,\bf{z}_t)   \\
        \gamma_t^z(\bf{x}_t,\bf{z}_t)
    \end{array} \!\!\right], 
\end{align}\ES
where $\gamma_t(\cdot)$, $\phi_t(\cdot)$ and $\psi_t(\cdot)$ process the  three  constraints $\Gamma$, $\Phi$ and $\Psi$ separately. We call \eqref{Eqn:NMIP} NMIP since both $\gamma_t(\cdot)$ $\phi_t(\cdot)$ and $\psi_t(\cdot)$ are memoryless, depending only on their current inputs $(\bf{x}_t, \bf{z}_t)$, $\bf{\bb{x}}_t$ and $\bf{\bb{z}}_t$, respectively. Furthermore, we assume that $\{\phi_t(\cdot)\}$ and $\{\psi_t(\cdot)\}$ are separable and Lipschitz continuous functions \cite{Berthier2017}. Let
\BS\begin{align}
    \bf{\bb{x}}_t&=\bf{x}+\bf{\bb{f}}_t,&\bf{\bb{z}}_t&=\bf{z}+\bf{\bb{s}}_t,\\
    \bf{x}_t&=\bf{x}+\bf{f}_t,
    &
    \bf{z}_t&=\bf{z}+\bf{s}_t,
\end{align}\ES
where $\bf{\bb{f}}_t$, $\bf{f}_t$, $\bf{\bb{s}}_t$ and $\bf{s}_t$ indicate the estimation errors with zero mean and variance:
\BS\begin{align}
    \bb{v}_t^x&=|\!\langle\bf{\bb{f}}_t,\bf{\bb{f}}_t\rangle\!|,&
    \bb{v}_t^z&=|\!\langle\bf{\bb{s}}_t,\bf{\bb{s}}_t\rangle\!|,\\
    v_t^x&=|\!\langle\bf{f}_t,\bf{f}_t\rangle\!|,&
    v_t^z&=|\!\langle\bf{s}_t,\bf{s}_t\rangle\!|.
\end{align}\ES

The orthogonality and asymptotic IID Gaussianity of NMIP was proved in \cite{Takeuchi2020, Rangan2019, pandit2020} based on the following error orthogonality.

\begin{lemma} [Orthogonality and Asymptotic IID Gaussianity]\label{Lem:IIDG}
Assume that $\bf{A}$ is unitarily invariant with $M, N\!\to\! \infty$  and the following orthogonality holds for NMIP: $\forall t\geq1$,
\BS\label{Eqn:error_orth}\begin{align}
    |\!\langle\bf{f}_t,\bf{\bb{f}}_t\rangle\!|&=0,&
    |\!\langle\bf{s}_t,\bf{\bb{s}}_t\rangle\!|&=0,\\
    |\!\langle\bf{\bb{f}}_{t+1},\bf{f}_t\rangle\!|&=0,&
    |\!\langle\bf{\bb{s}}_{t+1},\bf{s}_t\rangle\!|&=0,
\end{align}\ES 
Then, for $t\geq1$,
\begin{align} \label{Eqn:IIDG}
     v_{t}^{x}    &\overset{\rm a.s.}{=}  \tfrac{1}{N}  \|\phi_t(\bf{x}+ \sqrt{\bb{v}_t^x}\bf{\eta}^\bb{x}_t)-\bf{x}\|^2,\\
     \bb{v}_{t+1}^{x} &\overset{\rm a.s.}{=}  \tfrac{1}{N}  \| \gamma_t^x(\bf{x}_t,\bf{z}+ \sqrt{v_t^{z}}\bf{\eta}^z_t)-\bf{x}\|^2,\\
     \bb{v}_{t+1}^{z} &\overset{\rm a.s.}{=} \tfrac{1}{N}  \| \gamma_t^z(\bf{x}_t,\bf{z}+ \sqrt{v_t^{z}}\bf{\eta}^z_t)-\bf{z}\|^2,
\end{align} 
where $\bf{\eta}^\bb{x}_t\sim \mathcal{CN}(\bf{0},\bf{I}),\bf{\eta}^z_t\sim \mathcal{CN}(\bf{0},\bf{I})$, $\{\bf{\eta}^\bb{x}_t\}$ is independent of $\{\bf{\eta}^z_t\}$, and $\bf{\eta}^\bb{x}_t$ and $\bf{\eta}^z_t$ are independent of $\bf{x}$ and $\bf{z}$, respectively.  
\end{lemma}

\subsection{Review of GVAMP}
Throughout this paper, we assume that $\hat{\phi}_t(\cdot)$ and $\hat{\psi}(\cdot)$ are MMSE estimators:
\BS\begin{align} 
 \hat{\phi}_t(\bf{\bb{x}}_{t})&\equiv \mr{E}\{\bf{x}|\bf{\bb{x}}_{t},\Phi\},\\
  \hat{\psi}_t(\bf{\bb{z}}_{t})&\equiv \mr{E}\{\bf{z}|\bf{\bb{z}}_{t},\Psi\},
\end{align}\ES 
whose specific expressions are provided for the examples in section 7. 
As an instance of NMIP, GVAMP \cite{Schniter2016} is given in the following algorithm. The detailed derivation of GVAMP is provided in the next subsection.

\begin{framed}
 \emph{Generalized Vector AMP \cite{Schniter2016}:} Let $\rho_t=v_t^z/v_t^x$. Consider a LMMSE estimator of $\bf{x}$: 
\begin{align}\label{gamma_hat}
    \hat{\gamma}_t(\bf{x}_t)\equiv\bf{A}^{\mr H}(\rho_t\bf{I}+\bf{AA}^{\mr H})^{-1}(\bf{z}_t-\bf{Ax}_t).
\end{align}
 A GVAMP process is defined as: Starting with $t=1$, $v_1^{\bb x}=1$,$v_1^{\bb z}=\lambda_1$ and ${\bf{\bb x}}_{1}={\bf{\bb z}}_{1}=\bf{0}$, 
\BS\label{Eqn:GOAMP}
\begin{alignat}{2}
{\rm NLE:}\quad\quad  & \left[ \!\!\begin{array}{c}
         \bf{x}_t   \\
        \bf{z}_t
    \end{array} \!\!\right]  =\left[ \!\!\begin{array}{c}
        \phi_t(\bf{\bb{x}}_t)   \\
        \psi_t(\bf{\bb{z}}_t)
    \end{array} \!\!\right]=\left[ \!\!\begin{array}{c}
        \frac{1}{\epsilon^{\phi}_t}[\hat{\phi}_t(\bf{\bb{x}}_t)+(\epsilon^{\phi}_t-1)\bf{\bb{x}}_t] \\
        \frac{1}{\epsilon^{\psi}_t}[\hat{\psi}_t(\bf{\bb{z}}_t)+(\epsilon^{\psi}_t-1)\bf{\bb{z}}_t]
    \end{array} \!\!\right],\\
{\rm LE:}  \quad \quad & \left[ \!\!\begin{array}{c}
         \bf{\bb{x}}_{t+1}   \\
        \bf{\bb{z}}_{t+1} 
    \end{array} \!\!\right]
    = \left[ \!\!\begin{array}{c}
        \frac{1}{\delta\epsilon^\gamma_{t}} \hat{\gamma}_t(\bf{x}_t) + \bf{x}_t \\
        \tfrac{1}{1-{\epsilon}_{t}^\gamma}\left[  \bf{A}\big(\hat{\gamma}_t(\bf{x}_t) + \bf{x}_t\big) - {\epsilon}_{t}^\gamma \bf{z}_t \right]  
    \end{array}\!\! \right]
\end{alignat}\ES	 
with
\BS\label{Eqn:GOAMP-paras}\begin{align} 
    \epsilon^\phi_{t} &= 1-\tfrac{1}{N \bb{v}^{x}_t}\|\hat{\phi}_t(\bf{x}+\sqrt{\bb{v}^{x}_t} \bf{\eta} ) -\bf{x}\|^2, & {v}^x_{t} &=  \phi_{\mr{SE}}(\bb{v}^{x}_t) \equiv \big([\epsilon^\phi_{t}]^{-1} -1\big) \bb{v}^{x}_t,\\ 
    \epsilon^\psi_{t} &= 1-\tfrac{1}{M \bb{v}^{z}_t}\|\hat{\psi}_t(\bf{\bb{z}}_t) -\bf{z}\|^2, & {v}^z_{t} &=  \psi_{\mr{SE}}(\bb{v}^{z}_t) \equiv \big([\epsilon^\psi_{t}]^{-1} -1\big) \bb{v}^{z}_t,\\
    \epsilon^\gamma_{t}&=\tfrac{1}{M}{\rm tr}\{\bf{A}^{\mr H}(\rho_t\bf{I}+\bf{AA}^{\mr H})^{-1}\bf{A}\}, &  \left[ \!\!\!\begin{array}{c}
        \bb{v}_t^x   \\
        \bb{v}_t^z
    \end{array} \!\!\!\right] \!\! &=\gamma_{\mr{SE}}( v_t^{x},  v_t^{z}) \equiv \left[ \!\!\!\begin{array}{c}
       \big([\delta{\epsilon}_{t}^\gamma]^{-1} \!-\!1\big) v_t^{x}   \\
     \big([{\epsilon}_{t}^\gamma]^{-1} \!-\!1\big)^{-1}   v_t^{z}
    \end{array} \!\!\!\right],
\end{align}\ES
where $\bf{z}= \bf{\bb{z}}_t + \sqrt{\bb{v}^{z}_t} {\bf{\eta}}$, and $\bf{\eta}\sim {\cal{CN}}(\bf{0}, \bf{I})$ is independent of $\bf{x}$ and $\bf{\bb{z}}_t \sim {\cal{CN}}(\bf{0}, \sqrt{v_z-\bb{v}^z_t}\bf{I})$.  
\end{framed}

It is proved that GVAMP satisfies the orthogonality in \eqref{Eqn:error_orth}. Hence, the IID Gaussian property in \eqref{Eqn:IIDG} holds for GVAMP \cite{Takeuchi2020, Rangan2019, pandit2020}, which results in  the following state evolution. %{pandit2020,Reeves2017,Gabrie2018}

\emph{State Evolution:} The iterative performance can be tracked by the following state evolution: Starting with $t=1$, $\bb{v}_1^{x}=1$ and $\bb{v}_1^{z}=\lambda_1$,
\BS\begin{align}
     {\rm NLE:} \qquad\;\; (v_t^x, v_t^z) & = \big(\phi_{\rm SE}(\bb{v}_{t}^x), \psi_{\rm SE}(\bb{v}_{t}^z)\big),\\
    {\rm LE:} \quad  (\bb{v}_{t+1}^{x},\bb{v}_{t+1}^{z})&=\gamma_{\rm SE}(v_t^x,v_t^z).
\end{align}\ES
where $\gamma_{\rm SE}(\cdot)$, $\phi_{\rm SE}(\cdot)$ and $\psi_{\rm SE}(\cdot)$ are defined in \eqref{Eqn:GOAMP-paras}.

The following lemma was derived in \cite{pandit2020,Reeves2017,Gabrie2018}.

\begin{lemma}[Bayes Optimality]\label{Lem:optimality}
Assume that $M,\!N\!\to\!\infty$ with a fixed $\delta\!=\!\!M\!/\!N$, and GVAMP satisfies the unique fixed point condition. Then, GVAMP can achieve the minimum (i.e., Bayes-optimal) MSE as predicted by replica method for unitarily-invariant matrices.
\end{lemma}

In general, the NLE in GVAMP is a symbol-by-symbol estimator, whose time complexity is as low as ${\cal O}(N+M)$. The complexity of GVAMP is dominated by LMMSE-LE, which costs $\mathcal{O}(M^2N\!+\!M^3)$ time complexity per iteration for matrix multiplication and matrix inversion. Therefore, to reduce the complexity, it is desired to design a low-complexity Bayes-optimal LE for the message passing algorithm.

\subsection{Derivation of GVAMP Algorithm}
In \cite{Schniter2016},  the generalized linear model (GLM) was rewritten to a standard linear model (SLM) with $\bf{x}^{\rm new}\equiv[\bf{x}\ \bf{z}]^{\rm T}$, and then GVAMP was derived similarly as VAMP using vector composition and limitation. In contrast, we first estimate $\bf{x}$, and then find $\bf{z}$ based on the estimated $\bf{x}$ using a simple conversion expression between the LMMSE estimations of $\bf{x}$ and $\bf{z}$. Compared to previous work, our approach is more brief, direct and easy to understand. The specific derivations is as follows.

\subsubsection{Derivation of Orthogonal LMMSE-LE}
We derive the linear MMSE (LMMSE) estimation for the linear constraint $\bm{z} = \bm{Ax}$. We assume that the receiver knows the respective Gaussian observations ${\bm x}_t$ and ${\bm z}_t$ of $\bm{x}$ and $\bm{z}$. The problem is described as
\begin{align}
    \begin{cases}
            \bm{z} =\bm{Ax},\\
            {\bm z}_t={\bm z}  + {\bm s}_t,\quad\,
            {\bm s}_t\sim\mathcal{N}(\bf{0},v_t^z{\bm I}),\\
            {\bm x} =  {\bm x}_t + {\bm f}_t,\quad
            {\bm f}_t\sim\mathcal{N}(\bf{0},v_t^x{\bm I}).
    \end{cases}
\end{align}  
The goal is to estimate $P(\bm{x}|\bm{z}_t,\bm{x}_t)$ and $P(\bm{z}|\bm{z}_t,\bm{x}_t)$  using Bayes' rule:
\BS\begin{align}
    P(\bm{x}|\bm{z}_t,\bm{x}_t) = \frac{P(\bm{z}_t,\bm{x}_t|\bm{x})P(\bm{x})}{P(\bm{z}_t,\bm{x}_t)} \propto P(\bm{z}_t,\bm{x}_t|\bm{x})P(\bm{x}),\label{Eqn:x_post}\\
    P(\bm{z}|\bm{z}_t,\bm{x}_t) = \frac{P(\bm{z}_t,\bm{x}_t|\bm{z})P(\bm{z})}{P(\bm{z}_t,\bm{x}_t)} \propto P(\bm{z}_t,\bm{x}_t|\bm{z})P(\bf{z}).
\end{align}\ES

\emph{1) LMMSE Estimation of $\bf{x}$:} Since $\bm{z}_t - \bm{x} - \bm{x}_t$ is a Markov Chain, from \eqref{Eqn:x_post},  we have
\begin{align}
    P(\bm{x}|\bm{z}_t,\bm{x}_t)   P(\bm{x})\propto P(\bm{z}_t|\bm{x}) P(\bm{x}|\bm{x}_t). 
\end{align} 
where
\BS\begin{align}
    P({\bm z}_t|\bm{x}) &\propto e^{-\frac{ (\bm{z}_t-\bm{Ax})^{\rm H}(\bm{z}_t-\bm{Ax})}{2v_t^z}} \propto e^{-\frac{\bm{x}^{\rm H}\bm{A}^{\rm H}\!\bf{A}\bf{x}-2\bm{x}^{\rm H}\bf{A}^{\rm H}  \bf{z}_t}{2v_t^z}},\\
   P(\bm{x}|\bm{x}_t) &\propto e^{-\frac{\bm{(x}-\bf{x}_t)^{\rm H}(\bf{x}-\bf{x}_t)}{2v_t^x}} \propto e^{-\frac{\bm{x}^{\rm H}\bf{x} -2\bm{x}^{\rm H}\bf{x}_t}{2 v_t^x}}.
\end{align} \ES 
Therefore,
\begin{align}
   P(\bm{x}|\bm{z}_t,\bm{x}_t)  &\propto  e^{-\frac{\bm{x}^{\rm H}\bm{A}^{\rm H}\!\bf{A}\bf{x}-2\bm{x}^{\rm H}\bf{A}^{\rm H}  \bf{z}_t}{2v_t^z}} e^{-\frac{\bm{x}^{\rm H}\bf{x} -2\bm{x}^{\rm H} \bf{x}_t}{2 v_t^x}}\\\nonumber
    &= e^{-\frac{1}{2}\bm{x}^{\rm H}[(v_t^z)^{-1} \bm{A}^{\rm H} \!\!\bm{A} +(v_t^x)^{-1}\bf{I}]\bm{x} \,+\, \bm{x}^{\rm H}[(v_t^z)^{-1}{\bm{A}}^{\rm H}  \!\bm{z}_t + (v_t^x)^{-1}\bm{x}_t]},
\end{align} 
which follows a new Gaussian distribution: 
\begin{align}
    \!P(\bm{x}^{\gamma}_{\rm post}|\bm{x}) & \!\propto  \!e^{-\frac{(\bm{x} - \bm{x}^{\gamma}_{\rm post})^{\!\rm H}  \bf{V}_{x^{\gamma}_{\rm post}}^{-1}(\bm{x}-\bf{x}^{\gamma}_{\rm post})}{2}} \! \propto\! e^{-\frac{1}{2} \bm{x}^{\!\rm H}  \bf{V}_{x^{\gamma}_{\rm post}}^{-1}\! \bm{x} \,+\,  \bm{x}^{\!\rm H}   \bf{V}_{x^{\gamma}_{\rm post}}^{-1}\!\bm{x}^{\gamma}_{\rm post}}.
\end{align}
Therefore, we have  
\BS\begin{align} 
         \bf{V}_{x^{\gamma}_{\rm post}} &= [(v_t^z)^{-1}\bm{A}^{\rm H}\bm{A} + (v_t^x)^{-1}\bm{I}]^{-1}\\
         \bm{x}^{\gamma}_{\rm post} &= \bf{V}_{x^{\gamma}_{\rm post}}[(v_t^z)^{-1}\bm{A}^{\rm H}\bm{z}_t + (v_t^x)^{-1}\bm{x}_t]. 
\end{align}\ES
   
For over-load case, that $M<N$, the above expressions can be rewritten as
\begin{align}
    \bf{x}^{\gamma}_{\rm post}=\bf{x}_t + v_t^x\bf{A}^{\rm H}(v_t^z\bf{I} + v_t^x\bf{AA}^{\rm H})^{-1}(\bf{z}_t - \bf{Ax}_t).
\end{align}
Let define $\hat{\gamma}_t(\bf{x}_t)\equiv\bf{A}^{\mr H}(\rho_t\bf{I}+\bf{AA}^{\mr H})^{-1}(\bf{z}_t-\bf{Ax}_t)$ with $\rho_t=v_t^z/v_t^x$. We then output the orthogonal messages for $\bf{x}$:
\begin{align}
   \bf{\bb{x}}_{t+1} = \frac{1}{\delta\epsilon^\gamma_{t}} \hat{\gamma}_t(\bf{x}_t) + \bf{x}_t, 
\end{align}
where
\begin{align}
    \epsilon^\gamma_{t}=\tfrac{1}{M}{\rm tr}\{\bf{A}^{\mr H}(\rho_t\bf{I}+\bf{AA}^{\mr H})^{-1}\bf{A}\},
\end{align}
ensuring the trace of coefficient of $\bf{x}_t$ is equal to zero.

\emph{2) LMMSE Estimation of $\bf{z}$:} Following the linear constraint $\bf{z}=\bf{Ax}$, we have
\BS\begin{align} 
    {\bm z^{\gamma}_{\rm post}} &={\rm{E}} \{\bf{z} | \bm{z}_t,\bm{x}_t\}=\bf{A}{\rm{E}} \{\bf{x} | \bm{z}_t,\bm{x}_t\} =\bm{A}{\bm x^{\gamma}_{\rm post}},\\ 
    {\bf{V}}_{z^{\gamma}_{\rm post}} &\!=\!{\rm{Var}} \{\bf{z} | \bm{z}_t,\bm{x}_t\}\!=\!\bf{A}{\rm{Var}} \{\bf{x} | \bm{z}_t,\bm{x}_t\} \bm{A}^{\rm{H}}\!=\! \bm{A}{\bm{V}_{x^{\gamma}_{\rm post}}}\bm{A}^{\rm{H}}. 
\end{align}\ES

\emph{Proof:}
\BE
{\rm{E}} \{\bf{z} | \bm{z}_t,\bm{x}_t\}={\rm{E}} \{\bf{Ax} | \bm{z}_t,\bm{x}_t\}=\bf{A}{\rm{E}} \{\bf{x} | \bm{z}_t,\bm{x}_t\}=\bf{A}\bf{x}^{\gamma}_{\rm post},
\EE
\begin{align}
{\rm{Var}} \{\bf{z} | \bm{z}_t,\bm{x}_t\}&={\rm{E}} \{\bf{zz}^{\rm H} | \bm{z}_t,\bm{x}_t\} - {\rm{E}} \{\bf{z} | \bm{z}_t,\bm{x}_t\}{\rm{E}} \{\bf{z} | \bm{z}_t,\bm{x}_t\}^{\rm H}\nonumber\\
&={\rm{E}} \{\bf{Ax}\bf{x}^{\rm H}\bf{A}^{\rm H} | \bm{z}_t,\bm{x}_t\} - \bf{A}\bf{x}^{\gamma}_{\rm post}{\bf{x}^{\gamma}_{\rm post}}^{\rm H}\bf{A}^{\rm H}\nonumber\\
&=\bf{A}[{\rm{E}} \{\bf{x}\bf{x}^{\rm H} | \bm{z}_t,\bm{x}_t\} - \bf{x}^{\gamma}_{\rm post}{\bf{x}^{\gamma}_{\rm post}}^{\rm H}]\bf{A}^{\rm H}\nonumber\\
&=\bf{A}{\rm{Var}} \{\bf{x} | \bm{z}_t,\bm{x}_t\} \bm{A}^{\rm{H}}\nonumber\\
&=\bm{A}{\bm{V}_{x^{\gamma}_{\rm post}}}\bm{A}^{\rm{H}}.
\end{align}

Let define ${\bm z^{\gamma}_{\rm post}}=\bm{A}{\bm x^{\gamma}_{\rm post}}\equiv\bf{A}\big(\hat{\gamma}_t(\bf{x}_t) + \bf{x}_t\big)$. We then output the orthogonal messages for $\bf{z}$:
\begin{align}
    \bf{\bb{z}}_{t+1} = \tfrac{1}{1-{\epsilon}_{t}^\gamma}\left[  \bf{A}\big(\hat{\gamma}_t(\bf{x}_t) + \bf{x}_t\big) - {\epsilon}_{t}^\gamma \bf{z}_t \right].
\end{align}

\subsubsection{Derivation of Orthogonal NLE}
The following $\hat{\phi}_t(\cdot)$ and $\hat{\psi}(\cdot)$ are the MMSE NLEs of $\bf{x}$ and $\bf{z}$ respectively. Their specific expressions are provided for the examples in section 7.
\BS\begin{align} 
 \hat{\phi}_t(\bf{\bb{x}}_{t})&\equiv \mr{E}\{\bf{x}|\bf{\bb{x}}_{t},\Phi\},\\
  \hat{\psi}_t(\bf{\bb{z}}_{t})&\equiv \mr{E}\{\bf{z}|\bf{\bb{z}}_{t},\Psi\}.
\end{align}\ES 
\begin{itemize}
    \item The orthogonal NLE of $\bf{x}$ is given by orthogonalization:
\BS\begin{align}
    \bf{x}_t &= (v_{x_{\rm post}}^{-1} - v_{x_{\rm pri}}^{-1})^{-1} \big[v_{x_{\rm post}}^{-1}\hat{\phi}_t(\bb{\bf{x}}_t) - v_{x_{\rm pri}}^{-1}\bb{\bf{x}}_t\big]\\
    &= \frac{v_{x_{\rm pri}}}{v_{x_{\rm pri}} - v_{x_{\rm post}}} \big[\hat{\phi}_t(\bb{\bf{x}}_t) - \frac{v_{x_{\rm post}}}{v_{x_{\rm pri}}}\bb{\bf{x}}_t\big]\\
    &= \frac{1}{\epsilon^\phi_{t}}\big[\hat{\phi}_t(\bb{\bf{x}}_t) + (\epsilon^\phi_{t}-1)\bb{\bf{x}}_t\big],
\end{align}\ES
where
\begin{align}
    \epsilon^\phi_{t} = 1 - \frac{v_{x_{\rm post}}}{v_{x_{\rm pri}}},\hspace{0.3cm} v_{x_{\rm pri}} = \bb{v}_t^{x},\hspace{0.3cm} v_{x_{\rm post}} = \mr{var}\{\bf{x}|\bf{\bb{x}}^{t},\Phi\}=
    \tfrac{1}{N}\|\hat{\phi}_t(\bf{x}+\sqrt{\bb{v}^{x}_t} \bf{\eta} ) -\bf{x}\|^2
\end{align}
with $\bf{\eta}\sim {\cal{CN}}(\bf{0}, \bf{I})$ being independent of $\bf{x}$.

\item Similarly, the orthogonal NLE of $\bf{z}$ is given by
\begin{align}
    \bf{z}_t = \frac{1}{\epsilon^\psi_{t}}\big[\hat{\psi}_t(\bb{\bf{z}}_t) + (\epsilon^\psi_{t}-1)\bb{\bf{z}}_t\big]
\end{align}
with
\begin{align}
    \epsilon^\psi_{t} = 1 - \frac{v_{z_{\rm post}}}{v_{z_{\rm pri}}},\hspace{0.3cm} v_{z_{\rm pri}} = \bb{v}_t^{z},\hspace{0.3cm} v_{z_{\rm post}} = \mr{var}\{\bf{z}|\bf{\bb{z}}^{t},\Psi\}=
    \tfrac{1}{M}\|\hat{\psi}_t(\bf{\bb{z}}_t) -\bf{z}\|^2,
\end{align}
where $\bf{z}= \bf{\bb{z}}_t + \sqrt{\bb{v}^{z}_t} {\bf{\eta}}$, and $\bf{\eta}\sim {\cal{CN}}(\bf{0}, \bf{I})$ is independent of $\bf{\bb{z}}_t \sim {\cal{CN}}(\bf{0}, \sqrt{v_z-\bb{v}^{z}_t}\bf{I})$.
\end{itemize}

\section{Proof of Orthogonality and AIIDG of BO-GMAMP}%2
In this section, we first prove the orthogonality between estimation errors through equivalent transformation, and then derive the AIIDG of BO-GMAMP based on the Theorem 1 in main text. Let $\bf{W}_t=\bf{A}^{\rm H}\bf{B}^t\bf{A}$, $\widehat{\bf{W}}_t=\bf{AA}^{\rm H}\bf{B}^t$ and  $w_t  \equiv \tfrac{1}{M}{\rm tr}\{\bf{W}_t\}=\tfrac{1}{M}{\rm tr}\{\widehat{\bf{W}}_t\}$, we define
\BS\label{Eqn:orth_parameters}\begin{align}
    \vartheta_{t,i}&\equiv\xi_i\textstyle\prod_{\tau=i+1}^t\theta_\tau,&\Bar{w}_i&\equiv\lambda^{\dagger}w_{i}-w_{i+1},\\
   p_{t,i}& \equiv\vartheta_{t,i}w_{t-i},&\Bar{w}_{i,j}&\equiv \Bar{w}_{i+j} - w_iw_j,\\
   \bb{c}^x_t& =1/\textstyle\sum_{i=1}^t p_{t,i} , &\tilde{w}_{i,j}&\equiv   \Bar{w}_{i+j}/\delta - w_i w_j,\\ 
 \beta_t &= \tfrac{\xi_t}{\theta_t} - 1/\bb{c}^x_t, &  \bar{\Bar{w}}_{i}&\equiv\lambda^{\dagger}\Bar{w}_i- \Bar{w}_{i+1}. 
\end{align}\ES

\begin{proposition}\label{Pro:MLE_expand} 
The $\bf{\bb{x}}_{t+1}$ and its error can be expanded to
\BS\label{Eqn:MLE-x}\begin{align}
    \bf{\bb{x}}_{t+1} &= \bb{c}^x_t \left( \textstyle\sum_{i=1}^t  \bf{G}_{t,i}\bf{z}_i  - \textstyle\sum_{i=1}^t \bf{\bb{H}}_{t,i} \bf{x}_i \right), \label{Eqn:MLE-xa}\\
    \bf{\bb{f}}_{t+1} &=\bb{c}^x_t \left( \textstyle\sum_{i=1}^t  \bf{G}_{t,i}\bf{s}_i  - \textstyle\sum_{i=1}^t \bf{\bb{H}}_{t,i} \bf{f}_i \right),\label{Eqn:MLE-xb}
\end{align} %\xi_t (\theta^{-1}_t\bf{I}-\bf{A}\bf{A}^{\rm H})\bf{Ax}_t
where
\begin{align} 
    \bf{G}_{t,i}&\equiv\vartheta_{t,i}\delta^{-1}\bf{A}^{\rm H}\bf{B}^{t-i},\\ 
    \bf{\bb{H}}_{t,i}&\equiv\vartheta_{t,i}(\delta^{-1}\bf{W}_{t-i}- w_{t-i}\bf{I}), 
\end{align}\ES 
and $\bf{\bb{z}}_{t+1}$  and its error can be expanded to
\BS\label{Eqn:MLE_z}\begin{align}   
    \bf{\bb{z}}_{t+1}&= {\bb{c}^z_t}  \left( \beta_t \bf{z} + \tilde{\bf{\bb{s}}}_{t}  \right),  \label{Eqn:MLE_za} \\
    \bf{\bb{s}}_{t+1} &= ({\bb{c}^z _t}\beta_t -1)\bf{z} +{\bb{c}^z _t}\tilde{\bf{\bb{s}}}_{t},\label{Eqn:MLE_zb}
\end{align}
where $\bb{c}^z_t$ is optimized in next section, $\beta_t$ is defined in \eqref{Eqn:orth_parameters} and
\begin{align}
\tilde{\bf{\bb{s}}}_{t} &\equiv \textstyle\sum_{i=1}^t  \big(\bf{\bb{G}}_{t,i} \bf{s}_i -  \bf{H}_{t,i} \bf{f}_i\big) + \tfrac{\xi_t}{\theta_t}\bf{A}\bf{f}_t, \\ 
    \bf{\bb{G}}_{t,i}&\equiv \vartheta_{t,i}(\widehat{\bf{W}}_{t-i}-w_{t-i} \bf{I}),\\
     \bf{H}_{t,i}&\equiv \vartheta_{t,i}\bf{A}\bf{W}_{t-i}.%\\ 
   %\tilde{\bf{H}}_{t,t}&=   (\tfrac{\xi_t}{\theta_t}\bf{A}\bf{I}-\vartheta_{t,t}\bf{A}\bf{A}^{\rm H})\bf{A}.
\end{align}\ES   
\end{proposition}

See next subsection for further details.

\subsection{Proof of Proposition \ref{Pro:MLE_expand}}
According to the main text, a BO-GMAMP process is defined as: Starting with $t=1$ and $\hat{\bf{\bb{x}}}_{0}= {\bf{\bb{x}}}_{1}= \hat{\bf{\bb{z}}}_{1}= \bf{0}$, 
\BS\label{Eqn:GMAMP}
\begin{alignat}{2}
{\rm NLE:}&& \;\; \left[ \!\!\begin{array}{c}
         \bf{x}_{t}   \\
        \bf{z}_{t}
    \end{array} \!\!\right]  &=\left[ \!\!\begin{array}{c}
         \Bar{\phi}_t(\bf{\bb{x}}_t)   \\
        \Bar{\psi}_t(\bf{\bb{z}}_t)
    \end{array} \!\!\right]=\left[ \!\!\begin{array}{c}
         \scaleto{\zeta}{8pt}_{t, l_t} \phi_t(\bf{\bb{x}}_t)+\textstyle\sum_{i=1}^{l_t-1} \scaleto{\zeta}{8pt}_{t,i}\bf{x}_{t-l_t+i}, \vspace{2mm} \\
        \varrho_{t, l_t} \psi_t(\bf{\bb{z}}_t)+\textstyle\sum_{i=1}^{l_t-1} \varrho_{t,i}\bf{z}_{t-l_t+i}
    \end{array} \!\!\right],\label{Eqn:NLE}\\
{\rm MLE:} && \quad \left[ \!\!\begin{array}{c}
         \bf{\bb{x}}_{t+1}   \\
        \bf{\bb{z}}_{t+1}
    \end{array} \!\!\right]
    &\!=\! \gamma_t \big(\bf{X}_{1\rightarrow t}, \bf{Z}_{1\rightarrow t}\big) \!=\!   \left[ \!\!\!\begin{array}{c}
        \bb{c}^x_t\big(  \delta^{-1} \hat{\bf{\bb{x}}}_t  +  \textstyle\sum_{i=1}^t p_{t,i} \bf{x}_i  \big) \\
        \bb{c}^z_t\big[ \bf{A} \big( \hat{\bf{\bb{x}}}_t +\frac{\xi_t}{\theta_t}\bf{x}_t\big)   - \textstyle\sum_{i=1}^t p_{t,i} \bf{z}_i \big] 
    \end{array}\!\!\! \right], 
    \label{Eqn:MLE}
\end{alignat}\ES
where $\bf{X}_{1\rightarrow t}=[\bf{x}_1 \cdots \bf{x}_t]$ and $\bf{Z}_{1\rightarrow t}=[\bf{z}_1\cdots\bf{z}_t]$.

By expanding \eqref{Eqn:MLE} one by one, we can obtain \eqref{Eqn:MLE-xa} directly. Following \eqref{Eqn:MLE-xa}, the errors of  MLE are given by 
\BS\begin{align} 
     \bf{\bb{f}}_{t+1}&=\bb{c}^x_t \textstyle\sum\limits_{i=1}^t\big(   \bf{G}_{ti}\bf{z}_i-  \bf{\bb{H}}_{ti}\bf{x}_i \big)-\bf{x}\\
     &=   \bb{c}^x_t  \Big[   \textstyle\sum\limits_{i=1}^t\bf{G}_{ti}(\bf{Ax}+\bf{s}_i) -  \bf{\bb{H}}_{ti}(\bf{x}+\bf{f}_i) \Big]- \bf{x}\label{Eqn:error_b}\\
&=   \bb{c}^x_t    \Big[\!   \textstyle\sum\limits_{i=1}^t \big(  \bf{G}_{ti}{\bf{s}_i} - \bf{\bb{H}}_{t i} \bf{f}_i    \big) + \big[\textstyle\sum\limits_{i=1}^t (\bf{G}_{ti}\bf{A} -  \bf{\bb{H}}_{ti}) - \tfrac{1}{\bb{c}^x_t}\bf{I} \big] \bf{x} \Big]\label{Eqn:error_c}\\  
    &= \bb{c}^x_t\textstyle\sum\limits_{i=1}^t\big(\bf{G}_{ti}\bf{s}_i - \bf{\bb{H}}_{ti}\bf{f}_i\big). 
\end{align}\ES
Therefore, we obtain \eqref{Eqn:MLE-x}.  Furthermore,
\BS\begin{align}
    \bf{\bb{z}}_{t+1} &= \bb{c}^z_t \big[\textstyle\sum_{i=1}^t \big(\bf{\bb{G}}_{t,i}\bf{z}_i - \bf{H}_{t,i}\bf{x}_i\big) + \frac{\xi_t}{\theta_t}\bf{Ax}_t\big]\\
    &=   \bb{c}^z_t  \Big[   \textstyle\sum_{i=1}^t\big[\bf{\bb{G}}_{ti}(\bf{z}+\bf{s}_i) -  \bf{H}_{ti}(\bf{x}+\bf{f}_i) \big] + \frac{\xi_t}{\theta_t}\bf{A}(\bf{x}+\bf{f}_t) \Big]  \\
&=   \bb{c}^z_t    \Big[\!   \textstyle\sum_{i=1}^t \big(  \bf{\bb{G}}_{ti}{\bf{s}_i} - \bf{H}_{t i} \bf{f}_i    \big) + \frac{\xi_t}{\theta_t}\bf{Af}_t  +  \big[\underbrace{\textstyle\sum_{i=1}^t (\bf{\bb{G}}_{ti} -  \vartheta_{ti}\widehat{\bf{W}}_{t-i}) }_{=-1/\bb{c}^x_t\bf{I}}+ \frac{\xi_t}{\theta_t} \bf{I}\big] \bf{z} \Big] \\  
    &=\bb{c}^z_t\big[\textstyle\sum_{i=1}^t\big(\bf{\bb{G}}_{ti}\bf{s}_i - \bf{H}_{ti} \bf{f}_i\big) + \frac{\xi_t}{\theta_t}\bf{Af}_t+(\frac{\xi_t}{\theta_t}-1/\bb{c}^x_t)\bf{z}\big]\\  
    &={\bb{c}^z _t}  \left( \beta_t \bf{z} + \tilde{\bf{\bb{s}}}_{t}  \right).
\end{align}\ES
Thus, we have \eqref{Eqn:MLE_za}. In addition, \eqref{Eqn:MLE_zb} follows \eqref{Eqn:MLE_za}. Therefore, we obtain \eqref{Eqn:MLE_z}.

\subsection{Proof of Theorem 2 in Main Text}
$\bf{A}$ is unitarily invariant, i.e., $\bf{A}=\bf{U\Sigma V}^{\rm H}$, where $\bf{U}, \bf{\Sigma}$ and $\bf{V}$ are independent, and $\bf{U}$ and $\bf{V}$ are Haar distributed. Therefore, when $M, N\!\to\! \infty$, we have $\{\bf{x}, \bf{\bb x}_t, \bf{x}_t, \bf{\bb f}_t, \bf{f}_t \}$ is independent of $\{\bf{z}, \bf{\bb z}_t, \bf{z}_t, \bf{\bb s}_t, \bf{s}_t \}$, and they are column-wise IID.
 
It is easy to verify that, for $1\leq t'\leq t$,
\BS\label{Eqn:LE_orth}\begin{align}
    {\rm tr}\{\bf{\bb{H}}_{t,t'}\}&=0,\\
    {\rm tr}\{\bf{\bb{G}}_{t,t'}\}&=0. 
\end{align}\ES
In addition, $\{\bf{\bb f}_t, \bf{f}_t \}$ is independent of $\{\bf{\bb s}_t, \bf{s}_t \}$ and their rows are respectively IID. Therefore, we have the desired orthogonality, for $1\leq t'\leq t$,
 \begin{align}\label{Eqn:MLE_orth_proof}
    |\!\langle\bf{\bb{f}}_{t+1},\bf{f}_{t'}\rangle\!| =0,\qquad\qquad
    |\!\langle\bf{\bb{s}}_{t+1},\bf{s}_{t'}\rangle\!|=0.
\end{align} 
Second, since $\phi_t(\cdot)$ and $\psi_t(\cdot)$ are the same as GVAMP, we have: $ \forall t\geq 1 $,
\BE
 |\!\langle (\phi_t(\bf{\bb x}_t)-\bf{x}),\bf{\bb{f}}_t\rangle\!|  = 0, \qquad \qquad |\!\langle (\psi_t(\bf{\bb z}_t)-\bf{z}),\bf{\bb{s}}_t\rangle\!|  = 0.
\EE
Then, following \eqref{Eqn:LE_orth} in the previous iterations, we have: $t-l_t\le t'\le t$,
\BE
 |\!\langle\bf{f}_{t'},\bf{\bb{f}}_t \rangle\!|=0, \qquad\qquad |\!\langle\bf{s}_{t'},\bf{\bb{s}}_t \rangle\!|=0.
\EE
Hence, the following orthogonality holds.  
\BE\label{Eqn:NLE_orth_proof}
|\!\langle\bf{f}_t, \bf{\bb f}_{t} \rangle\!| = 0, \qquad\qquad |\!\langle\bf{s}_t, \bf{\bb s}_{t} \rangle\!| = 0.
\EE 
Therefore, we prove the orthogonality of BO-GMAMP, based on which the IID Gaussianity of BO-GMAMP can be obtained from Theorem 1 of the main text \cite{Takeuchi2020, Rangan2019, pandit2020}. Thus, we complete the proof of Theorem 2.

\section{Derivation of State Evolution of BO-GMAMP}%3
Let define the error covariance matrix of $\{\bf{x}_{t-l_t+1},...,\bf{x}_{t-1},\phi_t(\bf{\bb{x}}_t)\}$ (inputs of $\Bar{\phi}_t$) as
\begin{align}\label{Eqn:V_phi}
    \bf{V}_{t}^{\phi}\equiv
    \left[
    \begin{array}{cccc}
        v_{t-l_t+1, t-l_t+1}^{x} & \cdots & v_{t-l_t+1, t-1}^{x} & v_{t-l_t, t}^{\phi}\\
        \vdots & \ddots & \vdots & \vdots\\
        v_{t-1, t-l_t+1}^{x} & \cdots & v_{t-1 ,t-1}^{x} & v_{t-1, t}^{\phi}\\
        v_{t, t-l_t+1}^{\phi} & \cdots & v_{t, t-1}^{\phi} & v_{t, t}^{\phi}
    \end{array}
\right]_{l_t\times l_t},
\end{align}
and the error covariance matrix of $\{\bf{z}_{t-l_t+1},...,\bf{z}_{t-1},\psi_t(\bf{\bb{z}}_t)\}$ (inputs of $\Bar{\psi}_t$) as
\begin{align}\label{Eqn:V_psi}
    \bf{V}_{t}^{\psi}\equiv
    \left[
    \begin{array}{cccc}
        v_{t-l_t+1, t-l_t+1}^{z} & \cdots & v_{t-l_t+1, t-1}^{z} & v_{t-l_t+1, t}^{\psi}\\
        \vdots & \ddots & \vdots & \vdots\\
        v_{t-1, t-l_t+1}^{z} & \cdots & v_{t-1, t-1}^{z} & v_{t-1, t}^{\psi}\\
        v_{t, t-l_t+1}^{\psi} & \cdots & v_{t, t-1}^{\psi} & v_{t, t}^{\psi}
    \end{array}
\right]_{l_t\times l_t},
\end{align}
where
\BS\label{v_nodam}\begin{align}
    &v_{t,t'}^{\phi}\equiv|\!\langle[\phi_t(\bf{\bb{x}}_t)-\bf{x}],\bf{f}_{t'}\rangle\!|,\\
    &v_{t,t'}^{\psi}\equiv|\!\langle[\psi_t(\bf{\bb{z}}_t)-\bf{z}],\bf{s}_{t'}\rangle\!|
\end{align}\ES

Then, $\Bar{\phi}_{\rm SE}(\cdot)$, $\Bar{\psi}_{\rm SE}(\cdot)$ and $\gamma_{\rm SE}(\cdot)$ are given as below.

\begin{itemize}
\item $\Bar{\phi}_{\rm SE}(\cdot)$ is given by
    \BS\begin{align}
    v_{t, t'}^{x}=
    \begin{cases}
    \scaleto{\zeta}{8pt}_{t, l_t}v_{t,t'}^{\phi} + \sum_{i=1}^{l_t-1}\scaleto{\zeta}{8pt}_{t,i}v_{t-l_t+i, t'}^{x},\hspace{0.2cm}1\leq t' < t\\
    \scaleto{\boldsymbol{\zeta}}{8pt}_t^{\rm T}\bf{V}_{t}^{\phi}\scaleto{\boldsymbol{\zeta}}{8pt}_t,\hspace{3.3cm}t'=t
    \end{cases},
    \end{align}
    where $\bf{V}_{t}^{\phi}$ is defined in \eqref{Eqn:V_phi} and
    \BE\label{Eqn:SE_phi} 
    \bf{v}^{\phi}_{t} \equiv {\rm E}\big\{ [\phi_t(x\!+\!\eta_{t}) -x]^* [\tilde{\bf{x}}_{t}-x\bf{1}]\big\},  
    \EE\ES
    with $\tilde{\bf{x}}_{t}=[x_1 \cdots \; x_{t-1} \;  \phi_t(x\!+\!\eta_{t})]^{\rm T}$ and $[{\eta}_1 \cdots {\eta}_{t}]\sim \mathcal{CN}(\bf{0},\bf{V}_{t}^x)$ being independent of $x\sim P_X(x)$. The expectation can be evaluated by Monte Carlo method.
\item $\Bar{\psi}_{\rm SE}(\cdot)$ is given by
    \BS\begin{align}
    v_{t,  t'}^{z}=
    \begin{cases}
    \varrho_{t,  l_t}v_{t,  t'}^{\psi} + \sum_{i=1}^{l_t-1}\varrho_{t,i}v_{t-l_t+i,  t'}^{z},\hspace{0.2cm}1\leq t' < t\\
    \bf{\varrho}_t^{\rm T}\bf{V}_{t}^{\psi}\bf{\varrho}_t,\hspace{3.3cm}t'=t
    \end{cases},
    \end{align}
    where $\bf{V}_{t}^{\psi}$ is defined in \eqref{Eqn:V_psi} and
    \BE\label{Eqn:SE_psi} 
    \bf{v}^{\psi}_{t} \equiv {\rm E}\big\{ [\psi_t(\hat{z}_{t}) -z]^* [\tilde{\bf{z}}_{t}-z\bf{1}]\big\},  
    \EE\ES
    with $\tilde{\bf{z}}_{t}=[z_1 \cdots \; z_{t-1} \;  \psi_t(\hat{z}_{t})]^{\rm T}$, $z=\hat{z}_t+\eta_t$ and $[\eta_1 \cdots {\eta}_{t}]\sim \mathcal{CN}(\bf{0},\bf{V}_{t}^z)$ being independent of $[\hat{z}_1 \dots \hat{z}_t]$.
\item Substituting \eqref{Eqn:MLE-xb} and \eqref{Eqn:MLE_zb} into definition of errors, %\eqref{err_defi},
$\gamma_{\rm SE}(\cdot)$ is given by 
    \BE\label{Eqn:cov_x_MLE}
    \bb{v}_{t+1,t'+1}^x   =\bb{c}^x_t\bb{c}^x_{t'}\Big(\textstyle\sum\limits_{i=1}^t\textstyle\sum\limits_{j=1}^{t'}\vartheta_{t,i}\vartheta_{t'\!,j} (\tilde{w}_{t-i,t'-j} v_{i,j}^{x} + \delta^{-1}  w_{t+t'-i-j}  v_{i,j}^{z} ) \Big),
    \EE
    and 
    \BS\BE\label{Eqn:cov_z_MLE}
    \bb{v}_{t+1,t'+1}^z= ({\bb{c}^z _t}\beta_t -1) ({\bb{c}^z_{t'}}\beta_{t'} -1)w_0 +\bb{c}^z _t \bb{c}^z_{t'} v^{\tilde{\bb{s}}}_{t,t'}, 
    \EE
    where 
    \begin{align}
    v^{\tilde{\bb{s}}}_{t,t'}&\equiv|\!\langle \tilde{\bf{\bb{s}}}_t,\tilde{\bf{\bb{s}}}_{t'}\rangle\!| \\
    &=  \textstyle\sum\limits_{i=1}^t\textstyle\sum\limits_{j=1}^{t'}\vartheta_{t,i}\vartheta_{t'\!,j} \big( \bar{\Bar{w}}_{t+ t'-i -j}  v_{i,j}^{x} +  \Bar{w}_{t-i,t'-j}  v_{i,j}^{z} \big)  \nonumber\\
    &\ \ \ \ \ \ - \tfrac{\xi_{t'}}{ \theta_{t'}}\textstyle\sum\limits_{i=1}^t \vartheta_{t,i} \bar{w}_{t-i}v_{i,t'}^{x} - \tfrac{\xi_t}{\theta_{t}}\textstyle\sum\limits_{j=1}^{t'}  \vartheta_{t'\!,j}\bar{w}_{t'-j}v_{t,j}^{x} 
    + \tfrac{\xi_t\xi_{t'}}{ \theta_t\theta_{t'}}   w_0v_{t,t'}^{x}
    \label{Eqn:cov_s_tilde}\end{align}\ES 
    and $\bar{w}_i,\bar{w}_{i,j},\tilde{w}_{i,j}$ and $\bar{\bar{w}}_i$ are defined in \eqref{Eqn:orth_parameters}.
    
    See next subsections for further details.
\end{itemize}

\subsection{Derivation of $\bb{v}^{x}_{t+1,t'+1}$}
By substituting \eqref{Eqn:MLE-xb}  into definition of error, %\eqref{err_defi}, 
we have
\BS\begin{align}
    \bb{v}^{x}_{t+1,t'+1}&=|\!\langle\bf{\bb{f}}_{t+1},\bf{\bb{f}}_{t'+1} \rangle\!| \\
    &=\bb{c}^x_t\bb{c}^x_{t'}\Big(\textstyle\sum_{i=1}^t\textstyle\sum_{j=1}^{t'}\tfrac{1}{N}{\rm tr}\{\bf{G}_{t,i}^{\rm H}\bf{G}_{t',j}\}v_{i,j}^z + \tfrac{1}{N}{\rm tr}\{\bf{\bb{H}}_{t,i}^{\rm H}\bf{\bb{H}}_{t',j}\}v_{i,j}^x\Big) \\
    &=\bb{c}^x_t\bb{c}^x_{t'}\Big(\textstyle\sum_{i=1}^t\textstyle\sum_{j=1}^{t'}\vartheta_{t,i}\vartheta_{t',j}(\delta^{-1}w_{t+t'-i-j}v_{i,j}^z +  \tilde{w}_{t-i,t'-j}v_{i,j}^x)\Big),
\end{align}\ES
where
\BS\begin{align}
    \tfrac{1}{N}{\rm tr}\{\bf{G}_{t,i}^{\rm H}\bf{G}_{t',j}\}&=\vartheta_{t,i}\vartheta_{t',j}\delta^{-2}\tfrac{1}{N}{\rm tr}\{\bf{B}^{t-i}\bf{AA}^{\rm H}\bf{B}^{t'-j}\}  \\
    &=\vartheta_{t,i}\vartheta_{t',j}\delta^{-2}\tfrac{1}{N}{\rm tr}\{\bf{A}^{\rm H}\bf{B}^{t+t'-i-j}\bf{A}\} \\
    &=\vartheta_{t,i}\vartheta_{t',j}\delta^{-1}w_{t+t'-i-j}
\end{align}\ES
and
\BS\begin{align}
    \tfrac{1}{N}{\rm tr}\{\bf{\bb{H}}_{t,i}^{\rm H}\bf{\bb{H}}_{t',j}\}&=\vartheta_{t,i}\vartheta_{t',j}\tfrac{1}{N}{\rm tr}\{ (\delta^{-1}\bf{W}_{t-i}-w_{t-i}\bf{I})^{\rm H}(\delta^{-1}\bf{W}_{t'-j}-w_{t'-j}\bf{I})\} \\
    &=\vartheta_{t,i}\vartheta_{t',j}\tfrac{1}{N}{\rm tr}\{\delta^{-2}\bf{W}_{t-i}\bf{W}_{t'-j} - \delta^{-1}w_{t-i}\bf{W}_{t'-j} - \delta^{-1}w_{t'-j}\bf{W}_{t-i} + w_{t-i}w_{t'-j}\bf{I}\} \\
    &=\vartheta_{t,i}\vartheta_{t',j}\big(\tfrac{1}{N}{\rm tr}\{\delta^{-2}\bf{W}_{t-i}\bf{W}_{t'-j}\} - w_{t-i}w_{t'-j} - w_{t-i}w_{t'-j} + w_{t-i}w_{t'-j}\big) \\
    &=\vartheta_{t,i}\vartheta_{t',j}\big(\lambda^\dag\delta^{-1}w_{t+t'-i-j} - \delta^{-1}w_{t+t'-i-j+1} - w_{t-i}w_{t'-j}\big) \\
    &=\vartheta_{t,i}\vartheta_{t',j} \tilde{w}_{t-i,t'-j}
\end{align}\ES
with
\BS\begin{align}
    \tfrac{1}{N}{\rm tr}\{\delta^{-2}\bf{W}_{t-i}\bf{W}_{t'-j}\}&=\delta^{-2} \tfrac{1}{N}{\rm tr}\{\bf{A}^{\rm H}\bf{B}^{t-i}\bf{AA}^{\rm H}\bf{B}^{t'-j}\bf{A}\} \\
    &=\delta^{-2} \tfrac{1}{N}{\rm tr}\{\bf{A}^{\rm H}\bf{B}^{t-i}(\lambda^\dag\bf{I}-\bf{B})\bf{B}^{t'-j}\bf{A}\} \\
    &=\delta^{-2} \tfrac{1}{N}{\rm tr}\{\lambda^\dag\bf{A}^{\rm H}\bf{B}^{t+t'-i-j}\bf{A} - \bf{A}^{\rm H}\bf{B}^{t+t'-i-j+1}\bf{A}\} \\
    &=\delta^{-1}(\lambda^\dag w_{t+t'-i-j} - w_{t+t'-i-j+1}).
\end{align}\ES

\subsection{Derivation of $\bb{v}^{z}_{t+1,t'+1}$}
By substituting \eqref{Eqn:MLE_zb} into definition of error, %\eqref{err_defi}, 
we have
\BE
    \bb{v}^{z}_{t+1,t'+1}=|\!\langle\bf{\bb{s}}_{t+1},\bf{\bb{s}}_{t'+1} \rangle\!|= (\bb{c}^z_t\beta_t-1)(\bb{c}^z_{t'}\beta_{t'}-1)w_0 + \bb{c}^z_t\bb{c}^z_{t'}v_{t,t'}^{\tilde{\bf{\bb{s}}}},
\EE
where
\BS\begin{align}
    v^{\tilde{\bb{s}}}_{t,t'}&\equiv|\!\langle \tilde{\bf{\bb{s}}}_t,\tilde{\bf{\bb{s}}}_{t'} \rangle\!|  \\
    &=\textstyle\sum_{i=1}^t\textstyle\sum_{j=1}^{t'}\Big(\tfrac{1}{M}{\rm tr}\{\bf{\bb{G}}_{t,i}^{\rm H}\bf{\bb{G}}_{t',j}\}v_{i,j}^z + \tfrac{1}{M}{\rm tr}\{\bf{H}_{t,i}^{\rm H}\bf{H}_{t',j}\}v_{i,j}^x\Big)   \\
    &\hspace{0.4cm} -\textstyle\sum_{i=1}^t\tfrac{1}{M}{\rm tr}\{\bf{H}_{t,i}^{\rm H}(\tfrac{\xi_{t'}}{\theta_{t'}}\bf{A})\}v_{i,t'}^x -\textstyle\sum_{j=1}^{t'}\tfrac{1}{M}{\rm tr}\{(\tfrac{\xi_t}{\theta_t}\bf{A})^{\rm H}\bf{H}_{t',j}\}v_{t,j}^x + \tfrac{1}{M}{\rm tr}\{\tfrac{\xi_t}{\theta_t}\bf{A}^{\rm H}\tfrac{\xi_{t'}}{\theta_{t'}}\bf{A}\}v_{t,t'}^x \\
    &=
    \textstyle\sum_{i=1}^t\textstyle\sum_{j=1}^{t'}\vartheta_{t,i}\vartheta_{t',j}\big(\bar{w}_{t-i,t'-j}v_{i,j}^z + \bar{\bar{w}}_{t+t'-i-j}v_{i,j}^x\big)  \\
    &\hspace{2cm} -\tfrac{\xi_{t'}}{\theta_{t'}}\textstyle\sum_{i=1}^t\vartheta_{t,i}\bar{w}_{t-i}v_{i,t'}^x -\tfrac{\xi_t}{\theta_t}\textstyle\sum_{j=1}^{t'}\vartheta_{t',j}\bar{w}_{t'-j}v_{t,j}^x  + \tfrac{\xi_t\xi_{t'}}{\theta_t\theta_{t'}}w_0 v_{t,t'}^x \\
\end{align}\ES
with
\BS\begin{align}
    \tfrac{1}{M}{\rm tr}\{\bf{\bb{G}}_{t,i}^{\rm H}\bf{\bb{G}}_{t',j}\}&=\vartheta_{t,i}\vartheta_{t',j}\tfrac{1}{M}{\rm tr}\{(\widehat{\bf{W}}_{t-i}-w_{t-i}\bf{I})^{\rm H}(\widehat{\bf{W}}_{t'-j}-w_{t'-j}\bf{I})\} \\
    &=\vartheta_{t,i}\vartheta_{t',j}\tfrac{1}{M}{\rm tr}\{\widehat{\bf{W}}_{t-i}\widehat{\bf{W}}_{t'-j} + w_{t-i}w_{t'-j}\bf{I} - w_{t-i}\widehat{\bf{W}}_{t'-j} - w_{t'-j}\widehat{\bf{W}}_{t-i}\} \\
    &=\vartheta_{t,i}\vartheta_{t',j}(\lambda^\dag w_{t+t'-i-j} - w_{t+t'-i-j+1} -w_{t-i}w_{t'-j}) \\
    &=\vartheta_{t,i}\vartheta_{t',j}\bar{w}_{t-i,t'-j},
\end{align}\ES

\BS\begin{align}
    \tfrac{1}{M}{\rm tr}\{\bf{H}_{t,i}^{\rm H}\bf{H}_{t',j}\}&=\vartheta_{t,i}\vartheta_{t',j}\tfrac{1}{M}{\rm tr}\{\bf{A}^{\rm H}\bf{B}^{t-i}\bf{AA}^{\rm H}\bf{AA}^{\rm H}\bf{B}^{t'-j}\bf{A}\} \\
    &=\vartheta_{t,i}\vartheta_{t',j}\tfrac{1}{M}{\rm tr}\{\bf{A}^{\rm H}\bf{B}^{t-i}(\lambda^\dag\bf{I}-\bf{B})(\lambda^\dag\bf{I}-\bf{B})\bf{B}^{t'-j}\bf{A}\} \\
    &=\vartheta_{t,i}\vartheta_{t',j}\tfrac{1}{M}{\rm tr}\{{\lambda^\dag}^2\bf{W}_{t+t'-i-j} - 2\lambda^\dag\bf{W}_{t+t'-i-j+1} + \bf{W}_{t+t'-i-j+2}\} \\
    &=\vartheta_{t,i}\vartheta_{t',j}({\lambda^\dag}^2 w_{t+t'-i-j} - 2\lambda^\dag w_{t+t'-i-j+1} + w_{t+t'-i-j+2}) \\
    &=\vartheta_{t,i}\vartheta_{t',j}\bar{\bar{w}}_{t+t'-i-j},
\end{align}\ES

\BS\begin{align}
    \tfrac{1}{M}{\rm tr}\{\bf{H}_{t,i}^{\rm H}(\tfrac{\xi_{t'}}{\theta_{t'}}\bf{A})\}&=
    \vartheta_{t,i}\tfrac{1}{M}{\rm tr}\{\bf{A}^{\rm H}\bf{B}^{t-i}\bf{AA}^{\rm H}\tfrac{\xi_{t'}}{\theta_{t'}}\bf{A}\} \\
    &=\vartheta_{t,i}\tfrac{\xi_{t'}}{\theta_{t'}}\tfrac{1}{M}{\rm tr}\{\bf{A}^{\rm H}\bf{B}^{t-i}(\lambda^\dag\bf{I}-\bf{B})\bf{A}\} \\
    &=\vartheta_{t,i}\tfrac{\xi_{t'}}{\theta_{t'}}(\lambda^\dag w_{t-i} - w_{t-i+1}) \\
    &=\vartheta_{t,i}\tfrac{\xi_{t'}}{\theta_{t'}}\bar{w}_{t-i},
\end{align}\ES

\BS\begin{align}
    \tfrac{1}{M}{\rm tr}\{(\tfrac{\xi_t}{\theta_t}\bf{A})^{\rm H}\bf{H}_{t',j}\}&=\vartheta_{t',j}\tfrac{1}{M}{\rm tr}\{\tfrac{\xi_t}{\theta_t}\bf{A}^{\rm H}\bf{AA}^{\rm H}\bf{B}^{t'-j}\bf{A}\} \\
    &=\vartheta_{t',j}\tfrac{\xi_t}{\theta_t}\bar{w}_{t'-j}
\end{align}\ES
and
\BS\begin{align}
    \tfrac{1}{M}{\rm tr}\{\tfrac{\xi_t}{\theta_t}\bf{A}^{\rm H}\tfrac{\xi_{t'}}{\theta_{t'}}\bf{A}\}&=\tfrac{\xi_t\xi_{t'}}{\theta_t\theta_{t'}}\tfrac{1}{M}{\rm tr}\{\bf{A}^{\rm H}\bf{A}\} \\
    &=\tfrac{\xi_t\xi_{t'}}{\theta_t\theta_{t'}}w_0.
\end{align}\ES

\subsection{Approximation of $v^z_{t,t'}, v_{t,t'}^x$}%4
In practice, $\{v^{x}_{t,t'}, v^{z}_{t,t'}\}$ can be simply estimated using the following proposition.

\begin{proposition}\label{Pro:cov_app}
Define $v_x=|\!\langle\bf{x},\bf{x}\rangle\!|$ and $v_z=|\!\langle\bf{z},\bf{z}\rangle\!| $, and $\{v^{x}_{t,t'}, v^{z}_{t,t'}\}$ can be approximated by: For $ 1\leq{t'}\leq t$,
\BS\label{Eqn:v_approx}\begin{align} 
  v^{x}_{t,t'}  &\overset{\rm a.s.}{=} \tfrac{1}{N} \bf{x}_t^{\rm H}\bf{x}_{t'} +v^x_{t',t'}+v^x_{t,t}-v_x,\label{Eqn:v_approx_x}\\
  v^{z}_{t,t'}  &\overset{\rm a.s.}{=} \tfrac{1}{M}\bf{z}_t^{\rm H}\bf{z}_{t'} - v_z.\label{Eqn:v_approx_z}
\end{align}\ES 
\end{proposition}

The further details are given as follows. Note that MATLAB simulations may be unstable when the above approximate estimations are applied, so we instead adopt the more stable Monte Carlo statistical method\footnote{We run another offline program with known true $\bf{x}$ and $\bf{z}$. Specifically, for MLE $\Gamma$, with input $\{\bf{v}^x,\bf{v}^z\}$, the output estimation errors $\{\bb{\bf{v}}^x,\bb{\bf{v}}^z\}$ are obtained by the closed-form solutions derived in \eqref{Eqn:cov_x_MLE} and \eqref{Eqn:cov_z_MLE}. While for NLE $\Phi$ and $\Psi$, with input $\{\bb{\bf{v}}^x\}$ and $\{\bb{\bf{v}}^z\}$, the output estimation errors $\{\bf{v}^x\}$ and $\{\bf{v}^z\}$ are directly measured by the generated true $\bf{x}$ and $\bf{z}$. When the dimensions of $\bf{x}$ and $\bf{z}$, i.e., the number of random samples for statistical simulation, are large enough, the estimated $\{\bf{v}^x,\bf{v}^z\}$ in this program are consistent with that in practical systems.} to obtain $\{v^{x}_{t,t'}, v^{z}_{t,t'}\}$ in the experiments. How to accurately and efficiently estimate these two parameters is an interesting future work. 
\begin{itemize}
    \item Following Theorem 2 in main text, %\ref{The:IIDG_GMAMP}, 
    for $t\geq 1$, we have $\bf{x}_t=\bf{x}+\bf{f}_t$ and $\bf{f}_t$ is independent of $\bf{x}_t$. Hence,
    \BS \begin{align}
        \lim_{N\to\infty} \tfrac{1}{N}\bf{x}_t^{\rm H}\bf{x}_{t'} &= \tfrac{1}{N} (\bf{x}+\bf{f}_t)^{\rm H}(\bf{x}+\bf{f}_{t'}) \\
        &= \tfrac{1}{N}\big( \|\bf{x}\|^2 +   \bf{f}_t^{\rm H}\bf{f}_{t'}  +\bf{x}^{\rm H}\bf{f}_{t'}  +\bf{f}_t^{\rm H}\bf{x}  \big)\\
        &\overset{\rm a.s.}{=} v_x + v^x_{t,t'} +\tfrac{1}{N}\big[   (\bf{x}_{t'}-\bf{f}_{t'})^{\rm H}\bf{f}_{t'}  +\bf{f}_t^{\rm H}(\bf{x}_{t}-\bf{f}_{t}) \big]\\
        &\overset{\rm a.s.}{=} v_x + v^x_{t,t'} -  v^x_{t',t'} -  v^x_{t,t}.
    \end{align}\ES
    Thus, we have \eqref{Eqn:v_approx_x}.
    \item Following Theorem 2 in main text, %\ref{The:IIDG_GMAMP}, 
    for $t\geq 1$, $\bf{z}_t=\bf{z}+\bf{s}_t$ and $\bf{s}_t$ is independent of $\bf{z}$. Hence,
    \BS \begin{align}
        \lim_{N\to\infty} \tfrac{1}{M}\bf{z}_t^{\rm H}\bf{z}_{t'} &\overset{\rm a.s.}{=}\tfrac{1}{M} (\bf{z}+\bf{s}_t)^{\rm H}(\bf{z}+\bf{s}_{t'}) \\
        &\overset{\rm a.s.}{=}\tfrac{1}{M} \|\bf{z}\|^2 + \tfrac{1}{M} \bf{s}_t^{\rm H}\bf{s}_{t'}\\
        &= v_z + v^z_{t,t'}.
    \end{align}\ES
    Thus, we have \eqref{Eqn:v_approx_z}.    
\end{itemize}

\section{Proof of Convergence and Bayes-optimality of BO-GMAMP}%4
First, we prove the convergence of BO-GMAMP in subsection \ref{Converge} using the monotonically decreasing property of the optimized damping. Second, we prove that the fixed points of BO-GMAMP and GVAMP are the same in subsection \ref{FixPoint}. Therefore, BO-GMAMP converges to the same fixed point as GVAMP. Finally, following Lemma \ref{Lem:optimality}, we can say that the optimized BO-GMAMP can achieve the minimum (i.e., Bayes optimal) MSE as predicted by replica method for all unitarily-invariant transformation matrices if it has a unique fixed point.

\subsection{Convergence}\label{Converge}
Intuitively, the MSEs of $\{\bf{x}_{t+1},\bf{z}_{t+1}\}$ with optimized damping are not worse than that of $\{\bf{x}_t,\bf{z}_t\}$ in the previous iteration. That is, in the optimized BO-GMAMP, $\{v_{t,t}^x,v_{t,t}^z\}$ are monotonically decreasing sequences. Besides, $\{v_{t,t}^x,v_{t,t}^z\}$ both have the lower bound $0$. Therefore, sequences $\{v_{t,t}^x,v_{t,t}^z\}$ converge to the cetian value $\{v_{*}^x,v_{*}^z\}$, i.e., the convergence of the optimized BO-GMAMP is guaranteed.

\subsection{Fixed-Point Consistency of BO-GMAMP and GVAMP}\label{FixPoint}
The follow lemma gives a Taylor series expansion for the fixed point of GVAMP .
\begin{lemma}\label{fixlem}
Assume that $\{\bf{x}_t,\bf{\bb{x}}_t,\rho_t,\bf{z}_t,\bf{\bb{z}}_t\}$ converges to $\{\bf{x}_*,\bf{\bb{x}}_*,\rho_*,\bf{z}_*,\bf{\bb{z}}_*\}$. The fixed point of GVAMP is given by
\BS\begin{align}
    \bf{x}_*&=\phi_{\infty}(\bf{\bb{x}}_*),\\
    \bf{z}_*&=\psi_{\infty}(\bf{\bb{z}}_*),\\
    \bf{\bb{x}}_*&=\frac{1}{\delta\epsilon^{\gamma}_*}\bf{A}^{\rm H}\big(\textstyle\sum\nolimits_{i=0}^{\infty}\theta^i\bf{B}^i\big)(\bf{z}_*-\bf{Ax}_*)+\bf{x}_*,\\
    \bf{\bb{z}}_*&=\frac{1}{1-\epsilon^{\gamma}_*}\big[\bf{A}\big(\bf{A}^{\rm H}\big(\textstyle\sum\nolimits_{i=0}^{\infty}\theta^i\bf{B}^i\big)(\bf{z}_*-\bf{Ax}_*) + \bf{x}_*\big)-\epsilon^{\gamma}_*\bf{z}_*\big],
\end{align}\ES
where
\BS\begin{align}
    \theta&=(\lambda^{\dagger}+\rho_*)^{-1},\\
    \epsilon^{\gamma}_*&=\textstyle\sum\nolimits_{i=0}^{\infty}\theta^i(\lambda^{\dagger}b_i-b_{i+1}).
\end{align}\ES
\end{lemma}

See APPENDIX F-C in \cite{lei2020mamp} for further details.

Notice that $\theta=(\lambda^{\dagger}+\rho_*)^{-1}$ minimizes the spectral radius of $\bf{I}-\theta(\rho_*\bf{I}+\bf{AA}^{\rm H})$. Thus, we can ensure $\rho(\theta\bf{B})<1$. Supposing that $\{\bf{x}_t,\bf{\bb{x}}_t,\xi_t,\theta_t,\bf{z}_t,\bf{\bb{z}}_t\}$ in BO-GMAMP converges to $\{\bf{x}_*,\bf{\bb{x}}_*,\xi_*,\theta_*,\bf{z}_*,\bf{\bb{z}}_*\}$. We have
\BS\begin{align}
    &\lim_{t\to\infty}\vartheta_{ti}=\xi_*\theta^{t-i},\\
    &\lim_{t\to\infty}\bb{c}^x_t=\xi_*\varepsilon^{\gamma}_*,\\
    &\lim_{t\to\infty}\bf{G}_{ti}=\xi_*\theta^{t-i}\delta^{-1}\bf{A}^{\rm H}\bf{B}^{t-i},\\
    &\lim_{t\to\infty}\bf{\bb{G}}_{ti}=\xi_*\theta^{t-i}(\widehat{\bf{W}}_{t-i}-w_{t-i}\bf{I}),\\
    &\lim_{t\to\infty}\bf{H}_{ti}=\xi_*\theta^{t-i}\bf{A}\bf{W}_{t-i},\\
    &\lim_{t\to\infty}\bf{\bb{H}}_{ti}=\xi_*\theta^{t-i}(\delta^{-1}\bf{W}_{t-i}-w_{t-i}\bf{I}).
\end{align}\ES
According to (\ref{Eqn:NLE}), $\sum\nolimits_{i=1}^{l_t}\zeta_{ti}=1$ and $\sum\nolimits_{i=1}^{l_t}\varrho_{ti}=1$, we have
\BS\begin{align} 
    \bf{x}_*&=\phi_{\infty}(\bf{\bb{x}}_*),\\
    \bf{z}_*&=\psi_{\infty}(\bf{\bb{z}}_*),\\
    \bf{\bb{x}}_*&=\frac{1}{\varepsilon^{\gamma}_*}\textstyle\sum\nolimits_{i=0}^{\infty} \left[\theta^i\delta^{-1}\bf{A}^{\rm H}\bf{B}^i\bf{z}_* - \theta^i(\delta^{-1}\bf{W}_i-w_i\bf{I})\bf{x}_*\right],\\
    \bf{\bb{z}}_*&=\frac{1}{1-\varepsilon^\gamma_*}\left[\textstyle\sum\nolimits_{i=0}^{\infty}\big(\theta^i(\widehat{\bf{W}}_i-w_i\bf{I})\bf{z}_* - \theta^i\bf{A}\bf{W}_i\bf{x}_*\big) + \bf{Ax}_* \right].
\end{align}\ES
Since GVAMP and BO-GMAMP have the same $\phi_t(\cdot)$ and $\psi_t(\cdot)$, the fixed point of BO-GMAMP is the same as that of GVAMP in Lemma \ref{fixlem}.

\section{Parameter Optimization}%5
The optimization of $\{\bf{\zeta}_t,\bf{\varrho}_t,\theta_t\}$ have been discussed for MAMP in \cite{lei2020mamp}, which also applies to BO-GMAMP.

\subsection{Optimization of $\xi_t$}
An optimal $\xi_t$ that minimizes $\bb{v}_{t+1,t+1}^{x}$ is given by $\xi_1^{\rm opt}=1$ and for $t\geq2$,
\begin{align}\label{Eqn:xi}
    \xi_t^{\rm opt}=
    \begin{cases}
    \dfrac{c_{t2}c_{t0}+c_{t3}}{ c_{t1}c_{t0}+c_{t2}},\hspace{0.5cm}{\rm if}\ c_{t1}c_{t0}+c_{t2}\neq 0\vspace{2mm}\\
    +\infty,\hspace{1.5cm}{\rm otherwise}
    \end{cases},
\end{align}
where
\BS\label{c_defi}\begin{align}
    c_{t0}&=\sum_{i=1}^{t-1}p_{ti}/w_0,\\
    c_{t1}&=v_{tt}^x\delta\tilde{w}_{00}+v_{tt}^zw_0,\\
    c_{t2}&=-\sum_{i=1}^{t-1}\vartheta_{ti}(v_{ti}^x\delta\tilde{w}_{0t-i} + v_{ti}^z w_{t-i}),\\
    c_{t3}&=\sum_{i=1}^{t-1}\sum_{j=1}^{t-1}\vartheta_{ti}\vartheta_{tj}(v_{ij}^x\delta\tilde{w}_{t-it-j}+v_{ij}^z w_{2t-i-j}).
\end{align}\ES 
At this time, the minimum $\bb{v}_{t+1,t+1}^x$ is calculated as
\begin{align}
    \bb{v}_{t+1,t+1}^x(\xi_t^{\rm opt}) = \frac{c_{t1}{\xi_t^{\rm opt}}^2 - 2c_{t2}\xi_t^{\rm opt} + c_{t3}}{\delta w_0^2(\xi_t^{\rm opt}+c_{t0})^2}.
\end{align}

\subsection{Optimization of $\bb{c}^z_t$}
We calculate $ \bb{v}_{t+1,t+1}^z$ by
\BS\label{Eqn:v_z_bb_deri}\begin{align}
    \bb{v}_{t+1,t+1}^z &= |\!\langle\bf{\bb{s}}_{t},\bf{\bb{s}}_{t}\rangle\!|\\
    &= \tfrac{1}{M}{\rm E}\big\{\|({\bb{c}^z _t}\beta_t -1)\bf{z} +{\bb{c}^z _t}\tilde{\bf{\bb{s}}}_{t}\|^2\big\}\\
    &=({\bb{c}^z _t}\beta_t -1)^2 w_0 +{\bb{c}^z_t}^2 v^{\tilde{\bb{s}}}_{t,t}\\
    &=(\beta_t^2 w_0+ v^{\tilde{\bb{s}}}_{t,t}){\bb{c}^z_t}^2 -2w_0\beta_t\bb{c}^z _t + w_0.
\end{align}\ES

Since $\beta_t^2 w_0+ v^{\tilde{\bb{s}}}_{t,t}>0$, $\bb{v}_{t+1,t+1}^z$ convex of $\bb{c}^z _t$. Thus,  $\bb{v}_{t+1,t+1}^z$ is minimized by the solution of $\partial \bb{v}_{t+1,t+1}^z/ \partial \bb{c}^z _t =0$. Then, an optimal $\bb{c}^z_t$ that minimizes $\bb{v}_{t+1,t+1}^z$ is given by
\BS\label{Eqn:alpha_opt}\BE
    \bb{c}_t^{z,{\rm opt}} = \frac{\beta_t w_0}{\beta_t^2 w_0+v^{\tilde{\bb{s}}}_{t,t}},
\EE
where
\BE
    v^{\tilde{\bb{s}}}_{t,t} \equiv |\!\langle\tilde{\bf{\bb{s}}}_{t},\tilde{\bf{\bb{s}}}_{t}\rangle\!|.
\EE\ES
In this case, from \eqref{Eqn:alpha_opt} and \eqref{Eqn:v_z_bb_deri}, the minimum $\bb{v}_{t+1,t+1}^z$ is calculated as
\BE\label{Eqn:v_z_bb}
    \bb{v}_{t+1,t+1}^z(\bb{c}_t^{z,{\rm opt}}) = \frac{1}{w_0^{-1}+\beta_t^2/v^{\tilde{\bb{s}}}_{t,t}}.
\EE
Meanwhile, from \eqref{Eqn:MLE_z} and \eqref{Eqn:alpha_opt}, the following orthogonality holds
\BE\label{Eqn:z_bb_mmse_orth}
    |\!\langle\bf{\bf{\bb{z}}}_{t+1},\bf{\bf{\bb{s}}}_{t+1} \rangle\!|=0.
\EE

\section{Algorithm Summary}%6
We summarize BO-GMAMP and the state evolution of BO-GMAMP in the following Algorithm 1 and Algorithm 2.
\begin{algorithm}[hbt] 
\caption{BO-GMAMP}
\begin{algorithmic}
\STATE 
 \textbf{Input:} $\bf{A}, \bf{y}, \sigma^2, P_X(x), T, L, \{\hat{\phi}_t(\cdot), \hat{\psi}_t(\cdot)\}, \{\lambda_{\rm max}, \lambda_{\rm min}, \lambda_t\}$ \\
 \textbf{Initialization:} $\hat{\bf{\bb{x}}}=\hat{\bf{\bb{z}}}=\bf{\bb{x}}_1=\bf{\bb{z}}_1=\bf{0}, \bb{v}_{x}=\infty, \bb{v}_{z}=\lambda_1, \vartheta_1=\xi=1, \lambda^{\dagger}=(\lambda_{\rm max}+\lambda_{\rm min})/2$, $\delta=M/N$\\ 
  \hspace{2.3cm}  $\{w_i, \bar{\bar{w}}_i, 0\leq i<2T\}$ and $\{\bar{w}_{i,j}, \tilde{w}_{i,j},0\leq i,j\leq T\}$ by \eqref{Eqn:orth_parameters}.
  
\STATE
  \hspace{0.5cm}\textbf{for}\ {$t=1:T$}\ \textbf{do}\vspace{1.5mm}\\
  \hspace{1cm}$\left[ \!\!\!\begin{array}{c}
         (\hat{\bf{z}}_{t},\hat{v}_{t}^z)  \\
         (\hat{\bf{x}}_{t},\hat{v}_{t}^x)
    \end{array} \!\!\!\right] =\left[ \!\!\!\begin{array}{c}
         \hat{\psi}_t(\bf{\bb{z}},\bb{v}_z,\bf{y},\sigma^2)  \\
         \hat{\phi}_t(\bf{\bb{x}},\bb{v}_x)
     \end{array} \!\!\!\right]$, \hspace{1.2cm}$ \left[ \!\!\!\begin{array}{c}
       \bf{z}_{t} \\
       \bf{x}_{t}
     \end{array} \!\!\!\right] =\left[ \!\!\!\begin{array}{c} (\hat{\bf{z}}/\hat{v}_{t}^z - \bf{\bb{z}}/\bb{v}_z)/(1/\hat{v}_{t}^z - 1/\bb{v}_z)\\
     (\hat{\bf{x}}_{t}/\hat{v}_{t}^x \!-\! \bf{\bb{x}}/\bb{v}_x)/(1/\hat{v}_{t}^x\! - \!1/\bb{v}_x)\end{array} \!\!\!\right]$ \\\vspace{1mm} 
 
    \hspace{1cm}$\{v_{t,t'}^z\!=\!v_{t'\!, t}^z, t'\!\leq \!t\}$ obtained by SE,\hspace{0.8cm} $\!\{\!v_{t, t'}^x\!\!=\! v_{t'\!, t}^x, t'\!\!\leq\! t\}$ obtained by SE
    %$\{v_{t,t'}^z\!=\!v_{t'\!, t}^z\!=\!\tfrac{\bf{z}_{t}^{\rm H}\bf{z}_{t'}}{M}\!-\!1,   t'\!\leq \!t\}$,\hspace{1.5cm} $\!\{\!v_{t, t'}^x\!\!=\! v_{t'\!, t}^x\!\!=\!\!\tfrac{\bf{x}_{t}^{\rm H}\bf{x}_{t'}}{N}\!+\!v^x_{t'\!,t'}\!\!+\!v^x_{t,t}\!\!-\!1,  t'\!\!\leq\! t\}$
    \hspace{2.2cm}\% NLE  \\ \vspace{3mm}  
 
    \hspace{1.1cm}$l_t=\min\{L,t\}$, \hspace{3.5cm} $\left[ \!\!\!\begin{array}{c}  \{\bf{\varrho}=[\bf{V}_{t}^{\psi}]^{-1}\bf{1}, v_{\psi}=1/\bf{1}^{\rm T}\bf{\varrho}, \bf{\varrho}=v_{\psi}\bf{\varrho}\}\\
    \{\bf{\zeta}=[\bf{V}_{t}^{\phi}]^{-1}\bf{1}, v_{\phi}=1/\bf{1}^{\rm T}\bf{\zeta}, \bf{\zeta}=v_{\phi}\bf{\zeta}\} \end{array} \!\!\!\right]$ \\\vspace{1mm}
 
    \hspace{1cm}$\left[ \!\!\!\begin{array}{c}  
       \bf{z}_{t}\\
       \bf{x}_{t}
    \end{array} \!\!\!\right]\!=\!\sum\limits_{i=1}^{l_t} \left[ \!\!\!\begin{array}{c}  
    \varrho_i\bf{z}_{t-l_t+i}\\
    \zeta_i\bf{x}_{t-l_t+i}
    \end{array} \!\!\!\right] $, \hspace{2.4cm} $
    \left[ \!\!\!\begin{array}{c}  
    v_{t, t}^z\!=\!v_{\psi}, \{v_{t, t'}^z\!=\!v_{t'\!, t}^z\!=\!\!\sum_{i=1}^{l_t}\varrho_i v^z_{t-l_t+i, t'}\}\vspace{1mm}\\ 
    v_{t, t}^x\!=\!v_{\phi}, \{v_{t, t'}^x\!=\!v_{t'\!, t}^x\!=\!\!\sum_{i=1}^{l_t}\!\zeta_i v^x_{t-l_t+i, t'}\}
    \end{array} \!\!\!\right]$   \hspace{0.7cm} \% Damping \vspace{3mm} 

\STATE
   \hspace{1.1cm}$\theta=(\lambda^{\dagger}+v_{t,t}^z/v_{t,t}^x)^{-1}$,\hspace{2.7cm}$\{\vartheta_i=\theta\vartheta_i, p_i=\vartheta_i w_{t-i}, 1\leq i<t\}$\\\vspace{1mm}
 
  \hspace{1cm}$\{c_i,0\leq i\leq 3\}$ by \eqref{c_defi}, \hspace{2.3cm}$\{t\!\geq \!2: \xi\!=\!\vartheta_t\!=\!(c_2c_0\!+\!c_3)/(c_1c_0\!+\!c_2)\}$\hspace{1.5cm}  \% $\xi$ \\\vspace{1.5mm}
 
  \hspace{1cm}$\{p_t=\xi w_0, \tfrac{1}{\bb{c}^x}=p_t+w_0c_0, \beta\!=\!\tfrac{\xi}{\theta}\!-\!\tfrac{1}{\bb{c}^x}\}$, \hspace{0.06cm} 
  $\bb{v}_x\!=\!(c_1\xi^2\!-\!2c_2\xi\!+\!c_3){\bb{c}^x}^2\!/\delta$, $\bb{v}_z\!=\!1/\!\big(w_0^{-1}\!+\!\beta^2\!/v^{\tilde{\bb{s}}}_{t,t}\big)$ \\ \vspace{1mm}
  
  \hspace{1.05cm}$\hat{\bf{\bb{z}}}\!=\!\theta\lambda^{\dagger}\hat{\bf{\bb{z}}}\!+\!\xi \bf{z}_t \!-\!\bf{A}(\xi  \bf{x}_t \!+\! \theta\bf{A}^{\rm H}\hat{\bf{\bb{z}}})$, \hspace{1.05cm} $\hat{\bf{\bb{x}}}\!=\!\bf{A}^{\rm H}\hat{\bf{\bb{z}}},\{\bb{c}^z\!=\!\beta w_0/\!(\beta^2 w_0\!+\!v^{\tilde{\bb{s}}}_{t,t})\}$ 
  \\   \vspace{1.5mm}
  
  \hspace{1.05cm}$\bf{\bb{x}}\!=\!\bb{c}^x\big(\delta^{-1}\hat{\bf{\bb{x}}} + \sum_{i=1}^t p_i\bf{x}_i\big)$, \hspace{1.75cm} $\bf{\bb{z}}\!=\!\bb{c}^z\big[\bf{A}\big(\hat{\bf{\bb{x}}}+\xi/\theta\bf{x}_t) - \sum_{i=1}^t p_i\bf{z}_i\big]$  \hspace{1.85cm}\% MLE\\   \vspace{1.5mm}
  
  \hspace{0.5cm}\textbf{end for} \vspace{1mm} 
\STATE
  \textbf{Output:} $\{\hat{\bf{x}}_{t}, \hat{v}^x_{t}\}$
\end{algorithmic}
\end{algorithm}

\begin{algorithm}[htb] 
\caption{State Evolution (SE) of BO-GMAMP}
\begin{algorithmic}
\STATE
     \textbf{Input:} $\bf{A}$, $\sigma^{2}$, $T$, $L$, $\{\hat{\phi}_t(\cdot),\hat{\psi}_t(\cdot)\}, x\!\sim\!P_x, z\!\sim\!P_z, \{\lambda_{\min}, \lambda_{\max}, \lambda_t\}$. \\\vspace{1mm}
     \textbf{Initialization:} $\bb{v}^x=\infty, \bb{v}^z=\lambda_1, \bb{z}=\bf{0}, \vartheta_{1,1}\! = \!\xi\!\!=\!1$,   ${\lambda}^\dag \!\!=\! (\lambda_{\max}+ \lambda_{\min})/2$,  $\delta=M/N$, $\tilde{\bf{x}}_0=\tilde{\bf{z}}_0= {\rm null}$\\
    \hspace{2.3cm}    $\{w_i, \bar{\bar{w}}_i, 0\leq i<2T\}$ and $\{\bar{w}_{i,j}, \tilde{w}_{i,j},0\leq i,j\leq T\}$ by \eqref{Eqn:orth_parameters}.
    \vspace{2mm} 
 
    \textbf{for}\  {$t =1$ to $T$ }\ \textbf{do} \vspace{2mm} 
  
\STATE
   \hspace{1cm}$\big\{\eta^x_t$ {\rm by APPENDIX E in} \cite{lei2020mamp}, \; $\bb{x} = \bf{x} + \eta^x_t\big\}$\\\vspace{2mm}

   \hspace{1cm}$\left[ \!\!\!\begin{array}{c}
        (\hat{\bf{z}},\hat{v}_z)  \\
        (\hat{\bf{x}},\hat{v}_x)
    \end{array} \!\!\!\right] =\left[ \!\!\!\begin{array}{c}
        \hat{\psi}_t(\bf{\bb{z}},\bb{v}^z,Q(\bf{z}),\sigma^2)  \\
        \hat{\phi}_t(\bf{\bb{x}},\bb{v}^x)
    \end{array} \!\!\!\right]$, \hspace{0.95cm}$ \left[ \!\!\!\begin{array}{c}
      \bf{z}_{t} \\
      \bf{x}_{t}
    \end{array} \!\!\!\right] =\left[ \!\!\!\begin{array}{c} (\hat{\bf{z}}/\hat{v}_z - \bf{\bb{z}}/\bb{v}^z)/(1/\hat{v}_z - 1/\bb{v}^z)\\
    (\hat{\bf{x}}/\hat{v}_x \!-\! \bf{\bb{x}}/\bb{v}^x)/(1/\hat{v}_x\! - \!1/\bb{v}^x)\end{array} \!\!\!\right]$\\\vspace{2mm}

    \hspace{1.05cm}$\bf{v}^z_{t} \!\!=\! {\rm E}\big\{\! (z_{t}\!-\!{z})^* ([\tilde{\bf{z}}_{t-1} \;z_{t}]^\intercal\!-\!{z} \bf{1})\!\big\}$,\hspace{1.25cm}  $\bf{v}^x_{t} \!\!=\! {\rm E}\big\{\! (x_{t}\!-\!{x})^* ([\tilde{\bf{x}}_{t-1} \;x_{t}]^\intercal\!-\!{x} \bf{1})\!\big\}$  \hspace{2.35cm} \% NLE  \vspace{1mm}    
   
\STATE
    \hspace{1.05cm}$l_t=\min\{L,t\}$, \hspace{3.7cm} $\left[ \!\!\!\begin{array}{c}  
     \{\bf{\varrho}=[\bf{V}_{t}^{\psi}]^{-1}\bf{1}, v_{\psi}=1/\bf{1}^{\rm T}\bf{\varrho}, \bf{\varrho}=v_{\psi}\bf{\varrho}\}\\
     \{\bf{\zeta}=[\bf{V}_{t}^{\phi}]^{-1}\bf{1}, v_{\phi}=1/\bf{1}^{\rm T}\bf{\zeta}, \bf{\zeta}=v_{\phi}\bf{\zeta}\}
    \end{array} \!\!\!\right]$\\\vspace{1mm}
    
    \hspace{1cm}$\left[ \!\!\!\begin{array}{c}  
    \tilde{\bf{z}}_{t}\\
    \bf{x}_{t}
    \end{array} \!\!\!\right]\!=\! \left[ \!\!\!\begin{array}{c}  
    \tilde{\bf{x}}_{t-1}\ \sum_{i=1}^{l_t}\varrho_i\bf{z}_{t-l_t+i}\vspace{1mm}\\
    \tilde{\bf{z}}_{t-1}\ \sum_{i=1}^{l_t}\zeta_i\bf{x}_{t-l_t+i}
    \end{array} \!\!\!\right] $, \hspace{1.3cm} $\left[ \!\!\!\begin{array}{c}  
    v_{t, t}^z\!=\!v_{\psi}, \{v_{t, t'}^z\!=\!v_{t'\!, t}^z\!=\!\!\sum_{i=1}^{l_t}\varrho_i v^z_{t-l_t+i, t'}\}\vspace{1mm}\\ 
    v_{t, t}^x\!=\!v_{\phi}, \{v_{t, t'}^x\!=\!v_{t'\!, t}^x\!=\!\!\sum_{i=1}^{l_t}\!\zeta_i v^x_{t-l_t+i, t'}\}
    \end{array} \!\!\!\right]$   \hspace{1.1cm} \% Damping   \vspace{2mm} 

\STATE
   \hspace{1.05cm}$\theta_t=(\lambda^{\dagger}+v_{t,t}^z/v_{t,t}^x)^{-1}$,\hspace{2.8cm}$\{\vartheta_{t,i}=\theta_t\vartheta_{t-1,i}, p_i=\vartheta_{t,i}w_{t-i}, 1\leq i<t\}$\\\vspace{1mm}
 
   \hspace{1cm}$\{c_i,0\leq i\leq 3\}$ by \eqref{c_defi}, \hspace{2.5cm}$\{t\!\geq \!2: \xi_t\!=\!\vartheta_{t,t}\!=\!(c_2c_0\!+\!c_3)/(c_1c_0\!+\!c_2)\}$\hspace{1.5cm}  \% $\xi$ \\\vspace{1.5mm}
 
   \hspace{1cm}$\{1/\bb{c}^x_t=w_0(\xi_t + c_0), \beta_t\!=\!\xi_t/\theta_t\!-\!1/\bb{c}^x_t\}$, \hspace{0.2cm} 
   $\bb{v}^x_{t,t}\!=\!(c_1\xi^2\!-\!2c_2\xi\!+\!c_3){\bb{c}^x}^2\!/\delta$, $\bb{v}^z_{t,t}\!=\!1/\!\big(w_0^{-1}\!+\!\beta^2\!/v^{\tilde{\bb{s}}}_{t,t}\big)$ \\ \vspace{1mm}
  
   \hspace{1cm}$\big\{\bb{v}^x_{t,t'}=\bb{v}^x_{t',t}, 1\leq t'\leq t-1 \big\}$ {\rm by \eqref{Eqn:cov_x_MLE}}, \hspace{0.45cm} $v^{\tilde{\bb{s}}}_{t,t}$ {\rm by \eqref{Eqn:cov_s_tilde}}, $\bb{c}^z_t\!=\!\beta_t w_0/\!(\beta_t^2 w_0\!+\!v^{\tilde{\bb{s}}}_{t,t})$   \\   \vspace{1.5mm}
  
   \hspace{1cm}$\big\{\bb{v}^z_{t,t'}=\bb{v}^z_{t',t}, 1\leq t'\leq t-1\big\}$ {\rm by \eqref{Eqn:cov_z_MLE}},
   \hspace{0.35cm}$\big\{\tilde{\bb{s}}$ {\rm by APPENDIX E in \cite{lei2020mamp}}, \; $\bb{z}=\bb{c}^z_t(\beta_t\bf{z} + \tilde{\bb{s}}) \big\}$\hspace{0.25cm}\% MLE \\ \vspace{1.5mm}

   \hspace{0.5cm}\textbf{end for} \vspace{1mm} 
\STATE
   \textbf{Output:} $\hat{\bf{v}}_x$.
\end{algorithmic}
\end{algorithm}

\section{Supplementary Simulation Results}%7
In this section, we first give the expressions of MMSE NLE $\hat{\phi}_t(\cdot)$ for Bernoulli-Gaussian signaling and MMSE NLE  $\hat{\psi}(\cdot)$ for the clipping constraint. The details of Bernoulli-Gaussian signaling and clipping function $Q(\cdot)$ are set in main text.  Furthermore, we provide more simulation results of BO-GMAMP.

\subsection{Derivation of Bernoulli-Gaussian Demodulation Function $\hat{\phi}_t(\cdot)$}
We derive the MMSE estimation of a Bernoulli-Gaussian signal $\bf{x}_i \sim P_X(x)$ based on a Gaussian observation. The problem can be described as:
\begin{equation}\label{Eqn:subproblem 1}
\left\{
\begin{array}{l}
x = b \cdot g, \quad b \sim \mathcal{B}(\mu), \quad g \sim \mathcal{N}(u_g,v_g)\\
\bb{x}= x + \bb{n}^x, \quad
\bb{n}^x \sim \mathcal{N}(0,\bb{v}^x)
\end{array}
\right..
\end{equation}

First, we calculate the \emph{a-posteriori} probability of $b$ as follows.
\begin{eqnarray}
P(b=0|\bb{x}) \!\!&\propto&\!\!\! P(\bb{x}=b\cdot g+\bb{n}^x|b=0)P(b=0)\\\nonumber
&\propto&\!\!\! P(\bb{x}=\bb{n}^x|b=0)P(b=0)\\\nonumber
&\propto&\!\!\! \frac{1-p}{\sqrt{\bb{v}^x}}e^{-\frac{{\bb{x}}^2}{2 \bb{v}^x}},
\end{eqnarray}
and
\begin{eqnarray}
P(b=1|\bb{x}) \!\!&\propto&\!\!\! P(\bb{x}=b\cdot g + \bb{n}^x|b=1)P(b=1)\\\nonumber
&\propto&\!\!\! P(\bb{x}=g + \bb{n}^x|b=1)P(b=1)\\\nonumber
&\propto&\!\!\! \frac{p}{\sqrt{v_g + \bb{v}^x}}e^{-\frac{(\bb{x}-u_g)^2}{2(v_g + \bb{v}^x)}}.
\end{eqnarray}
Therefore, 
\begin{eqnarray}
p_{\rm{post}}\!\!&=&\!\!\!\frac{P(b=1|\bb{x})}{P(b=0|\bb{x})+P(b=1|\bb{x})}\nonumber\\
&=&\!\!\! \frac{\frac{p}{\sqrt{v_g + \bb{v}^x}}e^{-\frac{(\bb{x}-u_g)^2}{2(v_g + \bb{v}^x)}}}{\frac{1-p}{\sqrt{\bb{v}^x}}e^{-\frac{{\bb{x}}^2}{2\bb{v}^x}} + \frac{p}{\sqrt{v_g + \bb{v}^x}}e^{-\frac{(\bb{x}-u_g)^2}{2(v_g + \bb{v}^x)}}}\nonumber\\ 
&=&\!\!\! \frac{1}{  (p^{-1}-1)\sqrt{1+ v_g /\bb{v}^x}\ e^{\frac{u_g^2 \bb{v}^x - 2u_g \bb{x} \bb{v}^x - v_g {\bb{x}}^2}{2\bb{v}^x(v_g + \bb{v}^x)}} + 1}.
\end{eqnarray}

Next, we calculate the \emph{a-posteriori} probability of $g$ as follows.
\BE
P(g|b=1,\bb{x})\propto P(\bb{x}=g + \bb{n}^x|g)P(g)\propto e^{-\frac{(\bb{x}-g)^2}{2\bb{v}^x}} e^{-\frac{(g-u_g)^2}{2v_g}},
\EE

which follows a new Gaussian distribution $g\sim \mathcal{N}(u_{g_{\rm post}},v_{g_{\rm post}})$, where
\BS\begin{align}
    &v_{g_{\rm post}}= ( v_g^{-1} + {\bb{v}^x}^{-1} )^{-1},\\
    &u_{g_{\rm post}}=v_{g_{\rm post}} (v_g^{-1} u_g + {\bb{v}^x}^{-1}\bb{x} ).
\end{align}\ES

Then, the \emph{a-posteriori} expectation and variance of $x$ can be expressed as
\BS\begin{align}
    &\hat{x}={\rm{E}} \{x|\bb{x}\}=p_{\rm{post}} u_{g_{\rm post}},\\
    &\hat{v}_x={\rm{Var}} \{x|\bb{x}\}=p_{\rm{post}}(v_{g_{\rm post}} + u_{g_{\rm post}}^2) - \hat{x}^2.
\end{align}\ES

\subsection{Derivation of De-Clipping Function $\hat{\psi}_t(\cdot)$}
We derive the MMSE estimation for the following non-linear clip constraint.
\begin{align}
    \begin{cases} 
    y = Q(z) \equiv  {\rm Clip}(z) + n,\quad n\sim \mathcal{N}(0,\sigma^2)\\ 
    z = \bb{z}  + \bb{n}^z,\qquad \qquad\;\;\, \bb{n}^z\sim\mathcal{N}(0,\bb{v}^z)
    \end{cases},
\end{align}
where
\begin{align}
     {\rm Clip}(z)=
    \begin{cases}
         \mathfrak{c},\hspace{0.5cm} {\rm if}\; z\geq \mathfrak{c}\\
         z,\hspace{0.5cm} {\rm if} -\mathfrak{c}<z<\mathfrak{c}\\
         -\mathfrak{c},\hspace{0.2cm} {\rm if}\; z\leq -\mathfrak{c}
    \end{cases}.
\end{align}
The goal is estimating $P(z|\bb{z},y)$  using Bayes' rule:
\begin{align}
    P(z|\bb{z},y) &= \frac{P(\bb{z},y|z)P(z)}{P(\bb{z},y)}\propto
    P(\bb{z},y|z)P(z), \label{Eqn:z_post} 
\end{align} 
where $P(\bb{z},y)$ is fixed and can be treated as a normalization coefficient. We have no information about $P(z)$, which can be ignored or treated as constants.

Since $\bb{z} - z - y$ is a Markov Chain, from \eqref{Eqn:z_post}, we have
\begin{align}
     P(z|\bb{z},y)  
      = \frac{P(z|\bb{z})P(y|z)}{\int P(\bb{z}|z)P(y|z) dz},   
\end{align}

where $P(z|\bb{z})$ and $P(y|z)$ are respectively given by
\BS\begin{align}
    &P(z|\bb{z}) = \tfrac{1}{\sqrt{2\pi \bb{v}^z}}
    e^{-\frac{(z-\bb{z})^2}{2\bb{v}^z}},\\ 
    &P(y|z) =\begin{cases}
     \frac{1}{\sqrt{2\pi \sigma^2}}
    e^{-\frac{(\mathfrak{c}-y)^2}{2\sigma^2}},
    &{\rm if}\hspace{0.1cm} z\geq \mathfrak{c}\\
    \frac{1}{\sqrt{2\pi \sigma^2}}
    e^{-\frac{(z-y)^2}{2\sigma^2}},
    &{\rm if}\hspace{0.1cm} -\mathfrak{c}<z<\mathfrak{c}\\
    \frac{1}{\sqrt{2\pi \sigma^2}}
    e^{-\frac{(\mathfrak{c}+y)^2}{2\sigma^2}},
    &{\rm if}\hspace{0.1cm} z\leq -\mathfrak{c}
    \end{cases}.
\end{align}\ES

Thus,  we get the following \emph{a posteriori} estimation and variance of $z$.
\BS\begin{align}
    &  \hat{z}  = {\rm E}\{z|\bb{z},y\} = \int zP(z|\bb{z},y)dz, \\
    &\hat{v}_z = {\rm Var}\{z|\bb{z},y\} = \int z^2P(z|\bb{z},y)dz - (\hat{z}) ^2.
\end{align}\ES

\subsection{Simulation Results}
\begin{figure}[htb]
\centering
\begin{tabular}{ccc}
\includegraphics[width=0.32\textwidth]{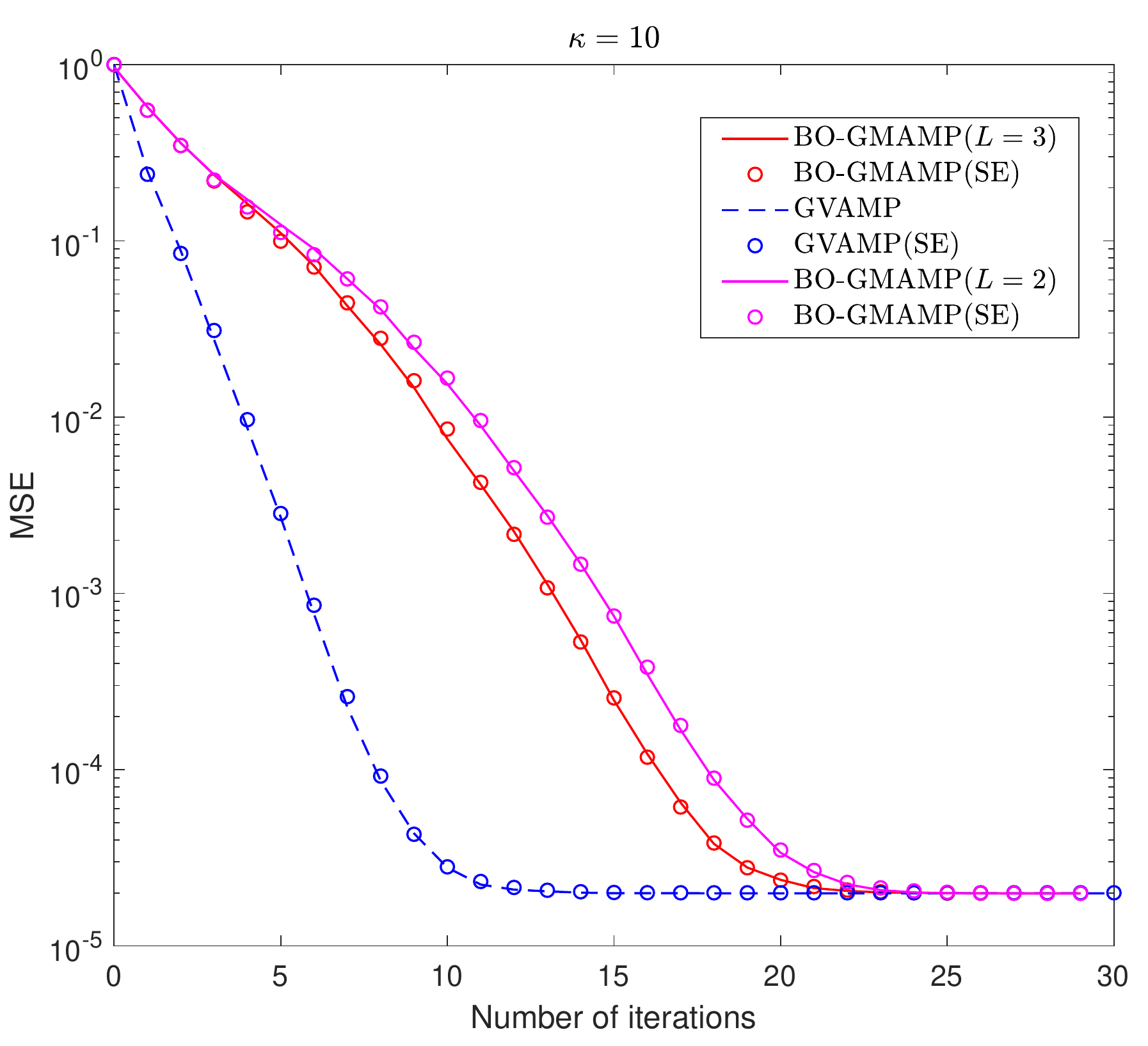}         &
\includegraphics[width=0.32\textwidth]{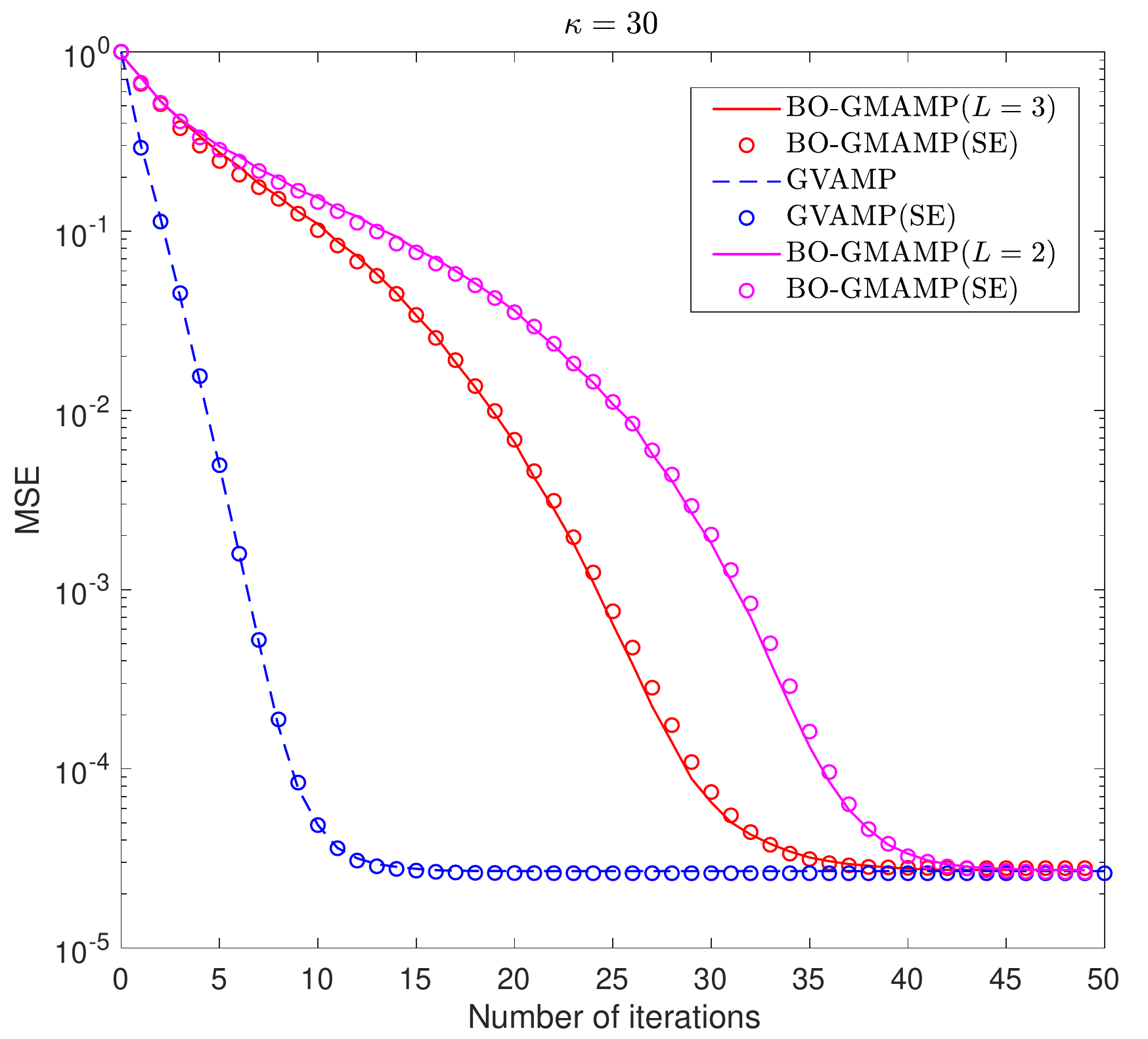}
& \includegraphics[width=0.32\textwidth]{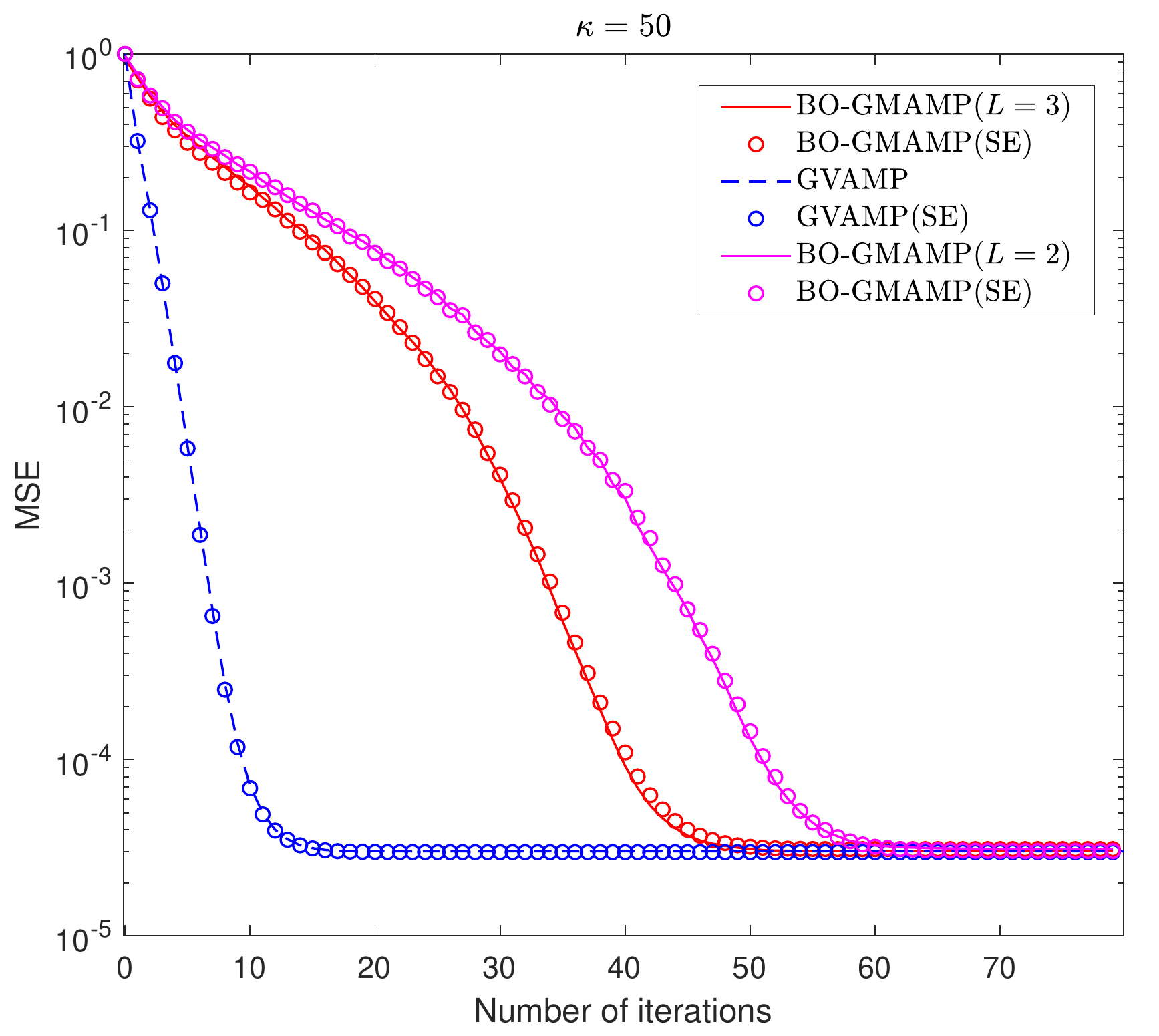} \\ 
\end{tabular}
\caption{\footnotesize MSE versus the number of iterations for GVAMP and BO-GMAMP. $\kappa=\{10,30,50\}$, $\mu=0.1$, $\mathfrak{c}=2$, $N=8192$, $\delta=0.5$, ${\rm SNR}=40{\rm dB}$ and damping length $L=3$ or $2$. "SE" denotes state evolution.} \label{fig:sim(kappa)}
\end{figure}

Next, we present some results about the influence of system parameters on the performance of BO-GMAMP. Fig. \ref{fig:sim(kappa)} shows the effect of condition number $\kappa$ and damping length $L$ on the speed of convergence of BO-GMAMP, where $\kappa\in\{10,30,50\}$ and $L\in\{2,3\}$. As we can see, with the increment of condition number $\kappa$, the speed of convergence slows down due to the instability of system. When $\kappa\in\{10,30,50\}$, BO-GMAMP converges within about 25, 45 and 60 iterations, respectively. Meanwhile, BO-GMAMP converges faster when damping length $L=3$ than when $L=2$.

\begin{figure}[htb]
\centering
\begin{tabular}{ccc}
\includegraphics[width=0.32\textwidth]{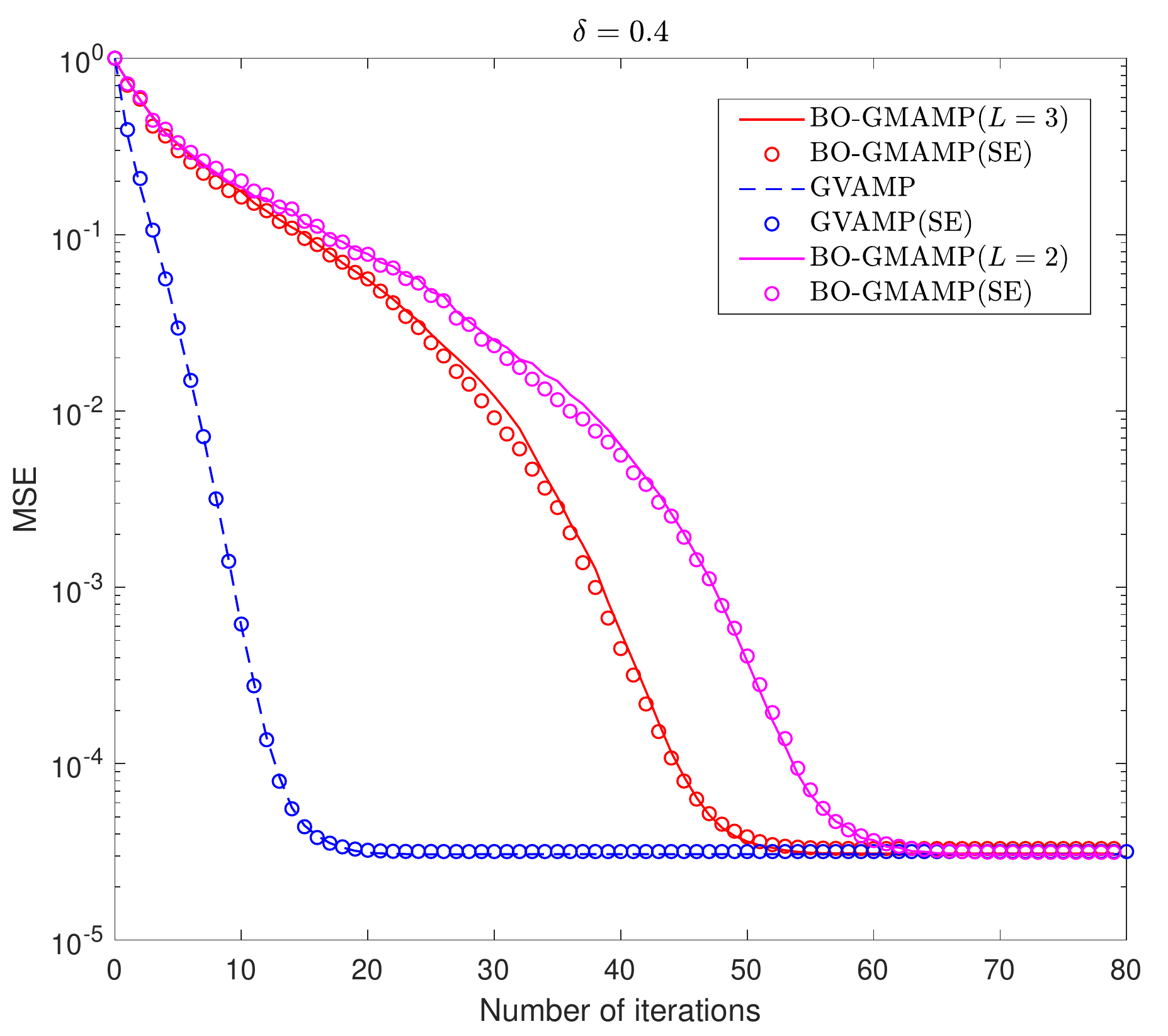}      &
\includegraphics[width=0.32\textwidth]{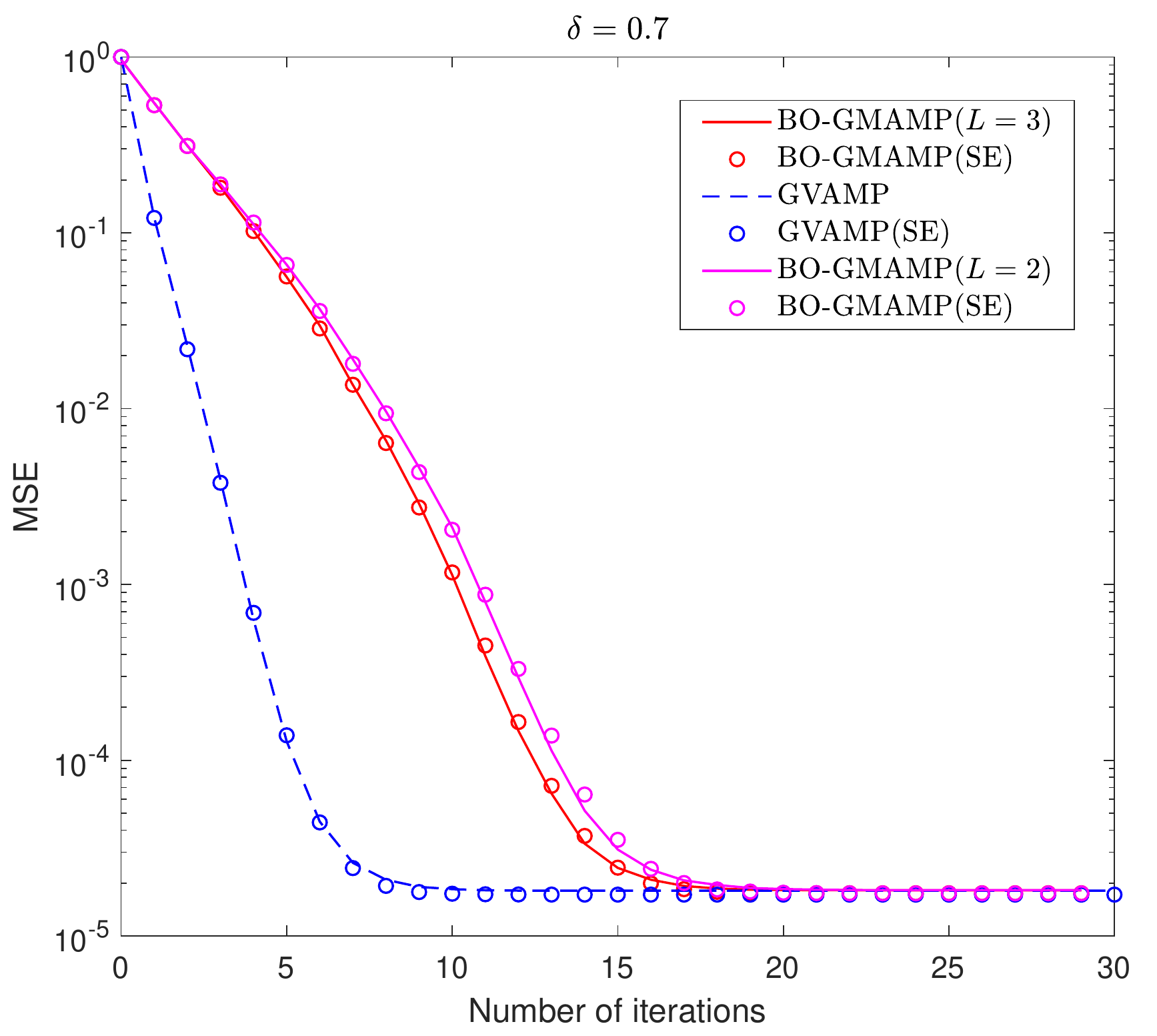}
& \includegraphics[width=0.32\textwidth]{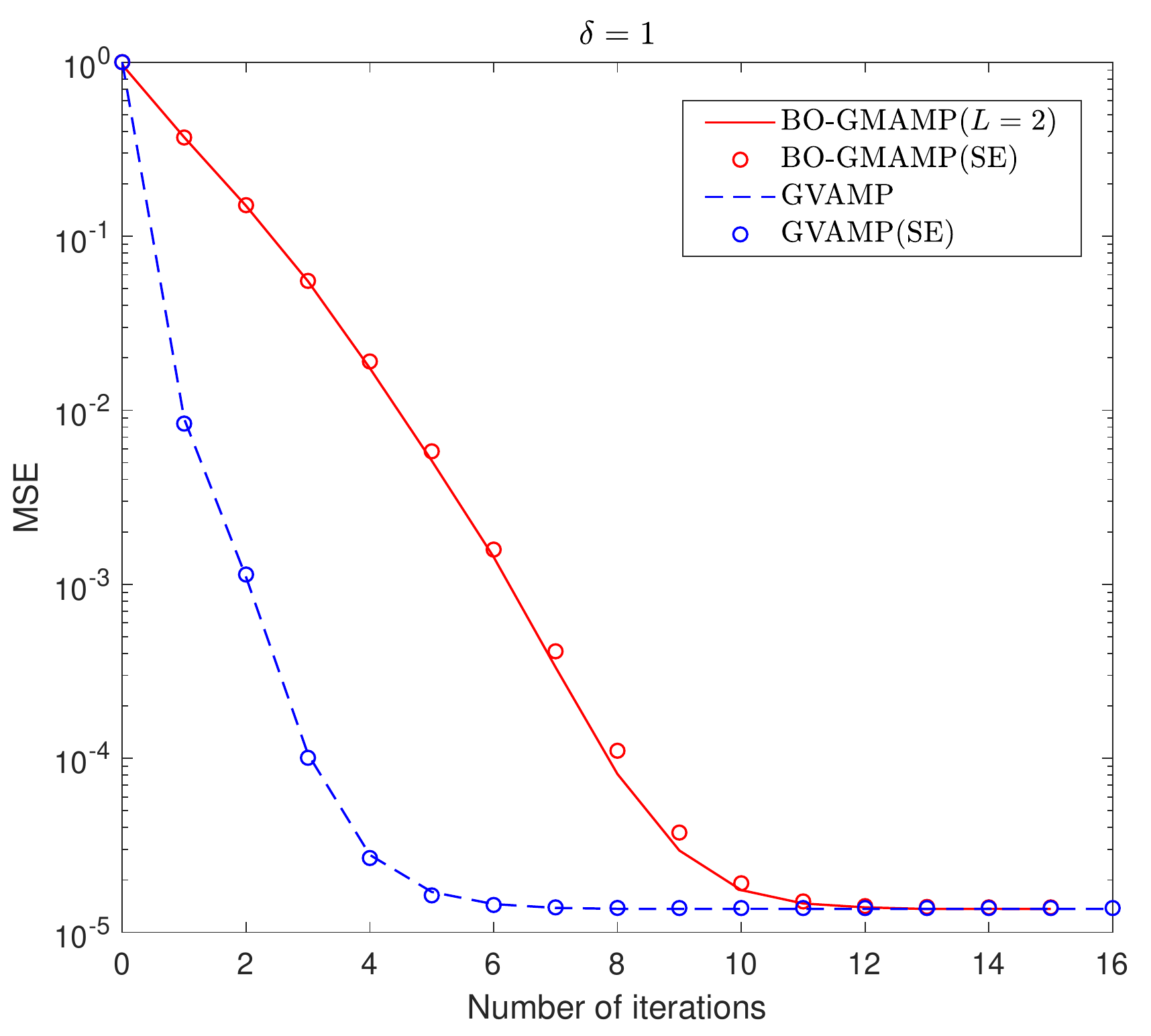}\\ 
\end{tabular}
\caption{\footnotesize MSE versus the number of iterations for GVAMP and BO-GMAMP. $\delta=\{0.4,0.7,1\}$, $\mu=0.1$, $\mathfrak{c}=2$, $N=8192$, $\kappa=20$, ${\rm SNR}=40{\rm dB}$ and damping length $L=3$ or $2$. "SE" denotes state evolution.} \label{fig:sim(delta)}
\end{figure}

Additionally, the influence of compression ratio $\delta$ and damping length $L$ on the performance of BO-GMAMP is shown in Fig. \ref{fig:sim(delta)}, where $\delta\in\{0.4,0.7,1\}$ and $L\in\{2,3\}$. Intuitively, as the compression ratio $\delta$ decreases, BO-GMAMP converges more and more slowly because of the increasing loss of information. When $\delta\in\{0.4,0.7,1\}$, BO-GMAMP converges within about 65, 20 and 12 iterations, respectively. Similarly, the proposed BO-GMAMP converges faster when damping length $L=3$ than when $L=2$. Specially, when $\delta=1$, BO-GMAMP curves at $L=2$ and $L=3$ almost coincide.

\end{document}